\documentclass[12pt]{article}

\newif\ifblind
\newif\ifarxiv

\blindfalse
\arxivtrue

\usepackage{clipboard}
\newclipboard{quotes}
\usepackage{hyperref}
\usepackage{graphicx}
\usepackage{amssymb,amsmath}
\usepackage{bm}
\usepackage{color}
\usepackage[authoryear,semicolon]{natbib}
\usepackage{multirow}
\usepackage{booktabs}
\usepackage{array}
\usepackage[normalem]{ulem}
\usepackage{tikz}
\usepackage{siunitx}
\usepackage{url}
\usepackage{standalone}
\usepackage{xr}

\newcolumntype{x}[1]{>{\centering\arraybackslash\hspace{0pt}}p{#1}}

\newcommand{\zerob} {{\bf 0}}

\newcommand{\thetab} {{\boldsymbol{\theta}}}

\newcommand{\nub} {{\boldsymbol{\nub}}}
\newcommand{\gammab} {{\boldsymbol{\gamma}}}

\newcommand{\etab} {{\boldsymbol{\eta}}}
\newcommand{\Thetab} {{\boldsymbol{\Theta}}}

\newcommand{\intd} {\textrm{d}}
\newcommand{\phib} {\boldsymbol{\phi}}

\newcommand{\Sigmamat} {{\bm \Sigma}}

\newcommand{\Lambdamat} {\mathbf{\Lambda}}

\newcommand{\Gammamat} {{\boldsymbol{\it \Gamma}}}
\renewcommand{\Gammamat} {{\boldsymbol{\Gamma}}}

\newcommand{\Amat} {\textbf{A}}
\newcommand{\Bmat} {\textbf{B}}

\newcommand{\Lmat} {\textbf{L}}
\newcommand{\Qmat} {\textbf{Q}}

\newcommand{\Wmat} {\textbf{W}}

\newcommand{\Mmat} {\textbf{M}}

\newcommand{\Cmat} {\mathbf{C}}

\newcommand{\Kmat} {\textbf{K}}
\newcommand{\Smat} {\textbf{S}}
\newcommand{\Zmat} {\textbf{Z}}

\newcommand{\Imat} {\textbf{I}}

\newcommand{\Hmat} {\textbf{H}}
\newcommand{\Fmat} {\textbf{F}}

\newcommand{\Vmat} {\textbf{V}}
\newcommand{\bvec} {\textbf{b}}

\newcommand{\avec} {\textbf{a}}
\newcommand{\evec} {\textbf{e}}
\newcommand{\hvec} {\textbf{h}}
\newcommand{\mvec} {\textbf{m}}

\newcommand{\wvec} {\textbf{w}}
\newcommand{\vvec} {\textbf{v}}
\newcommand{\svec} {\textbf{s}}
\newcommand{\uvec} {\textbf{u}}
\newcommand{\gvec} {\textbf{g}}
\newcommand{\fvec} {\textbf{f}}

\newcommand{\muvec} {\boldsymbol{\mu}}

\newcommand{\Phimat} {\mathbf{\Phi}}

\newcommand{\taub} {\boldsymbol {\tau}}

\renewcommand{\zerob}{\mathbf{0}}

\newcommand{\Yvec}{\mathbf{Y}}

\newcommand{\Zvec}{\mathbf{Z}}
\newcommand{\epsilonb}{\boldsymbol{\varepsilon}}

\newcommand{\Gau}{\mathrm{Gau}}

\DeclareMathOperator*{\argmax}{arg\,max}
\DeclareMathOperator*{\sig}{sig}
\newcommand{\obs}{\textrm{obs}}
\newcommand{\pred}{\textrm{pred}}
\newcommand{\thr}{\textrm{th}}

\let\originalleft\left
\let\originalright\right
\renewcommand{\left}{\mathopen{}\mathclose\bgroup\originalleft}
\renewcommand{\right}{\aftergroup\egroup\originalright}

\ifblind
\newcommand{\blind}{1}
\else
\newcommand{\blind}{0}
\fi

\newcommand\blfootnote[1]{%
  \begingroup
  \renewcommand\thefootnote{}\footnote{#1}%
  \addtocounter{footnote}{-1}%
  \endgroup
}

\newcommand{\myacktwo}{Andrew Zammit-Mangion's research was supported by an Australian Research Council (ARC) Discovery Early Career Research Award, DE180100203. Maurizio Filippone gratefully acknowledges support from the AXA Research Fund. The Aqua/MODIS Level 1B Calibrated Radiances (500m) data sets used in this work were acquired from the Level-1 and Atmosphere Archive \& Distribution System (LAADS) Distributed Active Archive Center (DAAC), located in the Goddard Space Flight Center in Greenbelt, Maryland (\url{https://ladsweb.nascom.nasa.gov/}). The authors wish to thank Simone Rossi for running the code for the random Fourier features model and Matt Moores, Aidan Sims and S{\'e}bastien Marmin for helpful discussions.}
\newcommand{\myack}{Andrew Zammit-Mangion is Senior Research Fellow in Statistics, School of Mathematics and Applied Statistics, Wollongong, Australia (e-mail: azm@uow.edu.au); Tin Lok James Ng is Lecturer in Statistics, School of Mathematics and Applied Statistics, Wollongong, Australia; Quan Vu is a PhD Student in Statistics, School of Mathematics and Applied Statistics, Wollongong, Australia; and Maurizio Filippone is Associate Professor of Data Science, Department of Data Science, EURECOM, Biot, France.~\myacktwo}

\begin{document}

\def\spacingset#1{\renewcommand{\baselinestretch}%
{#1}\small\normalsize} 

\spacingset{1.5}
\ifarxiv \spacingset{1} % DON'T change the spacing!
\else \fi
%%%%%%%%%%%%%%%%%%%%%%%%%%%%%%%%%%%%%%%%%%%%%%%%%%%%%%%%%%%%%%%%%%%%%%%%%%%%%%

\if0\blind
{
  \date{}
  \title{\bf Deep Compositional Spatial Models}
  \author{Andrew Zammit-Mangion \hspace{.2cm}\\
    School of Math.~and Applied Stat., University of Wollongong, Australia\\
     \\
    Tin Lok James Ng \\
    School of Math.~and Applied Stat., University of Wollongong, Australia\\
      \\
    Quan Vu \\
    School of Math.~and Applied Stat., University of Wollongong, Australia\\
      \\
    Maurizio Filippone\\
    Department of Data Science, EURECOM, France\ifarxiv\else\blfootnote{\myack}\fi}

} \fi

\if1\blind
{
  \bigskip
  \bigskip
  \bigskip
  \begin{center}
    {\LARGE\bf Deep Compositional Spatial Models}
\end{center}
  \medskip
} \fi

\ifblind\else\maketitle\fi

\bigskip
\begin{abstract}
Spatial processes with nonstationary and anisotropic covariance structure are often used when modelling, analysing and predicting complex environmental phenomena. Such processes may often be expressed as ones that have stationary and isotropic covariance structure on a warped spatial domain. However, the warping function is generally difficult to fit and not constrained to be injective, often resulting in `space-folding.' Here, we propose modelling an injective warping function through a composition of multiple elemental injective functions in a deep-learning framework. We consider two cases; first, when these functions are known up to some weights that need to be estimated, and, second,  when the weights in each layer are random. Inspired by recent methodological and technological advances in deep learning and deep Gaussian processes, we employ approximate Bayesian methods to make inference with these models using graphics processing units. Through simulation studies in one and two dimensions we show that the deep compositional spatial models are quick to fit, and are able to provide better predictions and uncertainty quantification than other deep stochastic models of similar complexity. We also show their remarkable capacity to model  nonstationary, anisotropic spatial data using radiances from the MODIS instrument aboard the Aqua satellite.
\end{abstract}
%Maximum 200 words

\noindent%
\vfill

\newpage

\section{Introduction}\label{sec:Intro}

Modelling nonstationary and anisotropic covariances in spatial processes is pivotal to obtaining reliable predictions and uncertainty quantification when analysing complex environmental phenomena. Several modelling classes have been established to model nonstationary covariances, ranging from convolution models to stochastic partial differential equation models, to mention a few \cite[e.g.,][]{Higdon_1999,Paciorek_2006,Fuglstad_2015}. %Most of these models  assume that the model used to induce nonstationarity is itself rather simple and can be described using what we call a one-layer, `shallow,' model.
Among the most well-known of these is the `warping approach' of \citet{Sampson_1992}. Consider a spatial process $Y(\cdot)$ on some spatial domain $G$, and assume that $Var(Y(\svec)) < \infty, \svec \in G$. In essence, Sampson and Guttorp proposed warping $G$ under a mapping $\fvec = \fvec_1 : G \rightarrow D_1$ such that the process has stationary and isotropic covariance structure on $D_1$. Their approach involves finding a multivariate function $\fvec_1(\cdot)$ such that the covariance function of $Y(\cdot)$ on the warped space, $C_{D_1}(\fvec_1(\cdot),\fvec_1(\cdot))$, is a univariate, positive-definite, function of distance, $C^o_{D_1}(h)$ say, where $h = \| \uvec_i - \uvec_j \|,~ \uvec_i,\uvec_j \in D_1$. In their case, $\fvec_1(\cdot)$ was determined using multi-dimensional scaling (MDS) and thin-plate splines. %As in many other works, the problem of non-injectivity in $\fvec_1$ (that results in space-folding) was not fully addressed. 

%Sampson and Guttorp's approach paved the path for a suite of spatial modelling tools, largely motivated by the intuitive results of \citet{Perrin_2000} who show that if $\fvec_1(\cdot)$ is bijective, and if both it and its inverse are differentiable, that then (under mild conditions) there is a one-to-one correspondence between  $\{C_{D_1}^o(\cdot), \fvec_1(\cdot)\}$ and the nonstationary covariance function on $G$. In this case $C_G(\cdot,\cdot)$ is said to be stationary isotropic reducible  \citep{Perrin_2000}.
\citet{Sampson_2001} provide a detailed review of  deformation methods up to the year 2001, and discuss two methods that are of particular relevance to this work. The first of these is the approach of \citet{Smith_1996}, in which the map $\fvec_1(\cdot)$ is modelled using a sum of radial basis functions derived from the thin-plate spline, and a likelihood-based approach is used for estimating the deformation. The second is that of \citet{Perrin_1999} who used compositions of radial basis function mappings to model $\fvec_1(\cdot)$. Bivariate Gaussian processes were first used to model the deformation in a Bayesian setting by  \citet{Schmidt_2003}, while Gaussian process deformations were also used in a state space framework by \citet{Morales_2013}. 

An interesting connection can be made between these warping models and feedforward neural nets, which have garnered much interest in the past decade \cite[e.g.,][]{LeCun_2015} and that express the latent function $\fvec(\cdot)$ through the composition of $n$  functions $\fvec_n\,\circ\,\fvec_{n-1}\,\circ\,\cdots\,\circ\,\fvec_1(\cdot)$. Indeed, the model of \cite{Smith_1996} has $n = 1$ hidden layer, while that of \citet{Perrin_1999} has $n > 1$ hidden layers. The model developed by \citet{Schmidt_2003} is a Gaussian process with one hidden layer, a special case of the general deep Gaussian process devised by \citet{Damianou_2013} and later extended by several authors including \citet{Hensman_2014} and \citet{Salimbeni_2017}. 

This connection begs the question as to whether spatial models can be made more representational of the data-generation process when using a warping function $\fvec(\cdot)$ which has $n$ hidden layers, where $n > 1$. This connection is all the more interesting given the recent interest in understanding the expressive power of deep neural networks \citep{Bengio_2011, Eldan_2016, Safran_2017}. In particular, it has been shown that deep networks are exponentially more efficient in function approximation than shallow networks \citep{Liang_2017, Arora_2018}. In Section 2 we review several models from both the statistical and machine-learning literature that can be used to construct nonstationry anisotropic covariance structure with deformations.% in conjunction with a comprehensive review that synergises the plethora of methods now available for modelling nonstationarity via deformation. This review is provided in Section 2.

%As convincingly demonstrated by \citet{Duvenaud_2014} and \citet{Dunlop_2017}, one issue that is not addressed by general deep structures is bijectivity of $\fvec$.
Injectivity of $\fvec(\cdot)$ has been an ongoing cause for concern for spatial statisticians: \citet{Sampson_1992} state that ``a mapping that folds usually results in a model that overfits the sample,'' while \citet{Schmidt_2003} assert that the folding is ``undesirable and/or implausible for environmental data.'' Due to the nature of the problems generally considered in their domain, for the machine-learning community the lack of a one-to-one mapping is largely considered a non-issue \citep[insofar as too much warping can lead to map degeneracy; see][]{Duvenaud_2014, Dunlop_2017}. On the other hand, various approaches have been used by spatial statisticians to enforce one-to-one correspondence, involving both soft \citep[e.g.,][]{Meiring_1997, Fouedjio_2015} and hard \citep[e.g.,][]{Iovleff_2004} constraints. While the latter generally lead to difficult optimisation problems, the former do not provide the injectivity guarantees we seek, and the cautious modeller will fit deformations that are generally too smooth. These difficulties have rendered other methods that model nonstationarity directly through the covariance function in a way that validity is guaranteed \citep[e.g.,][]{Paciorek_2006, Fuglstad_2015}, more attractive. Indeed, such models can capture stronger covariance nonstationarity than what is possible using simple mappings, such as the M{\"o}bius transformation that we consider in Section 3. However, in this work we show that compositions of multiple maps can yield the desired flexibility we seek; \Copy{Parameters}{further, at the cost of only a moderate amount of additional parameters when each individual map has a simple form.}

The main novel contribution of our work is the construction of a flexible \emph{deep compositional spatial model} in Section 3, which is built on the premise that a map constructed by composition of multiple injective maps is itself injective. We do not present the first instance of such a model: \citet{Perrin_1999} were probably the first to use multiple compositions of injective radial basis function mappings to model $\fvec(\cdot)$ in a spatial context. The flexible deep compositional spatial model we present extends their's on various fronts. First, inspired by multi-resolutional spatial modelling tools \citep[e.g.,][]{Nychka_2015, Zammit_2018}, we use multi-resolutional warpings that capture deformations at various scales. Second, in addition to the functions in \citet{Perrin_1999} for the multiresolution basis, we also consider warping layers with axial warping (scaling) units as well as large-scale M{\"o}bius transformations \citep[e.g.,][Section 11.3]{Dubrovin_1992}. %% The combination of a M{\"o}bius transformation and a nonlinear scaling is a powerful deformation tool in its own right:  As an example, in Figure~\ref{fig:Meobius_example}, left two panels, we illustrate the resulting warping that can be obtained when first carrying out a M{\"o}bius transformation, and then a simple axial warping of one coordinate axis. In the right panel we show a realisation (on $G$) of a Gaussian process that has a stationary, isotropic, exponential covariance function on the warped domain. The resulting compositional map, which is bijective, can be used to describe relatively complex spatially-varying scalings and orientations. Further, this compositional map is relatively simple, and consists of only 11 parameters in total (8 for the M{\"o}bius transformation and 3 for the axial warping).
Finally, our model is seated in a likelihood framework and is designed to take full advantage of the computational tools designed for deep neural networks, such as  stochastic gradient descent methods. This, in combination with the use of basis-function field representations for $Y(\cdot)$ \citep{Cressie_2008},  allows us to train relatively complex models in the presence of large data sets with relative ease.

In Section 4 we compare various approaches to doing spatial warping, including several types of deep Gaussian processes and our deep compositional spatial model, in a simulation experiment in one dimension. We also show the utility of our proposed model in two dimensions and on radiances using data from the MODIS instrument aboard the Aqua satellite. Section 5 concludes the work, and Appendix~A contains technical details on inference and prediction.

%% \begin{figure}
%%   \includegraphics[width=\textwidth]{2dMoebiusexample.png}
%%   \caption{Warping when constructing $\fvec$ through a composition of a M{\"o}bius transformation $\fvec_1$ and an axial warping $\fvec_2$. Left panel: A spatial grid pattern on $G$. Centre panel: The spatial grid depicted in the left panel plotted in $D_2$ after warping through $\fvec \equiv \fvec_2\,\circ\,\fvec_1$. Right panel: A single realisation from a Gaussian process with exponential, stationary, isotropic, covariance function simulated on $D_2$ and plotted on $G$. \label{fig:Meobius_example}}
%%   \end{figure}

\section{Background and related work}\label{sec:background}

In this section we review deep models that appear in both the statistical and the machine-learning literature and that are relevant to spatial deformation methods. In order to facilitate the review, we have classified the models into two groups: input-warped Gaussian processes and deep stochastic processes. The primary distinction between these two groups is that while models in the former class treat the warping as deterministic (resulting in covariances that can be evaluated deterministically for any two spatial locations), the latter treats the warping as a random process in itself (resulting in covariances that are themselves random). Throughout this section we only consider processes that are formed through function \emph{composition}: As shown by \citet{Dunlop_2017}, there are other ways to construct deep processes \citep[e.g., by extending the work of ][to multiple layers]{Paciorek_2006} but we do not consider those models here. %We also focus on warpings that are low-rank; we give a brief review of models with nonparametric warpings at the end of Sections \ref{sec:IWGP} and \ref{sec:DGP}.%We also focus on the model, and not the techniques used to fit the model; these can be found in the cited referenes.

\subsection{Input-warped Gaussian processes}\label{sec:IWGP}

Consider a Gaussian process $Y(\svec),\, \svec \in G,$ with covariance function $C_G(\cdot,\cdot)$. Input-warped Gaussian processes (IWGPs) are built on the premise that there exists a covariance function on some warped domain $D_n$, $C_{D_n}(\cdot, \cdot)$, that is simple enough to be expressed as a standard stationary, isotropic, covariance function $C_{D_n}^o(\cdot)$. The problem then reduces to finding the warping function $\fvec(\cdot)$. % and some parameters that define $C_{D_n}^o(\cdot)$ (although occasionally $C^o_{D_n}$ is also modelled nonparametrically).
In IWGP models, $\fvec(\cdot)$ is a deterministic (yet unknown) mapping constructed through composition. Specifically, $\fvec(\cdot) \equiv \fvec_n\,\circ\,\fvec_{n-1}\,\circ\,\cdots\,\circ\,\fvec_1(\cdot)$ and
$$
C_G(\svec, \uvec) \equiv C_{D_n}^o(\|(\fvec_{n}\,\circ\,\fvec_{n-1}\,\circ\,\cdots\,\circ\,\fvec_1(\svec)) - (\fvec_{n}\,\circ\,\fvec_{n-1}\,\circ\,\cdots\,\circ\,\fvec_1(\uvec))\|),\quad \svec,\uvec \in G.
$$
In the machine-learning literature, $C_{D_n}^o(\cdot)$ is often referred to as a deep kernel \citep[e.g.,][]{Wilson_2016} and the resulting process as a manifold GP \citep{Calandra_2016}. %We now further categorise the models seated in this framework according to whether $\fvec$ is modelled parametrically or otherwise.

The IWGPs most commonly used are structured feedforward neural nets. That is, for some $r_i$ basis functions at the $i$th layer, $\phib_i(\cdot\,; \Thetab_i) \equiv (\phi_{ij}(\cdot\,; \Thetab_i): j = 1,\dots,r_i)'$, and basis-function coefficients (or weights) at the $i$th layer $\Wmat_i \equiv (w_{ikj}: k = 1,\dots,d_i; j = 1,\dots,r_i)$ that map the input to a $d_i$-dimensional output,
\begin{align*}
%  \fvec_{1}(\svec) =& \Wmat_{1}(\Thetab_1)\phib_{1}(\svec; \Thetab_1), \quad \svec \in G,\label{eq:f1}\\
  % \fvec_{i}(\svec) =& \Wmat_{i}(\Thetab_i)\phib_{i}(\fvec_{i-1}(\svec); \Thetab_i),\quad \svec \in G; \quad i = 2,\dots,n,\label{eq:fi}
  \fvec_{i}(\svec) =& \Wmat_{i}\phib_{i}(\svec; \Thetab_i),\quad \svec \in D_{i-1}; \quad i = 1,\dots,n,\label{eq:fi}
\end{align*}
\noindent where $\fvec_i : D_{i-1}  \rightarrow D_i, i = 1,\dots, n, D_i \subset \mathbb{R}^{d_{i}}, D_0 \equiv G$, and $\Thetab_i$ are parameters appearing inside the $i$th layer that can be either fixed or estimated. %In most applications the weights $\Wmat_i$ are not parameterised in terms of the $\Thetab_i$ and these parameters only appear inside the basis functions, but as we show in Section \ref{sec:parDGPs} this is not always the case. %Throughout this article we omit the dependence on $\Thetab_i$ when we imply that the weights or basis functions do not contain any parameters that need to be estimated.
The basis functions used and the constraints imposed on the weights $\{\Wmat_i\}$ generally dictate the type of IWGP.

%% The motivation behind IWGPs for spatial deformation stems from the following simple observation derived from that of \citet{Cho_2009}. Let $K(\svec, \uvec)$ be a positive semi-definite bivariate function on $\svec,\uvec \in G$, such that $K(\svec,\svec) = K(\uvec, \uvec) = 1$ for all $\svec,\uvec$. Then, by Mercer's theorem, $K(\svec,\uvec) \approx \sum_{i=1}^n\psi_i(\svec)'\psi_i(\uvec)$ where $n\rightarrow\infty$ and the functions $\psib(\cdot) \equiv (\psi_i(\cdot): i = 1,\dots,n)'$ are orthogonal. Assume now that $\fvec_1(\cdot) \equiv \psib(\cdot)$; then it is straightforward to see that the squared Euclidean distance between two spatial locations, after warping, is given by
%% $$
%% \|\fvec_1(\svec) - \fvec_1(\uvec)\|^2 \propto 1 - K(\svec, \uvec);\quad \svec,\uvec \in G,
%% $$
%% which one can associate with the semivariogram, and thus a highly nonlinear  (and in this case bounded) function of displacement.  The 1-layer net is generally not constructed from orthogonal activation functions, but the effect is largely the same, that of changing the Euclidean distance metric on the original space to one that is sufficiently complex, through the learning of $\psib(\cdot)$ and thus, indirectly, the kernel $K(\cdot,\cdot)$. More layers effectively result in different kernels being applied recursively to the distance metric used to measure the distance between inputs in $G$.
%We now review some parametric IWGP architectures that could bear relevance to the deep spatial model.

The simplest $n=1$ low-rank IWGP is the single index linear model \citep{Choi_2011} where the input space is collapsed onto one dimension using a linear transformation. Specifically, $\phib_1(\svec) = \svec$, and therefore $f_1(\svec) = (w_{111}, w_{112},\dots)\svec$.  Multiple index linear models collapse the input space into one of a smaller dimension that is greater than one, and thus $\fvec_1(\svec) = \Wmat_1\svec$, where the number of rows in $\Wmat_1$ is less than the dimension of $\svec$. %Because these models reduce the dimension of the input space, they are also referred to as autoencoders.
\citet{Marmin_2018} augment the multiple index linear model by adding a second layer, with $\phib_2(\svec; \Thetab_2)$ set to Beta cumulative distribution functions (CDFs), which were also used for axial warping by \citet{Snoek_2014}. The (nonlinear) axial warping in the second layer is one-to-one, but the first, dimension reduction, map is generally not. Therefore, the warpings of the index linear models are not injective in general. %; as illustrated in Figure~\ref{fig:Meobius_example} the use of a M{\"o}bius transformation instead of a rigid transformation yields a class of models that is very flexible, while retaining bijectivity. 
Also, since spatial problems are low-dimensional problems, there is not much to be gained by using the encoding facility of index models. %and all subsequent methods we describe do not encode, and instead sometimes de-encode through dimension expansion, in the warping layers.

%The architecture proposed by \citet{Smith_1996} is motivated by its ability to fit thin-plate splines to the data.
\citet{Smith_1996} considered a spatial domain indexed by $\svec \in \mathbb{R}^2$, set $n = 1$, % $\phi_{11}(\svec) = s_1, \phi_{12}(\svec) =  s_2$ and $\phi_{1j}(\svec; \Thetab_1) = r(\svec; \gammab_{1j})^2\log{r(\svec; \gammab_{1j})}, j > 2,$ where $r(\svec; \gammab_{1j}) = \|\svec - \gammab_{1j} \|$ and $\gammab_{1j}$ are the centroids of the radial basis functions $\phi_{1j}, j > 2$.
and constructed $\phib_1(\cdot)$ using basis functions that reconstruct thin-plate splines. %Constraints were placed on $\Wmat_1$ to ensure identifiability.
As with the original MDS/thin-plate spline approach of \citet{Sampson_1992}, Smith's mapping is not injective in general. % and is structurally identical to traditional neural nets constructed using smooth basis-function activations, such as the $\tanh$ function or the sigmoid function. Both Mat{\'e}rn and a class of Bessel-function mixtures were considered for modelling $C^o_{D_1}(\cdot)$.
\citet{Perrin_1999} let $n > 1$, and constructed $\fvec(\cdot)$ from the  composition of radial-basis-function (RBF) deformations
%. The squared-exponential RBF at the $i$th layer takes the form
%\begin{equation}\label{eq:Perrin_original}
%\fvec_i(\svec) = \svec + w_i(\svec - \gammab_i)\exp(-a_i\|\svec - \gammab_i\|^2);\quad \svec \in D_{i-1} \subset \mathbb{R}^2,
%\end{equation}
%where $\gammab_i = (\gamma_{i1}, \gamma_{i2})'$  is the RBF centroid, $a_{i}$ is a scaling factor, $w_i$ controls the intensity of domain expansion/shinkage, and $\Thetab_{i} \equiv (\gammab'_{i}, a_{i})'$ is the parameter vector of the radial basis functions at the $i$th layer. In Section~\ref{sec:units} we show how this RBF can be re-expressed as a weighted sum of basis functions,
where the weights are constrained to ensure injectivity. %A squared exponential covariance function was used to model $C^o_{D_1}(\cdot)$.
\citet{Perrin_1999}'s work is the only one that we are aware of that uses parametric injective warpings other than axial warpings in a spatial modelling application. 

%The above mentioned IWGPs contain several parameters that are generally estimated in a maximum likelihood setting. In Section \ref{sec:DGP} we describe relatively newer models where the weights themselves are random, so that the warpings are stochastic processes in themselves. In this latter setting, approximations are generally needed to make inference on the latent warpings.

%\subsubsection{Nonparametric IWGPs}

%In Section \ref{sec:parIWGP} we focused on low-rank warping approaches, however there is also a sizeable portion of literature that focuses on
Sometimes nonparametric warping functions are used for $\fvec(\cdot)$. \citet{Sampson_1992} set $n = 1$, modelled $\fvec_1(\cdot) \in \mathbb{R}^2$ nonparametrically using thin-plate splines, and constructed  $C^o_{D_1}(\cdot)$  using a class of Gaussian probability mixtures. This same shallow kernel was used by \citet{Monestiez_1993}, \citet{Meiring_1997}, \citet{Zidek_2000}, \citet{Damian_2001}, and more recently by \citet{Kleiber_2016} for simulating nonstationary covariances. \citet{Bornn_2012} also used thin-plate splines but additionally considered dimension expansion (i.e., they let $d_1 > 2$) in a way that guarantees injectivity. Other related works include those of \citet{Iovleff_2004}, \citet{Anderes_2008}, \citet[][Section 3.10.3]{Gibbs_1998}, and \citet{Xiong_2007}.%We briefly discuss dimension-expansion approaches in Section \ref{sec:Conclusion}.

%% \citet{Iovleff_2004} set $n = 1$ and found $\fvec_1: \mathbb{R}^2 \longrightarrow \mathbb{R}^2$ at the observation locations using a constrained optimisation algorithm that enforces bijectivity and let $C^o_{D_1}(\cdot)$ be the power exponential.  \citet{Anderes_2008} deduced $\fvec_1 :  \mathbb{R}^2 \longrightarrow \mathbb{R}^2$ from local deformations estimated at a set of locations and then interpolated them over $G$. \citet[][Section 3.10.3]{Gibbs_1998} modelled $\fvec_1: \mathbb{R}^2 \longrightarrow \mathbb{R}^2$ by taking a path integral between any fixed point $\svec_0$ and $\svec$ of an arbitrary positive function (which is estimated for each component of $\fvec_1$). %Although claimed to be one-to-one, we note that it is easy to find many-to-one mappings with this approach: take the positive function to be identical for each component of $\fvec_1$ and a symmetric function about $\svec_0$. Then all points a distance $\|\svec_0 - \svec\|$ apart will map to the same point in the warped space.
%% \citet{Xiong_2007} simplifies Gibbs' approach by warping each axis separately; this ensures bijectivity but axial warpings have limited flexibility when used on their own.

\subsection{Deep stochastic processes}\label{sec:DGP}

In a deep stochastic process (DSP), the warping function is itself a stochastic process. By far the most common DSP is the deep Gaussian process (DGP) where the output of each layer is modelled as a GP. Note that DSPs are, in general, non-Gaussian processes (i.e., even when the DSP is a DGP). %% Despite the name and unlike the IWGP, the final layer in a DGP is not a Gaussian process over the inputs in $G$.
Both the finite-dimensional (low-rank) and the full-rank process representations of DSPs are of particular relevance to spatial deformation methods.

%\subsubsection{Low-rank DSPs}\label{sec:parDGPs}

In a low-rank DGP one equips each row in $\Wmat_i$ (i.e., $\wvec_i^{(k)}, k = 1,\dots,d_i$) with a multivariate Gaussian distribution with zero mean and covariance matrix $\Sigmamat_i^{(kk)}$, where $d_i$ is the output dimension of the $i$th layer. Then $f_{ik}(\cdot)$ is a zero mean Gaussian process with covariance function $C_{D_{i-1}}(\svec,\uvec) = \phib_i(\svec)'\Sigmamat_i^{(kk)}\phib_i(\uvec),\,\svec,\uvec\in D_{i-1}$. %More generally, if one equips $\textrm{vec}(\Wmat_i) \equiv (\wvec_i^{(k)'}: k = 1,\dots,d_i)'$ with a multivariate Gaussian distribution with mean zero, and valid block covariance matrix $\Sigmamat_i \equiv (\Sigmamat_i^{(kk')}: k,k' = 1,\dots,d_i)$, then $\fvec_i$ is a multivariate Gaussian process with cross-covariance function matrix $\Cmat_{D_{i-1}}(\svec,\uvec) = (\phib_i(\svec)'\Sigmamat_i^{(kk')}\phib_i(\uvec): k,k' = 1,\dots,d_i)$. Instead of leaving $\Sigmamat_i$ general positive-definite, one often gives it the form $\Sigmamat_i \equiv \Rmat_i \otimes \Sigmamat_i^{(11)}$ where $\otimes$ is the Kronecker product, and $\Rmat_i$ and $\Sigmamat_i^{(11)}$ are $d_i \times d_i$ and $r_i \times r_i$ positive-definite matrices, respectively. This modelling choice is usually made to reduce the number of parameters that need to be estimated and to take advantage of the computational benefits associated with the Kronecker product. %This would be familiar to many as a low-rank representation of a multivariate (infinite-dimensional) Gaussian process.
\citet{Cutajar_2016} consider a straightforward parameteric DGP where $Var(\wvec_i^{(k)}) = \Imat, k = 1,\dots,d_i$, and where $\phib_i(\cdot\,; \Thetab_i), i = 1,\dots,n$, is a Fourier basis. %Since the $\Thetab_i$ are also treated as random, the basis functions are referred to as random Fourier features, previously used for Gaussian process modelling by \citet{Quia_2010} and further investigated recently by \citet{Hensman_2017}.
The DGP of \citet{Damianou_2013} is based on compositions of sparse Gaussian processes (sparse GPs). A type of sparse GP, known as the subset-of-regressors approximation  \citep{Quinonero_2005} or the predictive process \citep{Banerjee_2008b}, can be written as a weighted sum of basis functions, and thus their process can be viewed as a low-rank variant; see also \citet{Hensman_2014, Dai_2016, Bui_2016}. %A class of sparse GPs, known as subset of regressors approximations, are known in the spatial statistics literature as predictive processes \citep{Banerjee_2008b}.% with the important distinction that the knot locations (or inducing points) used to construct the low-rank representation are generally assumed to be unknown in the former \citep[although knot prediction was introduced later in the latter, e.g., ][]{Katzfuss_2013}.
%As is well known \citep[e.g.,][]{Sang_2012} the sparse GP/predictive process is a finite-dimensional Nystr{\"o}m-approximation representation of a Gaussian process,
%% Sparse GPs are low-dimensional representations of GPs, and thus \citet{Damianou_2013}'s model, as well as all derivative works \citep[e.g.,][]{Hensman_2014, Dai_2016, Bui_2016} can be classed as low-rank DGPs. A type of sparse GP, known as the subset-of-regressors approximation  \citep{Quinonero_2005} or the predictive process \citep{Banerjee_2008b}, can be written as a weighted sum of basis functions. Specifically, let $k = 1$ and let the `parent' GP at the $i$th layer have covariance function $C_{D_{i-1}}(\cdot,\cdot)$. Further, consider a set of $m_i$ inducing points $\bar\Smat_{i-1} \equiv (\bar\svec_{i-1,1},\dots,\bar\svec_{i-1,m_i})$. Then,
%% $$
%% f_{i1}(\svec) = \wvec_i^{(1)}(\bar\Smat_{i-1},\Thetab_i)'\phib_i(\svec;\bar\Smat_{i-1}, \Thetab_i),
%% $$ 
%% where 
%% \begin{align*}
%% \phib_i(\svec; \bar\Smat_{i-1}, \Thetab_i) &\equiv (C_{D_{i-1}}(\svec, \bar\svec_{i-1,1}), \dots, C_{D_{i-1}}(\svec, \bar\svec_{i-1,m_i}))', \\
%% \wvec_i^{(1)}(\bar\Smat_{i-1}; \Thetab_i) &\sim \Gau(\zerob, C_{D_{i-1}}(\bar\Smat_{i-1},\bar\Smat_{i-1})^{-1}),
%% \end{align*}
%% where $C_{D_{i-1}}(\bar\Smat_{i-1},\bar\Smat_{i-1}) \equiv (C_{D_{i-1}}(\bar\svec_{i-1,k}, \bar\svec_{i-1,l}): k,l = 1,\dots, m_i)$.
In DGPs based on sparse GPs, inducing-point locations %$\bar\Smat_{i-1}$
as well as covariance-function parameters %$\Thetab_i$
generally need to be estimated. %Since these affect both the basis functions \emph{and} the weights,
Our experience (using the variational Bayes approximate inference scheme of \citet{Damianou_2013}) is that DGPs constructed by nesting sparse GPs are difficult to fit, even in the low-dimensional settings we consider.

%\subsubsection{Full-rank DSPs}

%% As with low-rank DSPs, the full-rank DGP is the only DSP that, to the best of our knowledge, has been developed and used to date, and its construction can be motivated from its low-rank sibling as follows: Let  $f_{ik}(\cdot)$ be Gaussian process with covariance function $C_{D_{i-1}}(\cdot,\cdot)$ corresponding to the $k$th output of the $i$th layer in a DGP. Then, from the Karhunen--Lo{\`e}ve theorem \citep[e.g.,][Chapter 4]{Cressie_2011}, we know that $f_{ik}(\cdot)$ admits the representation $f_{ik}(\cdot) = \sum_{j = 1}^\infty \omega_{ijk}\sqrt{\lambda_{ijk}}e_{ij}(\cdot)$, where $\omega_{ijk} \sim \Gau(0,1), j = 1,2,\dots$ and the $\{e_{ij}(\cdot)\}$ are orthogonal. We can therefore entertain the idea of representing $f_{ik}$ through an infinitely wide network by letting the basis functions $\phib_i(\cdot) = \evec_i(\cdot)$ and by letting  $w_{ijk} \sim \Gau(0,\lambda_{ijk})$. Alternatively, if we do not have knowledge of $C_{D_{i-1}}(\cdot,\cdot)$, we can use an infinitely wide network of basis functions $\phib_i(\cdot\,; \Thetab_i)$ where the parameters $\Thetab_i$ are random. As shown by \citet{Neal_1996}, Chapter 2, due to the Central Limit Theorem, under mild conditions this layer converges to a Gaussian process with covariance function $E(\phib(\cdot; \Thetab_i), \phib(\cdot; \Thetab_i))$ irrespective of whether our prior on $\Wmat_i$ is Gaussian or not.

 % Instead of resorting to infinitely-wide (in practice, very wide) layers, one may also ditch the notion of a neural net entirely and
In a full-rank DGP, each layer is defined to be a multivariate Gaussian process, that is, $  \fvec_{i}(\cdot) \sim MVGP(\muvec_i(\cdot), \Cmat_{D_{i-1}}(\cdot, \cdot); \Thetab_i),~1 \le i \le n,$ where $MVGP(\muvec_i(\cdot), \Cmat_{D_{i-1}}(\cdot, \cdot); \Thetab_i)$ is a multivariate Gaussian process with mean vector function $\muvec_i(\cdot)$ and cross-covariance matrix function $\Cmat_{D_{i-1}}(\cdot,\cdot)$. %Similar to the low-rank DGP, it is common to let $\Cmat_{D_{i-1}}(\cdot, \cdot) \equiv \Rmat_i\otimes C_{D_{i-1}}(\cdot, \cdot)$; such models are said to be \emph{separable}. There are connections between full-rank DGPs and low-rank DGPs where the number of hidden layers in the latter tends to infinity \citep[e.g.,][]{Neal_1996}.
Full-rank Gaussian processes are computationally burdensome to work with since estimation and inference algorithms with them will necessitate the decomposition of matrices of size $N \times N$, where $N$ is the number of data points. Yet, for moderately-sized problems, they are still computationally tractable, and were used in a spatial deformation context by \citet{Schmidt_2003} and \citet{Schmidt_2011}. %Since we envision deep compositional spatial models to be useful in large data settings, from now on we restrict our attention to low-rank IWGPs and DSPs.

Since the hidden functions in DGPs are multivariate Gaussian processes, they are not injective in general (in the sense that sample paths from the hidden functions will fold). Injective maps require the hidden layers to be non-Gaussian processes, and thus the model we require is a general deep stochastic process (DSP) that is non-Gaussian. Non-Gaussian DSPs can be highly complex processes, and to the best of our knowledge they have yet to be exploited for regression or classification tasks. In this article we present a deep compositional spatial process that is a non-Gaussian DSP. Specifically, it has the same structure as a low-rank DGP, but lets the weights in the hidden layers $(\Wmat_i, i = 1,\dots,n),$ be trans-Gaussian in order to ensure that sample paths at each of the hidden layers are injective maps.

\section{Deep Compositional Spatial Models}\label{sec:DCSMs}

In this section we introduce a class of flexible deep compositional spatial models where the geographic domain is warped through a composition of differentiable injective elemental warpings that we term \emph{units}. In Section \ref{sec:model} we give a general overview of the model; in Section \ref{sec:units} we describe the units; in Section \ref{sec:spatproc} we describe the spatial process at the top layer; and in Section~\ref{sec:inference-summary} we summarise inference and prediction strategies for when the model is a spatial IWGP (SIWGP) and a spatial DSP (SDSP), respectively.  Technical details are given in Appendix~\ref{sec:Inference}.

In this section and in Appendix~\ref{sec:Inference} we use the following notational convention. Consider an arbitrary vector $\bvec \in \mathbb{R}^{d_1}$ in $d_1$-dimensional space, and let $\Bmat \equiv (\bvec_1,\dots,\bvec_N)$ be a collection of $N$ vectors on the same space. Consider an arbitrary function $h: \mathbb{R}^{d_1} \rightarrow \mathbb{R}$, and let $\hvec(\cdot) \equiv (h_1(\cdot),\dots,h_{d_2}(\cdot))'$ be a vector of $d_2$ such mappings. We define $h(\Bmat) \equiv (h(\bvec_1),\dots,h(\bvec_N));  \hvec(\bvec) \equiv (h_1(\bvec), \dots, h_{d_2}(\bvec))';$ and $\hvec(\Bmat) \equiv (\hvec(\bvec_1), \dots, \hvec(\bvec_N)).$ That is, $h(\Bmat)$ returns a vector of size $1 \times N$ containing the evaluation of $h(\cdot)$ over the columns of $\Bmat$; $\hvec(\bvec)$ returns a vector of size $d_2 \times 1$ containing the evaluation of $h_i(\cdot)$ at $\bvec$, for $i = 1,\dots,d_2$; and $\hvec(\Bmat)$ returns a $d_2 \times N$ matrix containing the evaluations of $\hvec(\cdot)$ at all the column entries in $\Bmat$.

\subsection{Model overview}\label{sec:model}

The deep compositional spatial model we propose is constructed from several layers that (i) model the observed data conditional on the underlying process,  (ii) model the process on the warped domain, and (iii) injectively warp the geographic domain. For ease of exposition, we will focus on the ubiquitous Gaussian data model for the first layer, but the inferential frameworks we implement can, with some modification, accommodate other data models should this be needed. 

Let $Y(\cdot)$ be a Gaussian process on $G$ with covariance function $C_G(\svec,\uvec)$, $\svec, \uvec \in G$. Assume that we have access to $N$ noisy observations of $Y(\cdot)$, that is,
\begin{equation}\label{eq:obs}
Z_i = Y(\svec_i) + \epsilon_i, \quad i = 1,\dots, N,
\end{equation}
where $\epsilon_i \sim \Gau(0, \sigma^2_{\epsilon})$, $\sigma^2_{\epsilon}$ is the measurement-error variance, and $\svec_1,\dots,\svec_N \in G$ are the measurement locations.  We model the process $Y(\cdot)$  as a low-rank process using basis functions which, as is common in geostatistical applications, are assumed to be fixed and known a priori. The process model is thus given by $Y(\svec) = \wvec_{n+1}'\phib_{n+1}(\fvec(\svec)), \svec \in G$, where $\wvec_{n+1}$ are basis-function coefficients, $\phib_{n+1}(\cdot)$ are the basis functions, and $\fvec(\cdot)$ is the warping function comprised of $n$ compositional layers; we thus treat the process of interest as the output of the $(n+1)$th layer. We give more detail on the process layer in Section~\ref{sec:spatproc}.

We model the warping layers as low-rank processes and, as we do with the top layer, we use basis functions that are generally known up to a small number of parameters. We often  fix the basis functions, that is, we assume that the parameters are known. Such a choice simplifies the estimation problem considerably, but also introduces the requirement that the (input) domain at each layer is bounded and fixed a priori. Without loss of generality, we henceforth fix $D_i = [c_1, c_1 + 1]^{d_i}, i = 1,\dots, n,$ where $c_1 \in \mathbb{R}$. Our model thus differs slightly from those discussed in Section~\ref{sec:background} in that the outputs of each layer are (linearly) rescaled before being input into the following layer. Note that injectivity is retained under the individual rescaling of each output dimension. In practice these rescalings are done such that the $i$th warping of a set of $m$ input knots $\Fmat^\alpha_0 \equiv \Smat^\alpha$ (generally the set of unique observation locations), which we denote as $\Fmat_i^\alpha$, are interior or boundary points of $D_i$, for $i = 1,\dots, n$. 

Let $f^u_{ik}(\cdot) \equiv \wvec^{(k)'}_i\phib_i(\cdot)$ be the unscaled $k$th output of the $i$th layer, and let $g_{ik}({f}^u_{ik}(\cdot); {\Fmat}^\alpha_{i-1})$ denote the respective scaling function. The scaling function we use takes the form
\begin{equation}
g_{ik}({f}^u_{ik}(\cdot); \Fmat^\alpha_{i-1}) \equiv \frac{{f}^u_{ik}(\cdot) - \min(f^u_{ik}(\Fmat^\alpha_{i-1}))}{\max(f^u_{ik}(\Fmat^\alpha_{i-1})) - \min(f^u_{ik}(\Fmat^\alpha_{i-1}))} + c_1,\label{eq:scaling}
\end{equation}
where $f^u_{ik}(\Fmat^\alpha_{i-1}) = \wvec^{(k)'}_i\phib_i(\Fmat^\alpha_{i-1})$ and $\phib_i(\Fmat^\alpha_{i-1})$  is an $r_i \times m$ matrix of basis function evaluated at the knot locations. Hence, $f^u_{ik}(\Fmat^\alpha_{i-1})$ is the $k$th dimension of the knots' unscaled warped locations, and $\min(\cdot)$ and $\max(\cdot)$ return the minimum and maximum along this dimension, respectively. At each warping layer we collect the $d_i$ scaling functions into the  vector $\gvec_i(\fvec_i^u(\cdot); \Fmat^\alpha_{i-1}) \equiv (g_{ik}({f}^u_{ik}(\cdot);\Fmat^\alpha_{i-1}): k = 1,\dots,d_i)'$.

%% The input knots, $\widetilde\Smat$, should be representative of the spatial extent of the observed data on $G$, $\Smat \equiv (\svec_1,\dots,\svec_N)$. For simplicity (and this is what we do in our implementation) one can construct $\widetilde\Smat$ from the set of unique observation locations. Note that  $g_{ik}({f}^u_{ik}(\cdot); \Fmat^\alpha_{i-1})$ is invariant to linear scalings of ${f}^u_{ik}(\cdot)$, and hence of $\wvec_i^{(k)}$. Issues with non-identifiability of this kind can be easily resolved by using ridge regression when doing maximum-likelihood estimation, or through the use of appropriate prior distributions when doing Bayesian inference. %We did not experience identifiability issues in our implementations.

%% The form of $g_{ik}(\cdot)$ implies that we need to be careful about our choice of $\widetilde{\Smat}$. The choice of $\widetilde\Smat$ clearly should depend on the spatial arrangement of the observed data on $G$, $\Smat \equiv (\svec_1,\dots,\svec_N)$. However, for the partial derivatives of the minimum and maximum with respect to the weights to be defined at every layer, it is required that the columns of the warped locations at each layer are unique. Conveniently, under bijective warpings this is guaranteed if the chosen knots on $G$ (i.e., the columns of $\widetilde\Smat$) are unique. For simplicity (and this is what we do in our implementation) one can construct $\widetilde\Smat$ from the set of unique observation locations.

Our deep compositional spatial model has the following hierarchical structure:

\vspace{0.2in}

\begin{tabular}{rrll}
   \vspace{0.1in}
  Observation model: & $Z_i$ & $=$ & $Y(\svec_i) + \epsilon_i,\quad i = 1,\dots,N,$ \\ \vspace{0.1in}
  Top-layer process model: & $Y(\svec)$ & $=$ & $\wvec'_{n+1}\phib_{n+1}(\svec),\quad \svec \in D_n$ \\
  \multirow{3}{*}{Warping models:~~~$\begin{cases}~\\~ \end{cases}$ \hspace{-0.3in}} & $\fvec_{n}(\svec)$ & $=$ & $\gvec_n(\Wmat_{n}\phib_{n}(\svec; \Thetab_n); \Fmat^\alpha_{n-1}),\quad \svec \in D_{n-1},$\\
  &   $\vdots$ && \\
  &   $\fvec_{1}(\svec)$ & $=$ & $\gvec_1(\Wmat_{1}\phib_{1}(\svec; \Thetab_1); \Smat^\alpha), \quad \svec \in G,$
\end{tabular}
\vspace{0.1in}

\noindent where $\fvec_1 : G \rightarrow D_1$ and $\fvec_i : D_{i-1} \rightarrow D_i,\, i = 2,\dots, n$. In Section~\ref{sec:units} we describe specific forms of $\fvec^u_i(\cdot) \equiv ({f}^u_{ik}(\cdot): k = 1,\dots, d_i)', i = 1,\dots, n,$ that, through composition, can define flexible injective warpings.

\subsection{Warping units}\label{sec:units}

\subsubsection*{Axial warping units} 

An axial warping unit (AWU) is a nonlinear mapping of one of the input dimensions. The map is constrained to be strictly monotonic, and hence injective. The AWU at the $i$th layer has $d_{i-1}$ inputs and $d_i = d_{i-1}$ outputs. Only one of the inputs is warped, while the others are simply forwarded on to the following layer. In particular, we define an AWU that warps the $k$th input dimension as follows:
\begin{align*}
  f^u_{ik}(\svec) &= \wvec_i^{(k)'}\phib_i(\svec; \Thetab_i),\quad \svec \in D_{i-1},\\
  f_{ik'}^u(\svec) &= s_{k'},\quad k' \ne k,~ \svec \in D_{i-1},
\end{align*}
where $\phi_{i1}(\svec; \theta_{i1}) = s_k$, and $\phi_{ij}(\svec; \thetab_{ij}) = \sig(s_k; \thetab_{ij})$ for $j = 2,\dots,r_i$, with
\begin{equation}
 \sig(s; \thetab)  \equiv \frac{1}{1 + \exp(-\theta_{1}(s - \theta_{2}))}. \label{eq:sigmoid}
\end{equation}
The first basis function models a linear scaling, while the $\{\phi_{ij}(\,\cdot\,; \thetab_{ij}): j = 2,\dots,r_i\}$ are sigmoid functions that model nonlinear scaling. Strict monotonicity of the sigmoid functions ensures that if $\wvec_i^{(k)}$ is nonnegative, $f_{ik}^u(\svec)$, and hence  $\fvec_i^u(\svec)$, is injective. In the SIWGP, the nonnegativity can be guaranteed by estimating the transformed parameters  $\{\log(w_{ikj})\}$ and then transforming back through the exponential function. In the SDSP, injectivity of the sample paths can be ensured by letting the weights have a lognormal prior distribution, that is, by letting $\tilde{w}_{ijk} \equiv \log(w_{ikj}) \sim \Gau(\mu_{ikj}, \sigma^2_{ikj})$.

We fix the parameters $\Thetab_i \equiv (\theta_{i21}, \theta_{i22}, \theta_{i31}, \dots, \theta_{ir_i2})'$  such that $\{\phi_{ij}(\,\cdot\,; \thetab_{ij}): j = 2,\dots,r_i\}$ can reproduce a wide range of smooth warping functions over the entire input domain, $D_{i-1}$. This is feasible in the low dimensional settings of spatial applications, and results in a considerably streamlined inference problem with little loss in model representation.  The formulation is also intuitive: $f^u_{ik}(\cdot)$ results in little warping when all the unknown weights except the first are close to zero, while a large non-negative weight on one of the sigmoid functions will result in localised relative stretching  of the input domain. %We term the axial warping of one coordinate axis as a \emph{single axial warping unit} (SAWU), while that of all the axes as an \emph{axial warping unit} (AWU).

\begin{figure}[t!]
  \includegraphics[width=\textwidth]{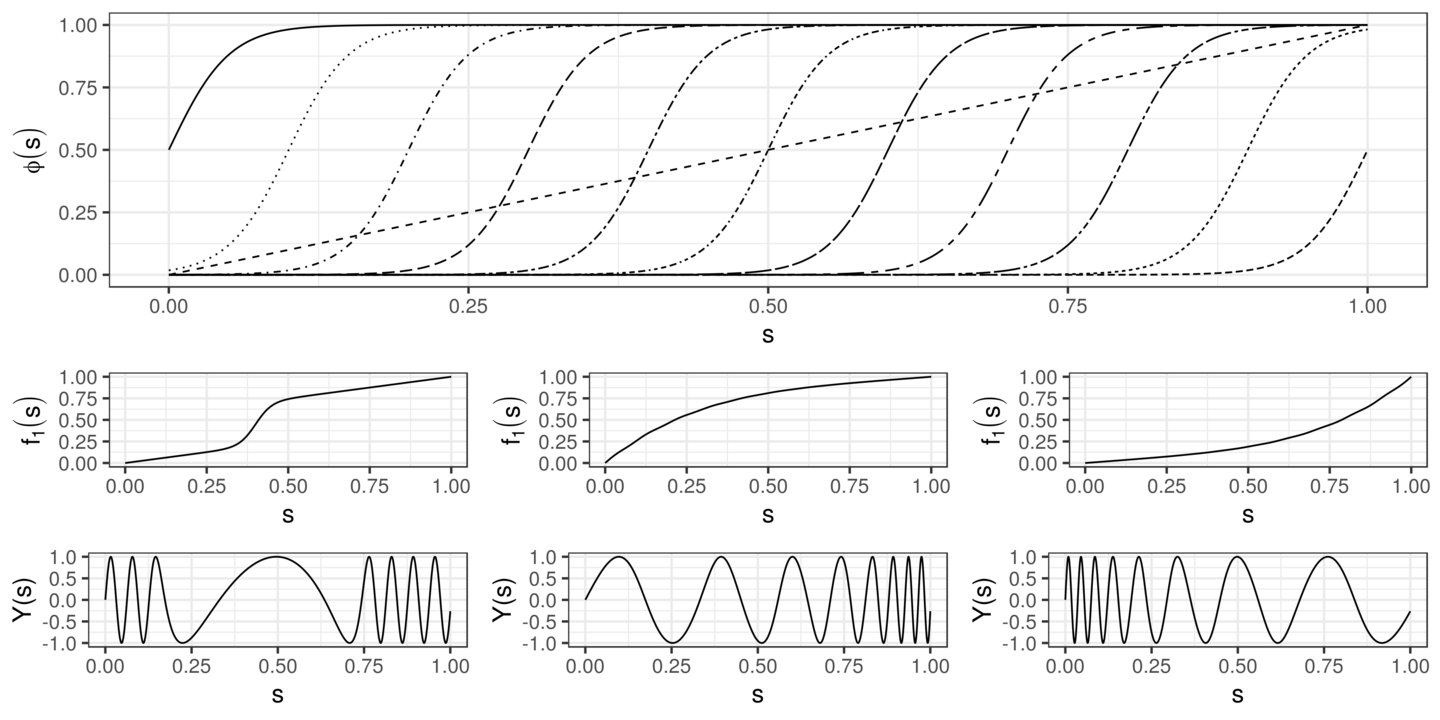}
  \caption{The axial warping unit. Top panel: The identity map and 11 sigmoid functions regularly spaced on the interval $[0, 1]$. Centre panels: Three different realisations of axial warping units reflective of different combinations of sigmoid-basis-function coefficients (see main text for details). Bottom panels: The output from an axial warping unit when the input is equal to $\sin(50s)$ for $s \in [0, 1]$. \label{fig:sigmoids}}
  \end{figure}

As an illustration of the AWU, consider the identity function and the 11 sigmoid functions in the interval $[0,1]$ shown in Figure~\ref{fig:sigmoids}, top panel. The bottom two panels of Figure~\ref{fig:sigmoids} show the warping function and its effect on an input signal equal to $\sin(50s)$ when (left panels) all basis-function coefficients are zero except for the fifth (from left to right), which is equal to 1, (middle panels) the basis-function coefficients increase cubicly (from left to right) from 0 to 1, and (right panels) the basis-function coefficients decrease cubicly (from left to right) from 1 to 0. In all cases, the output shown is that of the AWU rescaled to the interval $[0,1]$. An AWU formed from several basis functions can be much more flexible than one constructed using a Beta CDF \citep{Snoek_2014}; it is also likely easier to fit since the inference problem does not require basis-function parameter estimation, but only the estimation of a set of nonnegative weights which have local spatial scope.

\subsubsection*{RBF units} 

RBFs can be used to describe local expansions/contractions, and can warp at various resolutions. The RBF warping function is given by
%\begin{equation*}
%\fvec_i^u(\svec) = \svec + w_i(\svec - \gammab_i)\exp(-a_i\|\svec - \gammab_i\|^2);\quad \svec \in D_{i-1} \subset \mathbb{R}^2,
%\end{equation*}
%One can express Perrin and Monestiez's RBF warping as a weighted sum of basis functions. In particular,
$$
\fvec^u_i(\svec) = \Wmat_i\phib_i(\svec; \Thetab_i),\quad \svec \in D_{i-1},
$$
where $\phi_{i1}(\svec) = s_1, \phi_{i2}(\svec) = s_2$, $\phi_{i3}(\svec; \Thetab_i) = \psi_{i1}(\svec; \Thetab_i)$, and $\phi_{i4}(\svec; \Thetab_i) = \psi_{i2}(\svec; \Thetab_i)$, where $  \psi_{ij}(\svec; \Thetab_{i}) = (s_j - \gamma_{ij})\exp(-a_{i} \|\svec - \gammab_{i}\|^2),~ \svec \in D_{i-1},~~j = 1,2,$ and $\Thetab_i \equiv (\gammab_i', a_i)'$. The weight matrix has the form $\Wmat_i = [\Imat~w_i\Imat]$, so that only one weight $w_i$ needs to be estimated per layer (since, as with the AWUs, we fix $\Thetab_i$). Importantly, it is required that $-1 < w_i < \exp(3/2)/2$ for each $i$ to enforce injectivity \citep{Perrin_1999}. 

A single resolution RBF (SR-RBF) unit is formed from a composition of the RBFs of \citet{Perrin_1999}. The parameters $\Thetab_i$ are fixed in a way such that an SR-RBF unit can smoothly warp the entire domain, with higher resolutions able to provide more detailed and complex deformations. In our setup we let the $l$th resolution have the centroids of the RBFs arranged on a $3^l\times 3^l$ grid in $D_{i-1}$, so that at the $l$th resolution the SR-RBF unit has $3^{2l}$ layers. The scale parameter of the RBFs should increase with resolution. For $D_{i-1} = [0,1] \times [0,1]$ we set $a_i = 2\cdot (3^l - 1)^2$ (where $i$ is the layer corresponding to the RBF); this choice results in the $\exp(-1/2)$ contour lines of the squared-exponential components of the RBFs intersecting with those of their neighbours at a single point. 

\begin{figure}[t!]
  \includegraphics[width=0.9\textwidth]{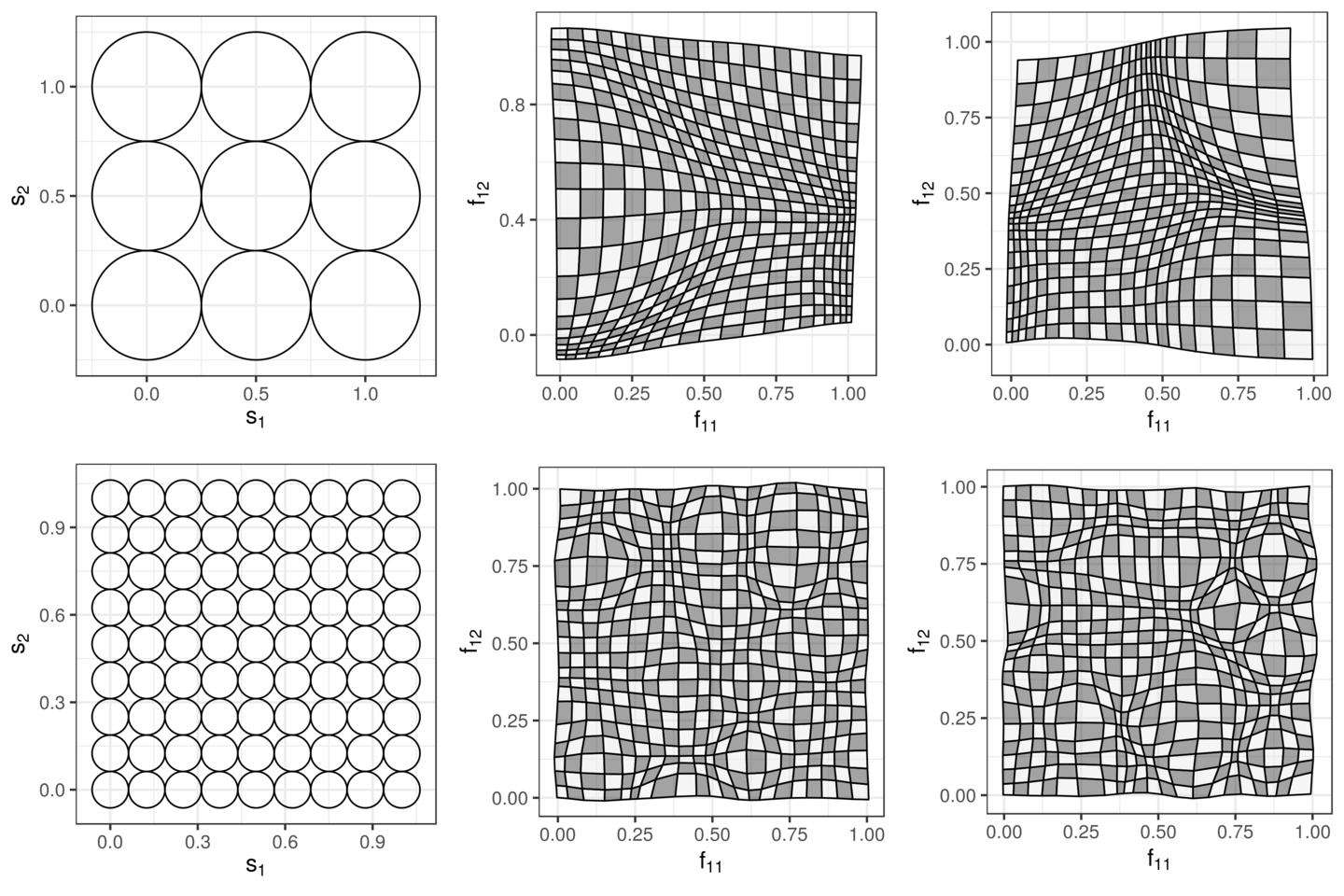}
  \caption{Single-resolution RBF (SR-RBF) units (left panels) and random realisations of warpings of regular chequered patterns on $[0,1]^2$ from these units (centre and right panels). Top panels: Resolution $l = 1$, where the RBFs are arranged on a $3 \times 3$ grid and $a_i = 8$. Bottom panels: Resolution $l = 2$, where the RBFs are arranged on a $9 \times 9$ grid and $a_i = 128$. \label{fig:RBFwarps}}
\end{figure}

In an SIWGP, the constraint on $w_i$ can be achieved by estimating the transformed parameter $\tilde w_i$ without any constraints, where $\tilde w_i = \textrm{logit}((1 + w_i)/(1 + \exp(3/2)/2)).$ In an SDSP we equip $\tilde w_i$ with a Gaussian distribution, that is, we let $\tilde w_i\sim \Gau(\mu_i, \sigma^2_i)$. Note that when $w_i = 0$  $(\tilde w_i \approx -0.8)$, $\fvec_i^u(\svec) = (s_1, s_2)', \svec \in D_{i-1}$, that is, the input to the layer is not warped. We therefore set $\mu_i = -0.8$.

In Figure~\ref{fig:RBFwarps} we show two resolutions of RBFs, and an example of warpings that can be generated using these basis functions, with the output rescaled to the unit square. \Copy{RBFcomposition}{We stress that unlike the AWUs, these RBFs are combined through composition and not summation in order to ensure injectivity of the composite map.} %Schematics showing the construction of the SR-RBF unit from elemental basis functions are given in Figure~\ref{fig:MRRBFs}.  
As in the case of spatial processes \citep[e.g.,][]{Cressie_2008, Nychka_2015}, we expect warpings to occur at various scales. We can model these multi-resolutional warpings through the composition of two or more SR-RBFs at different resolutions. We denote an SR-RBF at the $l$th resolution as SR-RBF($l$).

\subsubsection*{M{\"o}bius transformation units}

% M{\"o}bius transformations are bijective mappings from the complex plane, $\mathbb{C}$, to itself (which, for the purposes of this work, can be seen as a map from $\mathbb{R}^2$ to $\mathbb{R}^2$).
Define $z(\svec) \equiv s_1 + s_2\iota$, where $\iota \equiv \sqrt{-1}$, and let $\avec \in \mathbb{C}^4$. Then, the M{\"o}bius transformation is given by $\phi^m_i(z(\svec); \Thetab_i) = (a_1z(\svec) + a_2)/(a_3z(\svec) + a_4),$ where $\Thetab_i \equiv (a_1,\dots,a_4)'$. This warping unit contains 8 unknown parameters (the real and imaginary components of $\avec$) and all weights are fixed to one. That is,
\begin{align*}
  f_{i1}^u(\svec) &= Re(\phi^m_i(z(\svec); \Thetab_i)),\quad \svec \in D_{i-1}, \\
  f_{i2}^u(\svec) &= Im(\phi^m_i(z(\svec); \Thetab_i)),\quad \svec \in D_{i-1}, 
\end{align*}
\noindent where $Re(\cdot)$ and $Im(\cdot)$ return the real and imaginary components of their arguments, respectively. The M{\"o}bius transformation unit is distinct from the units considered so far, in that it does not have any weights that need to be estimated, but contains a set of parameters that do need to be. 

\begin{figure}
  \includegraphics[width=\textwidth]{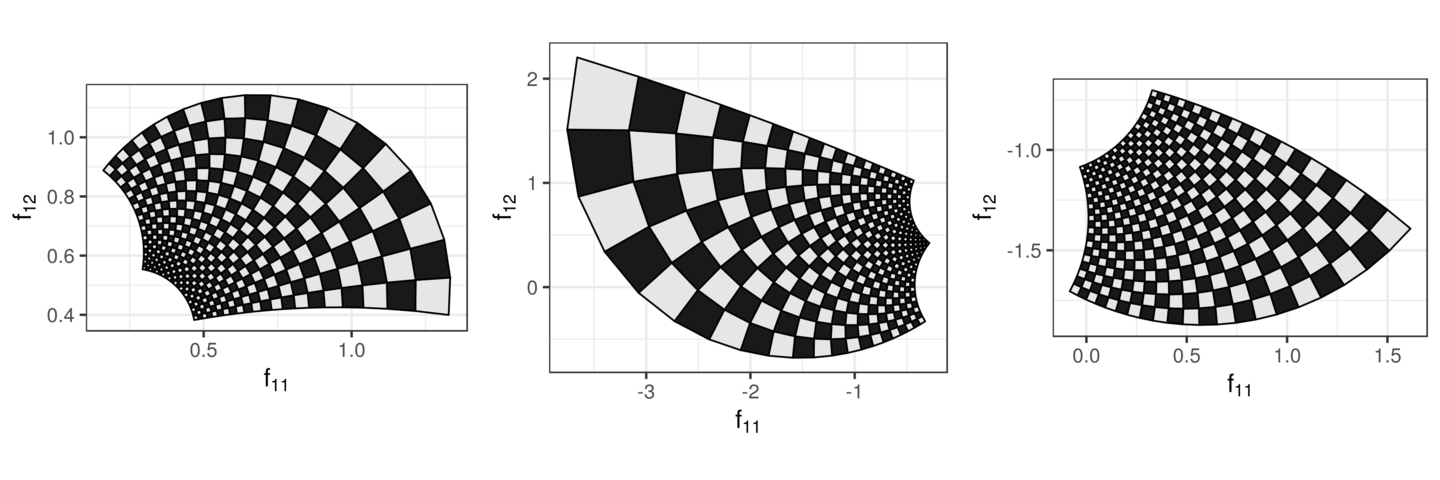}
  \caption{M{\"o}bius transformations of a regular chequered pattern on $D_{i-1} = [0, 1] \times [0,1]$, obtained by randomly generating the real and imaginary components of $\avec \in \mathbb{C}^4$ from a standard normal distribution subject to the constraint that $-a_4/a_3$ is not in the square enclosed by the points $0 + 0\iota, 0 + 1\iota, 1 + 1\iota, 1 + 0\iota$ on the complex plane. \label{fig:Moebiussims}} 
\end{figure}

One can show that a M{\"o}bius transformation of a M{\"o}bius transformation is itself a M{\"o}bius transformation and therefore there is no benefit in cascading more than one of these transformation units in immediate sequence \Copy{Mobius}{(conversely, a second M{\"o}bius transform separated from the first by other warping units does materially alter the warping function).} %It is also straightforward to show that the transformation itself is a composition of other elemental transformations, a linear scaling/rotation transformation ($\phi^m_i(z(\svec)) = a_1z(\svec)$), a translation ($\phi^m_i(z(\svec)) = z(\svec) + a_2$) and an inversion ($\phi^m_i(z(\svec)) = 1/z(\svec))$, and therefore no transformation units to this effect need to be used in conjunction with the M{\"o}bius transformation unit.
The unit maps $\svec$ to infinity for $z(\svec) = -a_4 / a_3$ and we therefore need to ensure that the spatial coordinates implied by the complex number $-a_4/ a_3$ are not in $D_{i-1}$. Assuming that $D_{i-1} = [0, 1]\times [0,1]$, this is equivalent to asserting that the real or imaginary component of $-a_4/a_3$ is outside of the interval $[0, 1]$, something that can be ensured when optimising $\Thetab_i$. Note that $a_2 = a_3 = 0$ and $a_1 = a_4 = 1$ implies no warping. Figure~\ref{fig:Moebiussims} shows three random M{\"o}bius transformations, where all components of $\avec$ were simulated from a standard normal distribution subject to the above constraint.

\subsection{The top-level spatial process}\label{sec:spatproc}

The output of the top, ($n + 1$)th, layer of the deep compositional spatial model is the process $Y(\cdot)$. To deal with moderately large datasets, we choose to have a low-rank representation for $Y(\cdot)$. Specifically, we let
$$
Y(\svec) = \wvec_{n+1}'\phib_{n+1}(\svec; \Thetab_{n+1}),\quad \svec \in D_n,
$$
where the random weights $\wvec_{n+1}$  have a Gaussian distribution with some mean $\muvec$ (which henceforth we take equal to $\zerob$ without loss of generality) and covariance matrix $\Sigmamat_{\tau_{n+1}}$, $\taub_{n+1}$ is a vector of unknown parameters appearing in $\Sigmamat_{\tau_{n+1}}$, and  the basis functions $\phib_{n+1}(\,\cdot\,;\Thetab_{n+1})$ take inputs in $D_n \subset \mathbb{R}^{d_n}$. Such a model is feasible when $d_n$ is small, say $d_n \le 4$, as is typical in spatial applications.

In our implementation we let $\phib_{n+1}(\,\cdot\,;\Thetab_{n+1})$ be a set of bisquare basis functions. That is, we let
$$
\phi_{n+1,j}(\svec; \thetab_{n+1,j}) \equiv \left\{\begin{array}{ll} \{1 - (\|\svec- \gammab_{n+1,j}\|/{\delta_{n+1,j}})^2\}^2; &\| \svec - \gammab_{n+1,j}  \| \le \delta_{n+1,j} \\
0; & \textrm{otherwise}, \end{array} \right.
$$
where the parameter vector $\thetab_{n+1,j} \equiv (\gammab_{n+1,j}', \delta_{n+1,j})'$ is made up of the centroid $\gammab_{n+1,j}$ and the aperture $\delta_{n+1,j}$. We let the centroids of the bisquare basis functions $\phib_{n+1}(\,\cdot\,;\Thetab_{n+1})$ be regularly spaced in $D_n$, and model the covariances of the weights as  $\Sigmamat_{\tau_{n+1}} = (\sigma^2 \exp(-\|\gammab_{n+1,j} - \gammab_{n+1,j'}\| / l): j,j' = 1,\dots,r_{n+1})$ \citep[see][for more details]{Zammit_2018} where $\taub_{n+1} = (\sigma^2, l)'$.  Note that this top-level spatial process does not induce a stationary isotropic covariance structure, but one that is usually able to approximate stationary and isotropic covariances reasonably well. 

\Copy{Sigma}{Since $\Sigmamat_{\tau_{n+1}}$ is constructed using only the distances between the basis-function centroids, it is reasonable to regularly place the basis-function centroids in $D_n$. The number of basis functions to use is a design choice. Exploratory analysis of the data's spectral composition may be used as a guide \citep[e.g.,][]{Zammit_2012b}, but for many applications several hundred is reasonable. As discussed in Section~\ref{sec:Complexity}, the user is limited to a few thousand in practice.}  Similar models that can deal with a larger number of basis functions, and where instead a sparse precision matrix $\Qmat_{\tau_{n+1}}$ is modelled \citep[e.g.,][]{Lindgren_2011, Nychka_2015}, are left for future consideration (see Section~\ref{sec:Conclusion}).

\subsection{Inference and prediction}\label{sec:inference-summary}

In SIWGPs and SDSPs, inference needs to be made on both the weights and the parameters. In order the render the optimisation problem on the weights unconstrained, we transform the weights using a transformation that is specific to the layer type. Recall that the transformation is a $\log$ function when the layer is an AWU, and a $\textrm{logit}$ function when the layer is an RBF. Denote the transformation functions as $h_i(\cdot)$, $i = 1,\dots, n$, which are strictly monotonic and invertible. Then, inference on the weights is done by first making inference on the transformed weights $\widetilde\Wmat_{i} = h_i(\Wmat_i), i = 1,\dots, n$.

Parameters and weights in the SIWGP can be estimated in a straightforward manner using maximum likelihood. With our Gaussian data model, one can also make use of the integrated likelihood, where the weights at the top layer, $\wvec_{n+1}$ (which are not transformed), are integrated out from the likelihood function. In our implementation, gradients were found using automatic differentiation (AD), where the gradients of the integrated likelihood with respect to the unknown weights and parameters are computed during runtime using, for example, back propagation \citep[][Section 6.5]{Goodfellow_2016}. AD obviates the need for analytical gradient computation, and has gained considerable interest in recent years through its use in the popular statistical modelling and fitting packages \texttt{Stan}, \texttt{greta}, and \texttt{TMB}. For this work we employed the AD functionality in the library \texttt{TensorFlow} \citep{Tensorflow} through \texttt{R} \citep{R, TensorflowR}. Once the parameters and weights are estimated, prediction can be done using standard Gaussian conditioning; details are given in Appendix~\ref{sec:SIWGP}.

In SDSPs, the transformed weight vectors in the $i$th layer, $\{\tilde \wvec_i^{(k)}: k = 1,\dots,d_i\}$, are equipped with multivariate Gaussian distributions with means $\muvec_i$ such that they are reflective of no, or little, warping, and covariance matrix $\sigma^2_i\Imat$. In our implementation, we fixed $\sigma^2_i$ to a large value for each $i$ to keep these prior distributions diffuse. The parameters $\{\sigma^2_i\}$ could instead be estimated or fixed to small values to add soft limits to the intensity of the warpings in the hidden layers.

%As with DGPs, inference with the SDSP is generally a difficult problem since $\log p(\Zvec \mid \widetilde\Wmat_{1:n}, \Thetab_{1:n}, \taub_{n+1}, \sigma^2_\epsilon)$ is a highly nonlinear function of \red{These are not defined now} $\widetilde\Wmat_{1:n}$. Hence, marginalisation of $\widetilde\Wmat_{1:n}$ is not possible analytically, and one cannot easily compute expectations with respect to $\widetilde\Wmat_{1:n} \mid \Zvec, \Thetab_{1:n}, \taub_{n+1}, \sigma^2_\epsilon$. This renders the use of expectation-maximisation inference frameworks difficult to implement for SDSP models.

As with DGPs, inference with the SDSP is generally a difficult problem since the integrated likelihood is a highly nonlinear function of the transformed weights. Hence, marginalisation of the weights $\widetilde\Wmat_{i}, i = 1,\dots, n,$ is not possible analytically, and one cannot easily compute expectations with respect to $\widetilde\Wmat_{i} \mid \Zvec, \Thetab_{1:n}, \taub_{n+1}, \sigma^2_\epsilon$, say, where $\Thetab_{1:n}\equiv\{\Thetab_1,\dots,\Thetab_n\}$. %This renders the use of expectation-maximisation inference frameworks difficult to implement for SDSP models.
We therefore use an approximate Bayesian approach, variational Bayes (VB), to make inference on the latent weights and unknown parameters. For excellent introductions to the VB approach to approximating intractable posterior distributions, see  \citet[][Chapter 2]{Beal_2003}, \citet[][Chapter 10]{Bishop_2006}, and \citet{Blei_2017}.

Briefly, the VB approach we adopt aims to maximise a lower-bound of the integrated likelihood. It involves specifying a variational distribution $q(\cdot)$, which we factorise across the layers and outputs, Specifically, we let  $q(\widetilde\Wmat_i) = \prod_{k = 1}^{d_i} q(\tilde\wvec_i^{(k)})$, where $q(\tilde\wvec_i^{(k)}) = \Gau(\mvec_i^{(k)}, \Vmat_i^{(k)}(\etab_i^{(k)})),~i = 1,\dots,n$, $\mvec_i^{(k)}$ is the variational expectation of the weights associated with the $k$th output dimension in the $i$th layer, and $\Vmat_i^{(k)}(\etab_i^{(k)})$ the corresponding covariance matrix, parameterised through parameters $\etab_i^{(k)}$. Estimating the variational expectations and the covariance-matrix parameters requires computing intractable expectations, which we instead approximate via Monte Carlo. Predictions from the variational posterior distributions are also done via Monte Carlo; in Appendix~\ref{sec:SDSP} we show that these predictions take the form of a Gaussian mixture and thus, unlike those from the SIWGP, can be highly non-Gaussian. \Copy{Bayesian}{Although other Bayesian computation methods, such as Markov chain Monte Carlo (MCMC), can be used instead,  it is not clear whether our inferences and predictions would benefit from the more exact approaches. First, variational posterior distributions have attractive asymptotic consistency properties \citep{Wang_2019}. Second, as we show in Section~\ref{sec:1D} through a simple one-dimensional example, classical MCMC algorithms are likely to be infeasible for large problems. Third, approximations to MCMC algorithms, such as the suite of stochastic gradient MCMC algorithms, are feasible, but require design choices that could easily compromise validity and convergence \citep{Teh_2016}. Finally,  we see a negligible difference between the predictions done using VB and those done via MCMC in our example of Section~\ref{sec:1D}. } For full technical details on the VB approach we implement see Appendix~\ref{sec:SDSP}. We discuss computational properties of our algorithms in Appendix~\ref{sec:Complexity}.

\section{Experiments}\label{sec:Experiments}

We assessed the SIWGP and the SDSP on both simulated data and real data, and compared their predictive performance to those of various models. We consider two simple one-dimensional examples in Section~\ref{sec:1D}, a two-dimensional example in Section \ref{sec:2D}, and images of cloud and ice taken from the MODIS instrument aboard the Aqua satellite in Section \ref{sec:L1}. In all cases, predictive performance was assessed by evaluating diagnostics on data not used for model fitting, namely the mean absolute prediction error (MAPE), the root mean-squared-prediction-error (RMSPE), the continuous ranked probability score (CRPS), and the interval score of the 95\% prediction interval (IS); see \citet{Gneiting_2007}. All experiments were carried out on a high-end desktop computer with 32GB of RAM, an Intel\textregistered~Core\textsuperscript{TM} i9-7900X CPU, and an NVIDIA\textregistered~1080 Ti graphics processing unit (GPU). Data and reproducible code for all experiments is provided as supplementary material.

\subsection{1D simulations}\label{sec:1D}

The first experiment  assesses the suitability of the SIWGP and SDSP in modelling nonstationary processes, and compares them to other existing deep and shallow models, on simple case studies. We consider two cases on $G = [-0.5, 0.5]$, where the underlying processes are
\begin{equation}\label{eq:1Dprocesses}
Y^{(1,1)}(s) = \begin{cases}
  -0.5& |s| > 0.2 \\
  0.5 & \textrm{otherwise},
\end{cases}
\qquad
 Y^{(1,2)}(s) = \begin{cases}
 \exp\left(4 + \frac{5}{2s(10s + 5)}\right) & -0.5 < s < 0 \\
 1 &  0.2 \le s \le 0.3 \\
 -1 & 0.3 < s \le 0.4 \\
 0 & \textrm{otherwise}.
  \end{cases}
\end{equation}
 The  process $Y^{(1,1)}(\cdot)$ is a rectangular function while the process $Y^{(1,2)}(\cdot)$ is the same, up to a scaling of $s$, as that considered by \citet{Monterrubio_2020}, and contains a smooth component and a discontinuous component. For both cases we used 300 spatial points randomly generated on $[-0.5, 0.5]$ as our observation locations, and generated data at these 300 locations by adding Gaussian measurement error with zero mean and variance $\sigma^2_\epsilon = 0.01$ to the process values at these locations. We used a fine grid of 1001 points on the interval $[-0.5, 0.5]$ as our prediction domain; this grid was used for computing the validation diagnostics outlined in the introduction to this section.

The SIWGP and SDSP were configured to have a single warping layer, specifically an AWU with 50 sigmoid functions, with the basis functions' points of inflection regularly spaced on $[-0.5, 0.5]$, and with the steepness parameter $\theta_{1} = 200$ in \eqref{eq:sigmoid}. The second top-level (process) layer was configured to have 50 bisquare functions with centroids regularly spaced on $[-0.5, 0.5]$. For the SDSP, the covariance matrix of the variational distribution, $\Vmat_1^{(1)}(\etab_1^{(1)}),$ was constrained to be diagonal, and the number of Monte Carlo samples we used in the VB algorithm was fixed to 10. An Adam optimiser \citep{Kingma_2014b} was used to optimise the process and variational parameters. In both experiments we proceeded in three stages: We first optimised the transformed weights $\tilde\wvec_1^{(1)}$ (in case of the SIWGP) or the variational means $\mvec_1^{(1)}$  (in case of the SDSP) using 100 gradient steps, then all the other parameters with $\tilde\wvec_1^{(1)}$ or $\mvec_1^{(1)}$ fixed for another 100 steps, and then all parameters simultaneously for a final 100 steps. Convergence of the likelihood (SIWGP) and variational lower bound (SDSP) was monitored for each case. \Copy{Easy}{In this and the following experiments, we found that obtaining reasonable parameter estimates was always straightforward. We suspect that this is due to the fact that our models are quite parsimonious overall (we have considerably less parameters than data points) and that the basis functions we use have fixed, local spatial scope.}

Although there are a wide range of spatial models that use a large number of basis functions (e.g., wavelets or Wendland functions) that might also be appropriate for modelling these data, here we focus on comparing the SIWGP and SDSP to other models that also use warping functions. We do this since, while high-rank models tend to give a good fit to the data, their ability to model nonstationarity, and hence their ability to capture the true data-generating process, tends to be limited.  We therefore used the generated data to compare the deep spatial models to the full-rank $n = 1$ DGP (DGPfull) of \citet{Schmidt_2003}, the Random Fourier Features DGP of \citet{Cutajar_2016} with $n = 2$ (DGPRFF), the sparse DGP of \citet{Damianou_2013} with one hidden layer (DGPsparse), and a (shallow) GP with a Mat{\'e}rn covariance function with smoothness parameter $\nu = 3/2$. The former three models were constructed from GPs with squared-exponential covariance functions. Details on the implementation of these alternative models are available in Section \ref{sec:1D-details} \ifarxiv (supplementary material)\else of the online supplementary material\fi. To check the validity of the approximate variational inferences in this simple setting we also ran MCMC on the weights of the SDSP after fixing the parameters at the top (process) layer to those estimated by VB, $\widehat\taub_{n+1}$ (SDSP-MCMC).  Table~\ref{tab:1D} summarises the models we considered, the associated inference method, the software used, and the hardware used. Software packages used include \texttt{TensorFlow} \citep{TensorflowR}, \texttt{PyTorch},\footnote{\url{https://pytorch.org/}} \texttt{GPflow} \citep{Matthews_2017}, and \texttt{Stan} \citep{Carpenter_2017}. 

\begin{table}[t!]
  \caption{Inference Method, Software, and Hardware Used for the Models Considered in the 1D Simulation Experiments.%, namely the full-rank $n = 1$ DGP (DGPfull), a random Fourier features approximation to a $n = 2$ DGP (DGPRFF),  the sparse deep Gaussian process (DGPsparse), a shallow Gaussian process  with a Mat{\'e}rn 3/2 covariance function (GP), the spatial deep stochastic process fitted using VB (SDSP), the SDSP with the weights re-inferred using MCMC (SDSP-MCMC), and the spatial input-warped Gaussian process (SIWGP). For more details see Appendix A in the Supplementary Material.
    \vspace{0.1in}}\label{tab:1D}

  \small
  \begin{center}
\begin{tabular}{llll}
Model & Inference method & Software & Hardware \\ \hline
DGPfull & Elliptical slice sampling & \texttt{Python} & CPU \\
DGPRFF & Stochastic variational inference & \texttt{Python} and \texttt{PyTorch} & GPU \\
DGPsparse & Variational inference & \texttt{R} and \texttt{TensorFlow} & GPU \\
GP & Maximum likelihood & \texttt{Python} and \texttt{TensorFlow} & GPU \\
SDSP & Stochastic variational inference & \texttt{R} and \texttt{TensorFlow} & GPU \\
SDSP-MCMC & Hamiltonian Monte Carlo & \texttt{R} and \texttt{Stan} & CPU \\
SIWGP & Maximum likelihood & \texttt{R} and \texttt{TensorFlow} & GPU 
\end{tabular}
\end{center}
\end{table}

\begin{figure}
  \ifarxiv
  \includegraphics[width=0.9\textwidth]{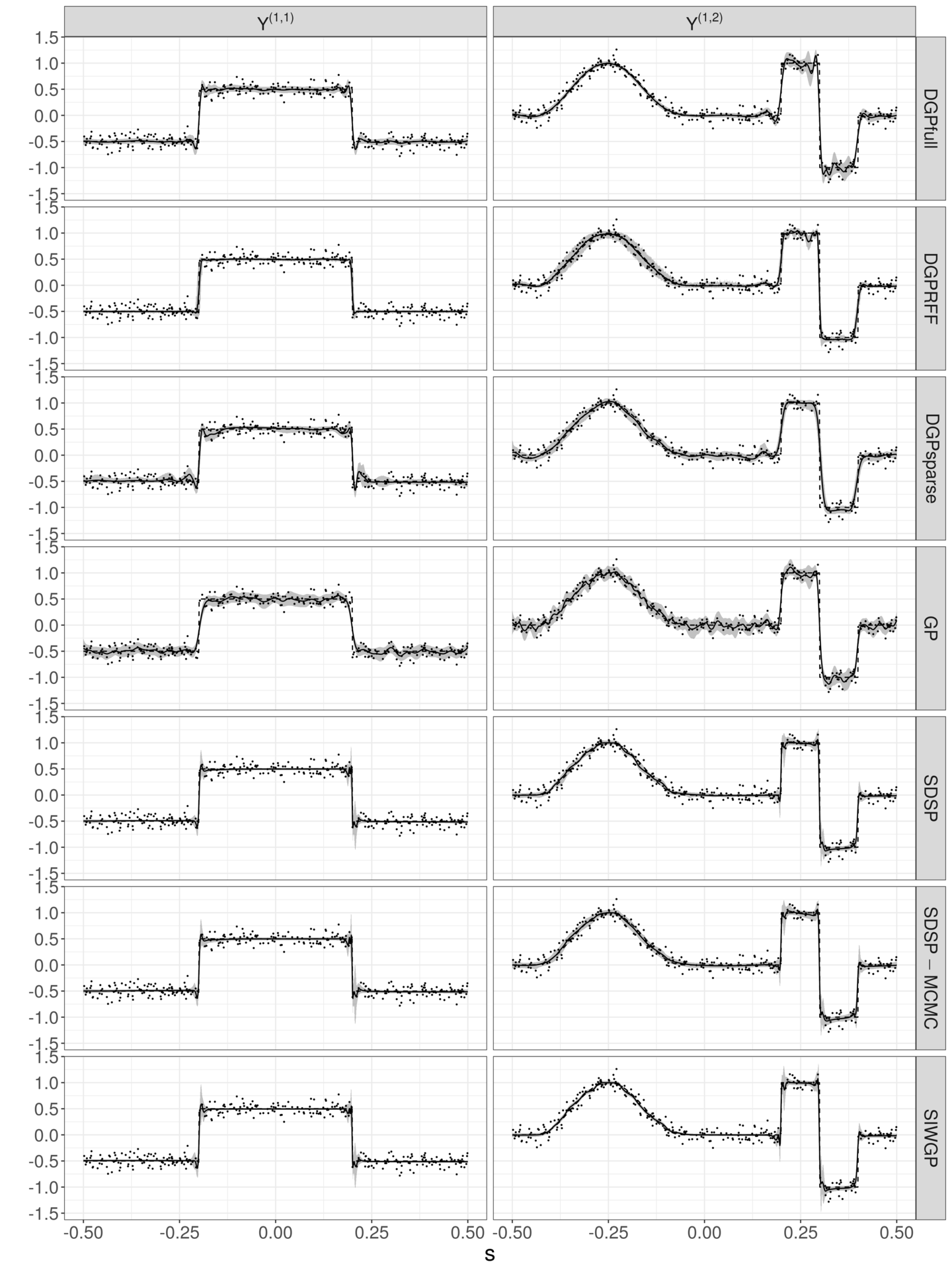}
  \else
  \includegraphics[width=0.9\textwidth]{../deep_spat_src_blinded_for_submission/1Dtests/img/1Dresults.png}
  \fi
  \caption{Data (dots), true process (dashed line), prediction (solid line) and 95\% prediction interval (grey shading) for the two processes defined in \eqref{eq:1Dprocesses} and the various models/inferential methods listed in Table~\ref{tab:1D}. \label{fig:1D}}
  \end{figure}

For both processes, the predictions, and 95\% prediction intervals, as well as the data points used for simulation, are depicted in the panels of Figure~\ref{fig:1D}.  The stationarity assumption of the Mat{\'e}rn GP is inadequate for these processes, and thus the GP inadequately predicts signal variability in regions where there is not any, and step transitions that are too smooth. Although all deep variants considered here contain only one or two warping layers, they are mostly able to adequately distinguish between signal `dead zones' and regions of signal variability, although not all are able to capture the steep step transitions. Note that with the SIWGP and SDSP the prediction uncertainty increases at the step transitions; this is to be expected since in these regions the domain is `stretched out,' resulting in relative local data paucity. Increased uncertainty in (relatively) poorly observed regions of high signal variability is an attractive feature of these deep models.

From Table~\ref{tab:1Dresults} we see that the SIWGP and SDSP outperform the other models with $n \in \{0, 1\}$, both in terms of prediction and uncertainty quantification. The SDSP's performance is comparable to that of the DGPRFF with $n = 2$ hidden layers for the step function (we obtained worse predictions with the DGPRFF for $n = 1$). %Interestingly, we do not see a notable difference between the SIWGP and the SDSP in this simple example (we do notice a difference in performance on the more complicated models considered in later sections). %the SIWGP faired slightly better on uncertainty quantification than the SDSP counterpart, but only marginally so. (This better performance was not there with the more complicated models considered in later sections.) The slightly lower scores for the SDSP are due to marginally wider prediction intervals as well as slightly worse predictions.
We do not observe a notable difference between the SIWGP and the SDSP in this simple example and, reassuringly, SDSP-MCMC provided very similar results to the SDSP fitted using VB.

  \begin{table}[t!]
  \caption{Diagnostic Results for $Y^{(1,1)}(\cdot)$ and $Y^{(1,2)}(\cdot)$.%Diagnostic results for $Y^{(1,1)}$, namely the mean absolute prediction error (MAPE), root-mean-squared prediction error (RMSPE), continuous-ranked probability score (CRPS) and interval score at the 5\% level (IS) of the various models listed in Table~\ref{tab:1D}. The diagnostics are computed from the true process values and the predictive distributions on a fine gridding of $[-0.5, 0.5]$.
  }\label{tab:1Dresults}
  \begin{center}
  {\small 
    \begin{tabular}{lc|cccc|cccc}
     \multicolumn{2}{c}{}   &   \multicolumn{4}{c}{$Y^{(1,1)}(\cdot)$} & \multicolumn{4}{c}{$Y^{(1,2)}(\cdot)$} \\
  Model &$n$& MAPE & RMSPE & CRPS & IS & MAPE & RMSPE & CRPS & IS\\   \hline
DGPfull &1& 0.0184 & 0.0474 & 0.0150 & 0.2413 & 0.0381 & 0.0892 & 0.0292 & 0.5155 \\ 
  DGPRFF &2& 0.0093 & 0.0393 & 0.0080 & 0.0824 & 0.0337 & 0.0821 & 0.0252 & 0.3089 \\ 
  DGPsparse &1& 0.0294 & 0.0566 & 0.0212 & 0.2050 & 0.0484 & 0.1108 & 0.0365 & 0.5650 \\ 
  GP &0& 0.0381 & 0.0712 & 0.0296 & 0.5058 & 0.0516 & 0.0958 & 0.0395 & 0.6329 \\ 
  SDSP &1& 0.0119 & 0.0323 & 0.0086 & 0.1037 & 0.0260 & 0.0660 & 0.0190 & 0.2143 \\ 
  SDSP-MCMC &1& 0.0119 & 0.0324 & 0.0092 & 0.1083 & 0.0264 & 0.0660 & 0.0195 & 0.2307 \\ 
  SIWGP &1& 0.0119 & 0.0316 & 0.0082 & 0.0890 & 0.0253 & 0.0673 & 0.0189 & 0.2246 \\ 
\end{tabular}}
\end{center}
\end{table}

  While it is difficult to compare the computational demands of the various models and inferential methods considered, it is worth noting that fitting and prediction with the SIWGP required only 4 s, while with the SDSP only 8 s. On the other hand it took over one hour to generate 10,000 samples using \texttt{Stan} with the SDSP, first because the computations were done on a CPU and, second, because a minimum number of Monte Carlo iterations is required to assess convergence; this number greatly exceeds the number of gradient ascent steps needed when doing maximum likelihood or variational inference. It also took a few hours to obtain a good fit with the DGPRFF model despite the use of a GPU, largely because very small step sizes were needed to ensure the warping (which is not constrained to be injective) did not exhibit the degeneracy discussed in \citet{Duvenaud_2014}. Finally, it took a few days to obtain useful MCMC traces from an elliptical slice sampler on the DGPfull model, where each sample took more than a minute on a CPU and where traces tended to be highly auto-correlated. The considerable computational advantages of the SIWGP and the SDSP stem from both the parsimonious representation of the injective warpings and the approximate inference schemes used.

In Section \ref{sec:1D-stat-experiment} \ifarxiv (supplementary material) \else of the online supplementary material \fi  we also show that the deep compositional spatial models are able to perform reasonably well even when the underlying process has a stationary covariance function. This experiment also shows the advantages of the SDSP over the SIWGP, which begins to over-fit as the number of basis functions used in the AWU is increased.

  %It is thus clear that approximate inference methods have an important role to play in the analysis of spatial DSPs.

\subsection{2D simulations}\label{sec:2D}

The second experiment serves two purposes; first, to assess whether the fitting mechanisms we employ are able to recover the underlying warping function when data are generated from an SIWGP with known architecture and, second, to assess what the impact is of using a different number of warping functions, or basis functions in the top layer, to what is used in the true model.

\begin{figure}[t!]
  \begin{center}
    \ifarxiv
    \includegraphics[width=0.9\textwidth]{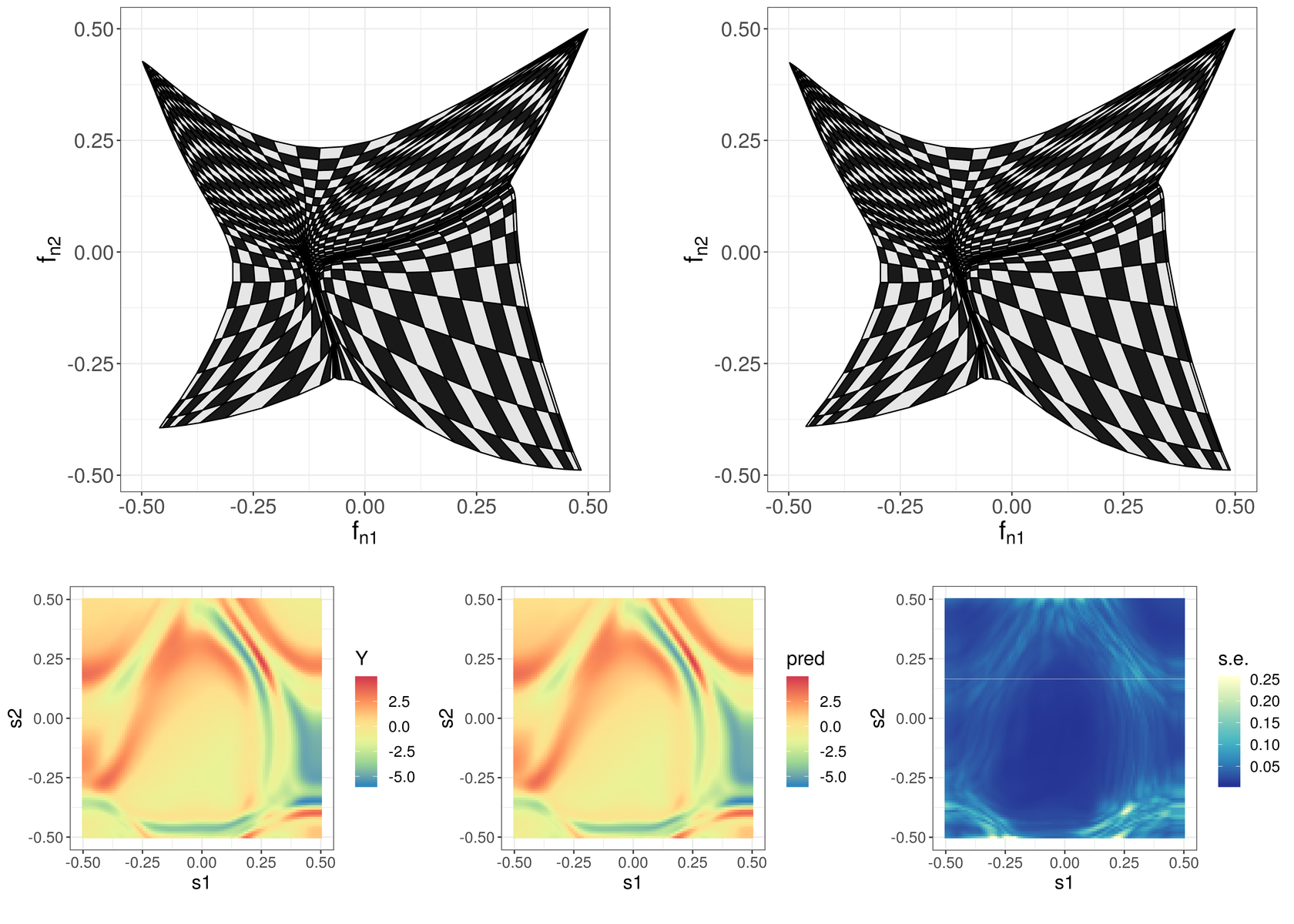}
    \else
    \includegraphics[width=0.9\textwidth]{../deep_spat_src_blinded_for_submission/2Dtests/img/AWU_RBF_2D.png}
    \fi
    \end{center}
  \caption{(Top-left) True warping function for the case $Y^{(2,1)}(\cdot)$, depicted here through its action on a regular chequered pattern on $G$ and (top-right) the SDSP variational posterior mean of the warping following variational inference. (Bottom-left) the simulated process, (bottom-centre) the prediction and (bottom-right) the prediction standard error.  \label{fig:2DA}}
  \end{figure}

\begin{figure}[t!]
  \begin{center}
    \ifarxiv
    \includegraphics[width=0.9\textwidth]{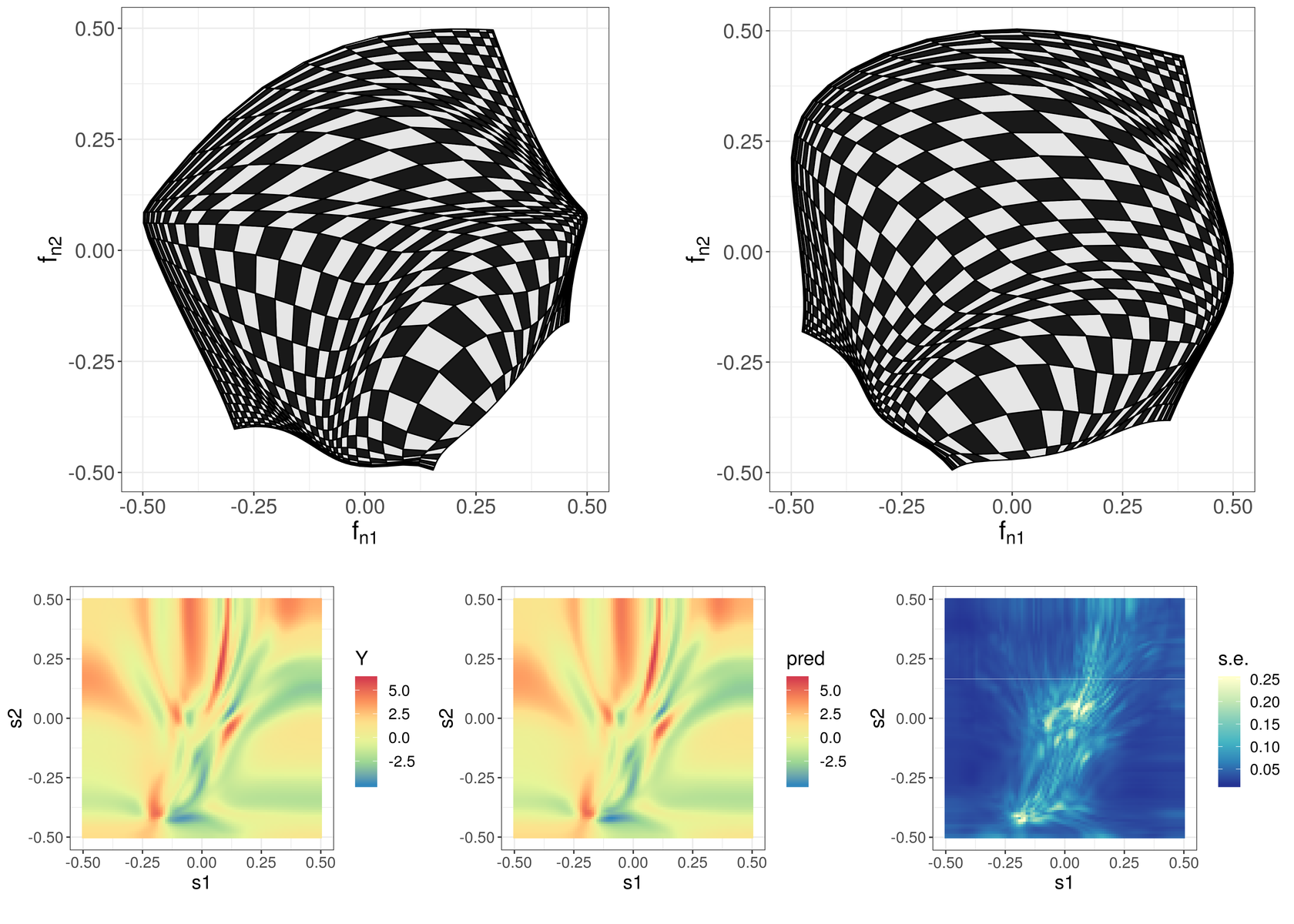}
    \else
    \includegraphics[width=0.9\textwidth]{../deep_spat_src_blinded_for_submission/2Dtests/img/AWU_RBF_LFT_2D.png}
    \fi
      \end{center}
  \caption{Same as Figure~\ref{fig:2DA} but for the case $Y^{(2,2)}(\cdot)$. \label{fig:2DB}}
  \end{figure}

We simulated data in two dimensions from two underlying SIWGPs on $G = [-0.5, 0.5]^2$, $Y^{(2,1)}(\cdot)$ and $Y^{(2,2)}(\cdot)$. The first SIWGP, $Y^{(2,1)}(\cdot)$, was constructed from an AWU comprising 50 sigmoid functions in each dimension, and an SR-RBF(1) unit (hence $n = 11$), while the second SIWGP, $Y^{(2,2)}(\cdot)$, was constructed using the same warping functions as $Y^{(2,1)}(\cdot)$ with a M{\"o}bius transformation added to the last layer ($n = 12$). For both cases, the steepness parameters in the AWUs were set to $\theta_1 = 200$ in \eqref{eq:sigmoid}. The warping functions associated with the two SIWGPs are shown in the top-left panels of Figures~\ref{fig:2DA} and \ref{fig:2DB}, respectively.

Both $Y^{(2,1)}(\cdot)$ and $Y^{(2,2)}(\cdot)$ were modelled using $r_{n+1} = 400$ bisquare basis functions, $\sigma^2$ was set to 1, and $l$ to $0.04$. Two thousand points were then uniformly sampled from $G$ and used as observation locatons. Gaussian measurement error with variance $\sigma^2_\epsilon = 0.01$ was then added to the process at these locations to yield two simulation data sets with which to fit an SIWGP and SDSP. The simulated processes are shown in the bottom-left panels of Figures~\ref{fig:2DA} and \ref{fig:2DB}, respectively.

\subsubsection{Recovery of the underlying warping function}\label{sec:recovery}

We first fitted both an SIWGP and an SDSP to $Y^{(2,1)}(\cdot)$ and $Y^{(2,2)}(\cdot)$ using the same model that was used to generate the data. We used the same optimisation strategy as in Section~\ref{sec:1D} to estimate the weights and the parameters, but this time 400 steps in each stage were required to ensure convergence. Fitting and predicting with the SIWGP took around 1 minute, while the SDSP took around 4 minutes.

The SDSP variational posterior mean of the warping functions $\fvec(\cdot)$ for both case studies are shown in the top-right panels of Figures~\ref{fig:2DA} and \ref{fig:2DB}, respectively. In both cases the estimated warpings are, up to a rotation, remarkably similar. The predictions faithfully reproduce the true process, although this is somewhat expected from 2000 points in two dimensions. What is strikingly different from standard Gaussian-process regression (even when anisotropic covariance functions are used) is the prediction standard error map, where the uncertainty is high in areas of high process variability, and where the uncertainty `contours' follow those of the underlying process. Indeed, the deep spatial models thus seem to provide a better representation of the underlying data-generating process. We will also observe this when we analyise radiances from the MODIS instrument in Section \ref{sec:L1}. As in Section~\ref{sec:1D}, we did not observe a material difference between the out-of-sample predictions of the SIWGP and the SDSP in terms of MAPE, RMSPE, CRPS, and IS.

\subsubsection{Impact of using different model architectures}\label{sec:different-architectures}

\begin{figure}[t!]
    \ifarxiv
    \includegraphics[width = \textwidth]{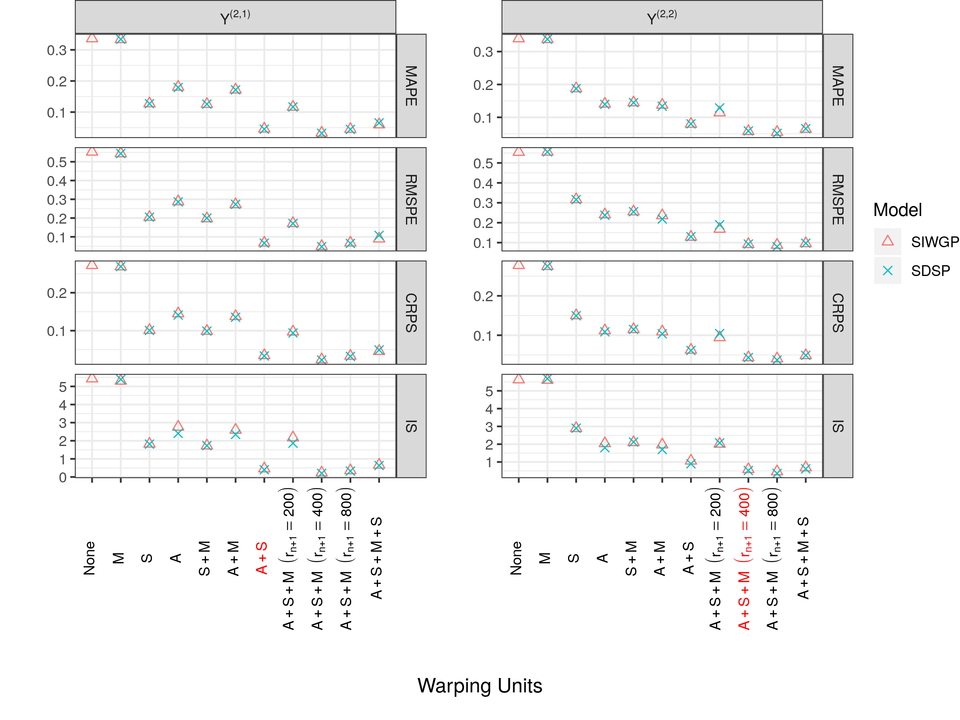}
    \else
    \includegraphics[width = \textwidth]{../deep_spat_src_blinded_for_submission/2Dtests/img/Exp2_all_results.png}
    \fi
  \caption{Diagnostic results from using various models to predict $Y^{(2,1)}(\cdot)$ (left panel) and $Y^{(2,2)}(\cdot)$ (right panel) from noisy data. The different models have different warping units (M -- M\"obius, S -- SR-RBF(1), A -- two AWUs, one for each dimension) and, unless otherwise stated, $r_{n+1} = 400$. For example, M\,+\,S, has a warping function composed of a M\"obius transformation and an SR-RBF(1) unit. The true model for each experiment is identified using a red font colour.}\label{fig:warping_models}
 \end{figure}

We next fitted SIWGPs and SDSPs with different architectures (i.e., different warping units and different number of layers and basis functions) to $Y^{(2,1)}(\cdot)$ and $Y^{(2,2)}(\cdot)$. The models we considered together with the corresponding prediction diagnostics are summarised in Figure~\ref{fig:warping_models}, from which we note the following. First, it is apparent that, for both data sets, using an underlying warping function generally improves the predictive performance, sometimes considerably so. Second, some warping units can be more important than others; for example, the improvement obtained from using a M{\"o}bius transformation is relatively small for both these data sets. Third, extra model complexity, both in terms of the number of warping units and $r_{n+1}$, does not substantially compromise the prediction performance. Unlike what we show in our one-dimensional example of Section~\ref{sec:1D-stat-experiment} \ifarxiv (supplementary material)\else of the online supplementary material\fi, the SIWGP does not over-fit here, although we expect that it will do so if the complexity of the warping function is increased further. Finally, a value for $r_{n+1}$ that is too small can adversely affect the predictions. This is a result of over-smoothing, a problem commonly observed in these low-rank models \citep[e.g.,][]{Zammit_2018}.

These results suggest that the problem of model (or architecture) selection can be approached by fitting several models with different warping functions and number of basis functions in the top layer, and choosing the one that gives the best out-of-sample performance in terms of some model selection criterion. Such a strategy is feasible in practice since the above models each only took between one and a few minutes to fit and predict with. We provide more discussion on the issue of architecture choice in Section~\ref{sec:Conclusion}.

\subsection{Experiment using MODIS L1B radiances}\label{sec:L1}

This experiment assesses the utility of the SDSP in an applied setting. Data for this experiment were obtained from spatial calibrated L1B radiances at a 500 m resolution from the Moderate Resolution Imaging Spectroradiometer (MODIS) instrument aboard the Aqua satellite \citep{MODIS}. In the product, radiances in units of W/m$^2$/\SI{}{\micro\metre}/st are provided for 36 bands in the \SI{0.4}{\micro\metre} to \SI{14.4}{\micro\metre} region of the electromagnetic spectrum. Here we consider the third of these bands, ranging from 0.459 \SI{}{\micro\metre} to 0.479 \SI{}{\micro\metre}, which is within the visible spectrum.  %https://modis.gsfc.nasa.gov/about/specifications.php
\Copy{Motivation}{The data we use have high signal-to-noise ratio and are complete. Therefore, although spatial prediction is not required for gap-filling, the data are ideally suited for comparatively validating models in a realistic setting.}

The L1B product is composed of several scenes that are of size 2708 $\times$ 4060 pixels. Since these scenes are at a very high resolution we first regridded them into scenes of size 136 $\times$ 203 that are at a 10 km resolution. From these we then sampled 4000 grid cells at random to make up our observed data set; the other $23,608$ grid cells were left for out-of-sample validation. The goal is to assess the performance of the deep models in predicting these out-of-sample data from the 4000 `observed' data.

\begin{figure}[t!]
  \includegraphics[width=0.49\textwidth]{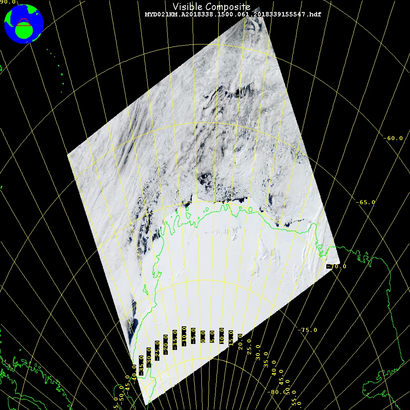}
  \includegraphics[width=0.49\textwidth]{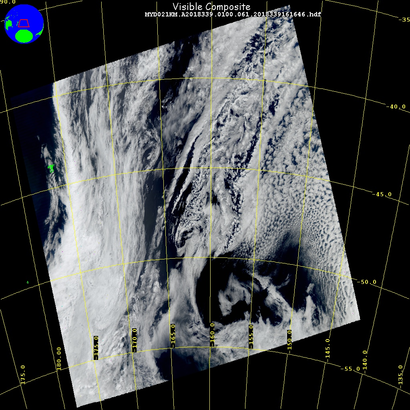}
  \caption{Calibrated L1B radiances in the \SI{0.459}{\micro\metre} to \SI{0.479}{\micro\metre} band from the MODIS instrument aboard the Aqua satellite. Longitudinal and latitudinal lines are shown in yellow while land boundaries are shown in green. The inset in the top-left corner of each image shows the location of the scene on the globe. The images were taken on 04 December 2018 15:00 UTC (left panel) and 05 December 2018 01:00 UTC (right panel), respectively. The images are courtesy of NASA (Source: \url{https://ladsweb.modaps.eosdis.nasa.gov}).}\label{fig:L1}
 \end{figure}

In the course of our study we found that the SDSPs perform as well as stationary Gaussian-process models when there is no clear covariance nonstationarity, or only mild covariance nonstationarity, in the visible image. On the other hand, improvement in predictive performance could be achieved when there was clear covariance nonstationarity. In this section we present two scenes, shown in Figure~\ref{fig:L1}, where we found that catering for a high degree of covariance nonstationarity proved beneficial. The first scene is from 04 December 2018 15:00 UTC over Antarctica, with radiances being detected from both ice and clouds. While radiances from the ice regime are almost spatially constant, those from the clouds are clearly more variable, and exhibit spatially-varying anisotropy. The second scene is from 05 December 2018 01:00 UTC in the South Pacific, just east of New Zealand. Here the radiances are predominantly from clouds that exhibit complex covariance nonstationarity.

\begin{figure}[t!]
  \begin{center}
  \ifarxiv  
  \includegraphics[width=0.47\textwidth]{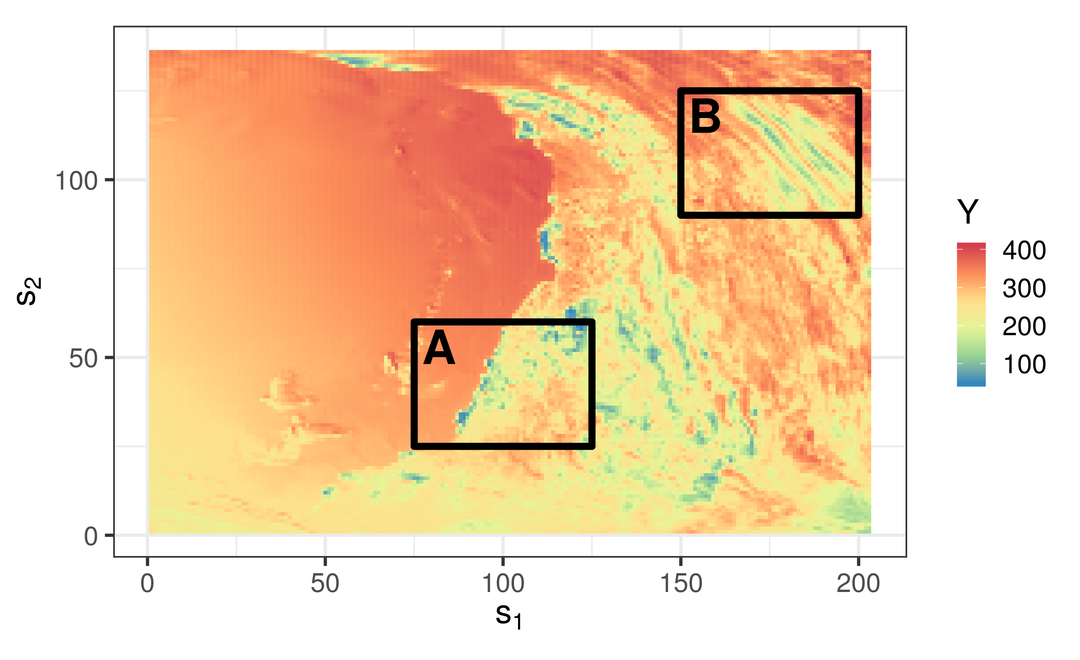}
  \includegraphics[width=0.47\textwidth]{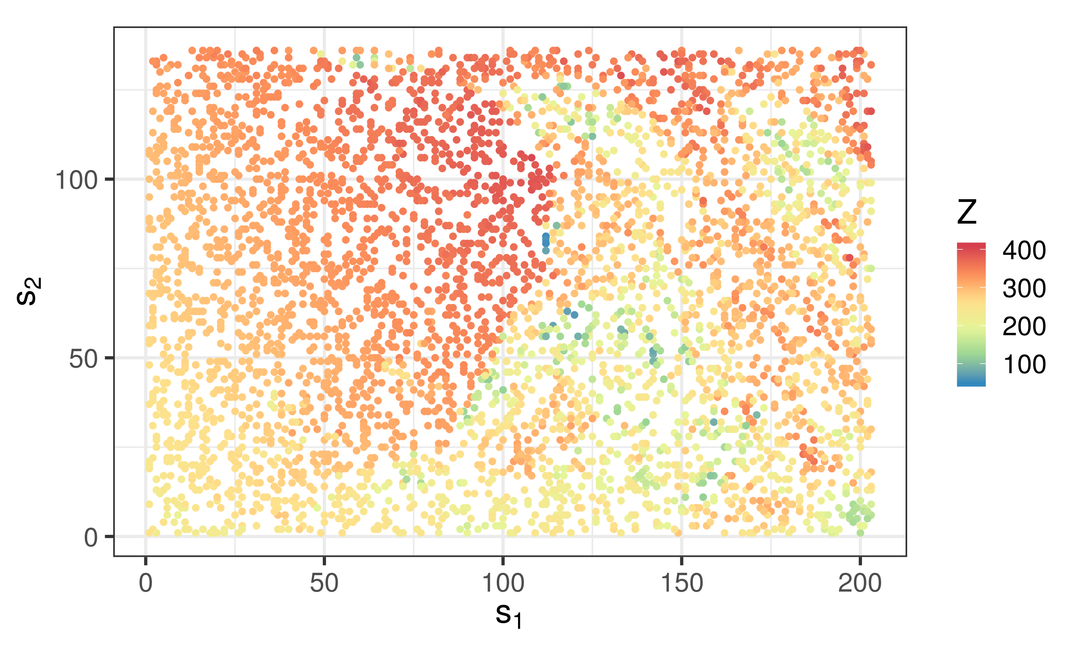}\\
  \includegraphics[width=0.47\textwidth]{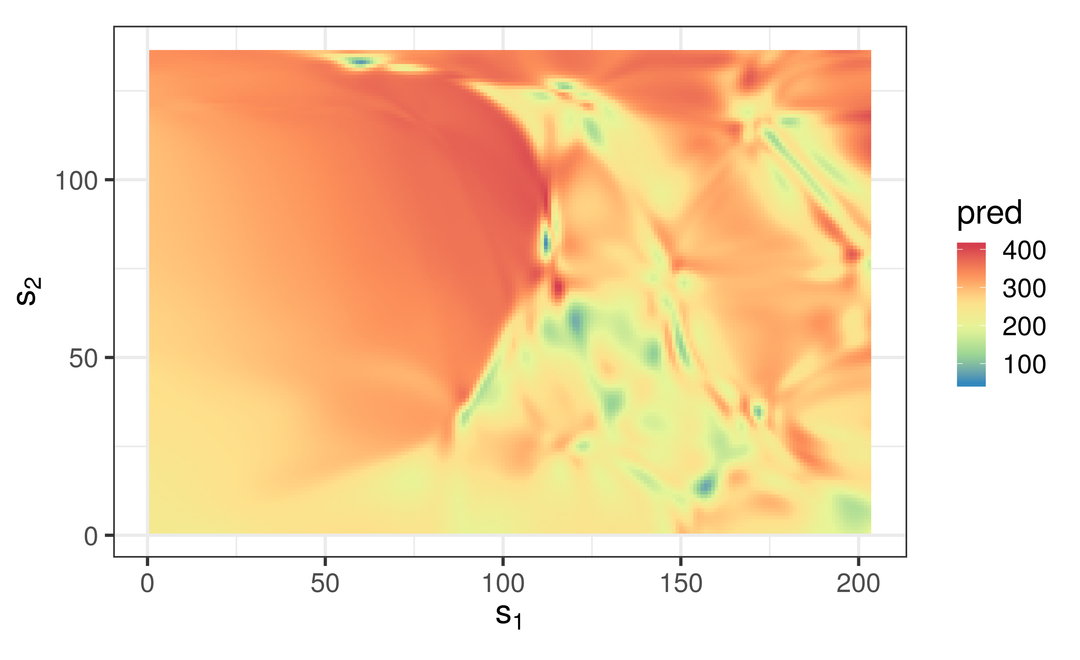}
  \includegraphics[width=0.47\textwidth]{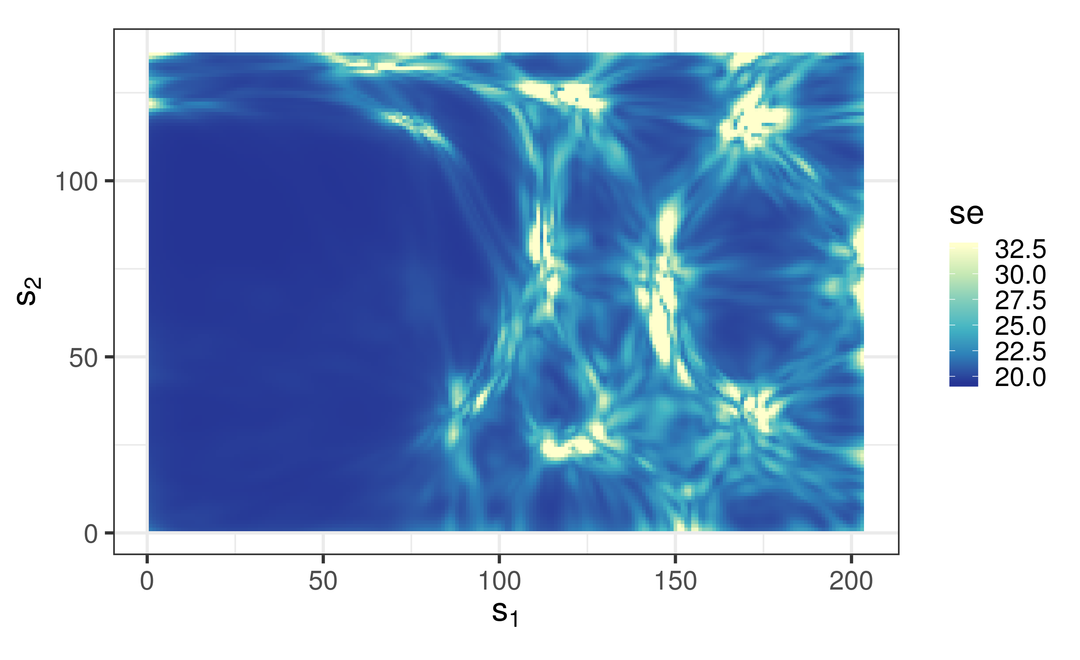}\\
  \includegraphics[width=0.47\textwidth]{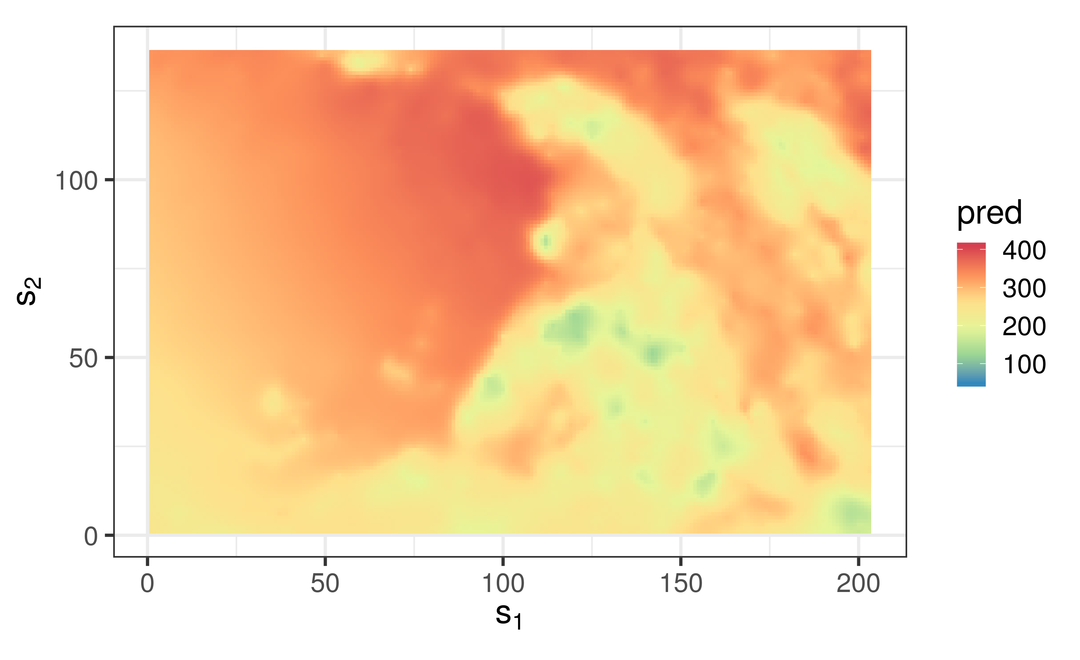}
  \includegraphics[width=0.47\textwidth]{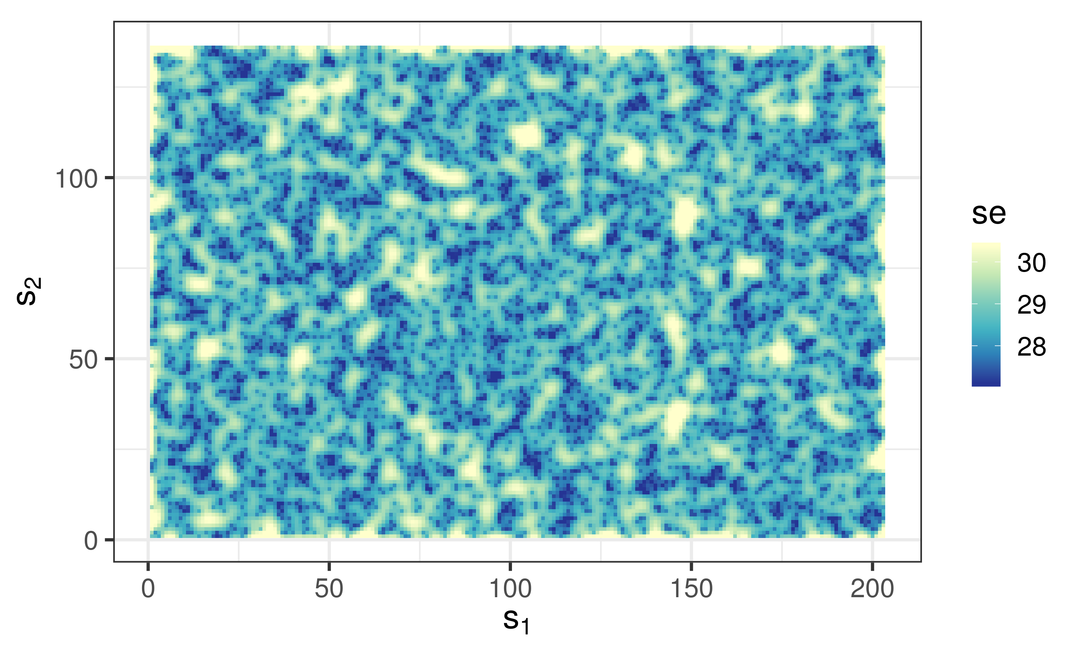}
  \else
  \includegraphics[width=0.47\textwidth]{../deep_spat_src_blinded_for_submission/MODIStest/img/MODIS1truth.png}
  \includegraphics[width=0.47\textwidth]{../deep_spat_src_blinded_for_submission/MODIStest/img/MODIS1obs.png}\\
  \includegraphics[width=0.47\textwidth]{../deep_spat_src_blinded_for_submission/MODIStest/img/MODIS1pred_SDSP.png}
  \includegraphics[width=0.47\textwidth]{../deep_spat_src_blinded_for_submission/MODIStest/img/MODIS1se_SDSP.png}\\
  \includegraphics[width=0.47\textwidth]{../deep_spat_src_blinded_for_submission/MODIStest/img/MODIS1pred_krige.png}
  \includegraphics[width=0.47\textwidth]{../deep_spat_src_blinded_for_submission/MODIStest/img/MODIS1se_krige.png}
  \fi
  \end{center}
  \caption{Results corresponding to the scene shown in the left panel of Figure~\ref{fig:L1}. (Top-left) The scene regridded to a lower resolution, with black squares marking insets A and B shown in more detail in the top two rows of Figure~\ref{fig:insets}. (Top-right) The observations, uniformly sampled from the gridded data. (Centre) The prediction (left) and prediction standard error (right) when modelling using the SDSP. (Bottom) The prediction (left) and prediction standard error (right) when using Gaussian-process regression with stationary, isotropic exponential covariance function. All values shown are in units of W/m$^2$/\SI{}{\micro\metre}/st.}\label{fig:MODIS1}
 \end{figure}

For this study we considered an SDSP with two AWUs (one for each spatial dimension) composed of 50 sigmoid functions each (two layers), a multi-resolution RBF unit consisting of SR-RBF(1) and SR-RBF(2) units (9 and 81 layers, respectively), and a M{\"o}bius transformation (one layer) for a total of $n = 93$ layers. We used $r_{n+1} = 1600$ bisquare basis functions in the process layer. It took approximately 14 minutes to fit and predict with the SDSP.  We compared the SDSP to standard Gaussian process regression with a stationary, isotropic exponential covariance function, where the covariance function was fitted using standard variogram techniques. Fitting and prediction with \texttt{gstat} \citep{Pebesma_2004} required approximately three minutes. We also compared it to a shallower version of the SDSP containing only an SR-RBF(1) unit (SDSP-RBF), and to an $n = 0$ shallow model where $r_1 = 1600$ bisquare basis functions. The latter $n = 0$ model is identical to that considered by \cite{Zammit_2018}, and we thus denote it as the `FRK' (short for Fixed Rank Kriging) model. We employ the FRK model so that we can assess the benefit of including a warping function when the number of basis functions in the top layer is fixed by design (e.g., because of computational requirements). The SDSP-RBF and FRK models required approximately eight minutes and one minute, respectively, to fit and predict with.

\begin{figure}[t!]
  \begin{center}
    \ifarxiv
    \includegraphics[width=0.47\textwidth]{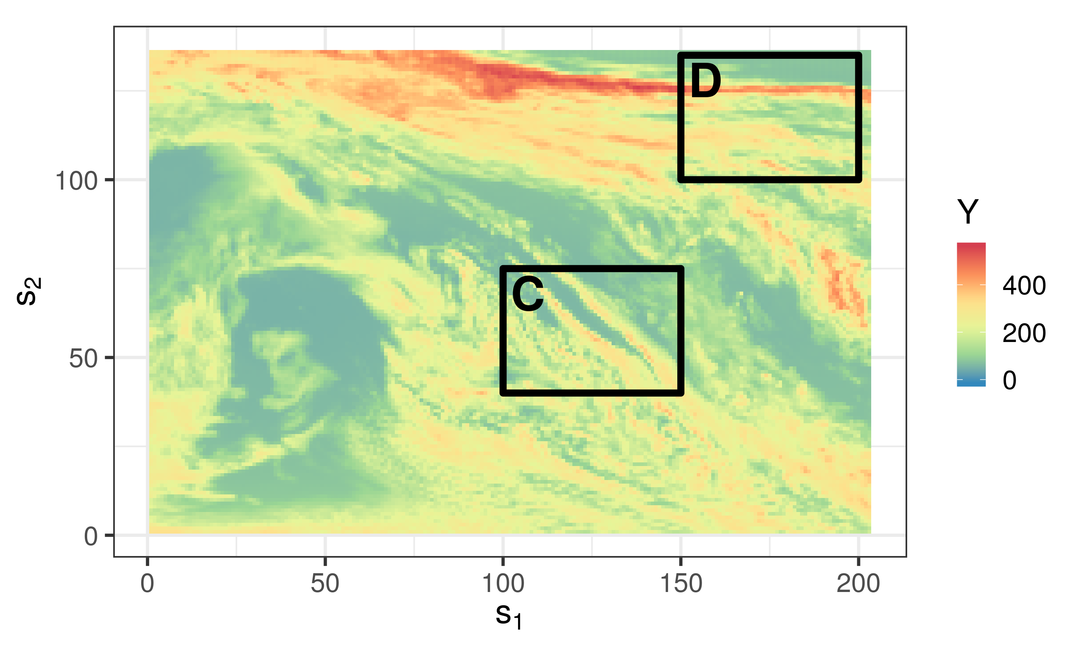}
  \includegraphics[width=0.47\textwidth]{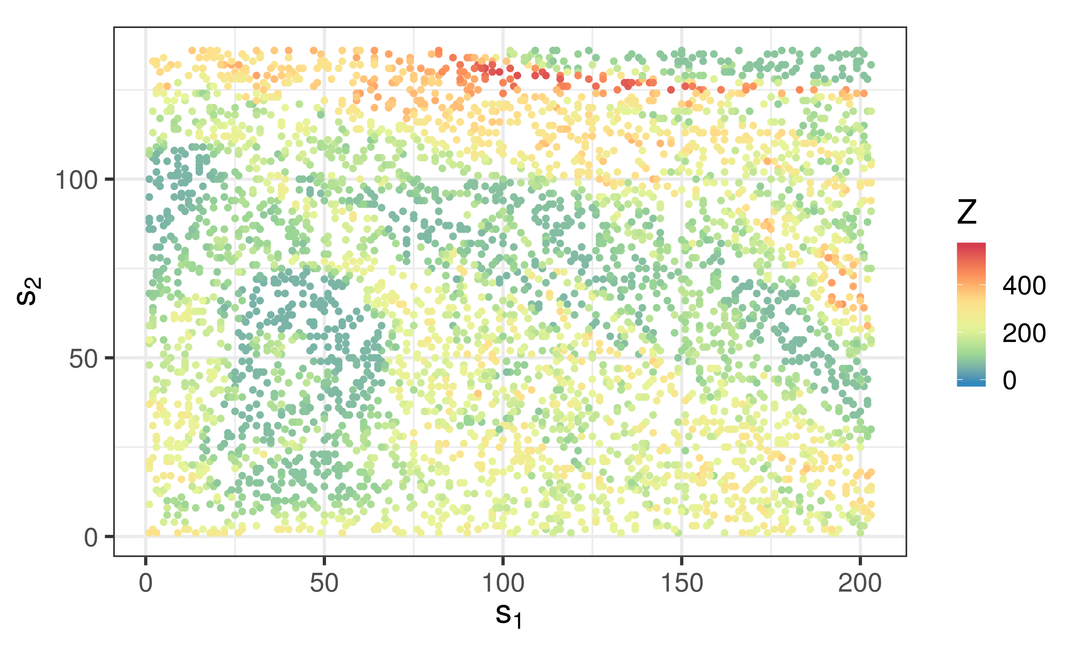}\\
  \includegraphics[width=0.47\textwidth]{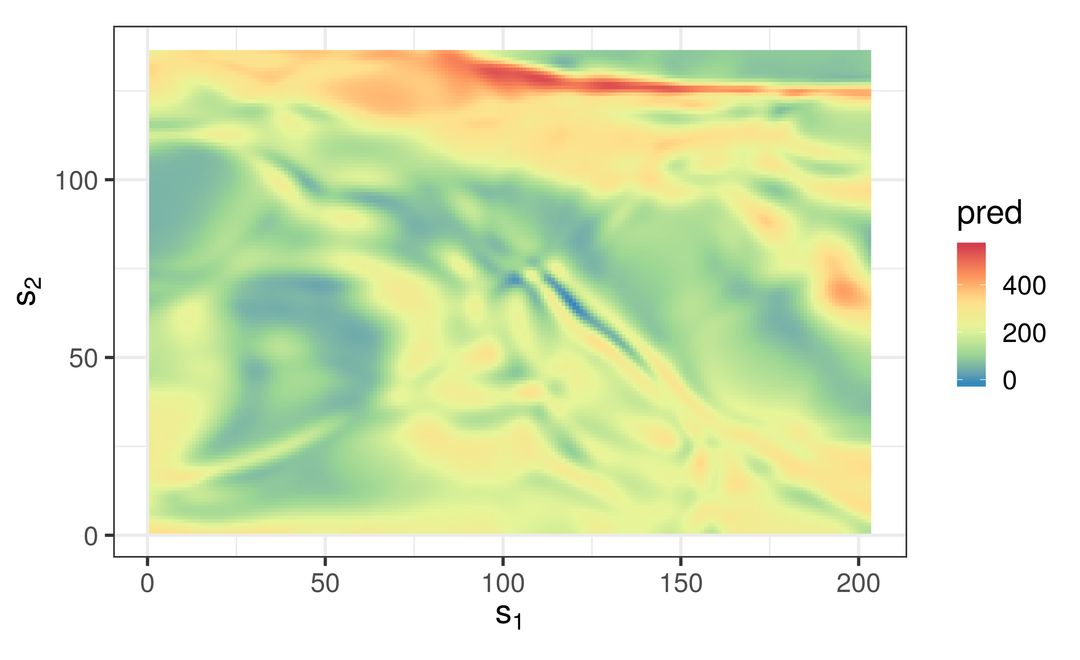}
  \includegraphics[width=0.47\textwidth]{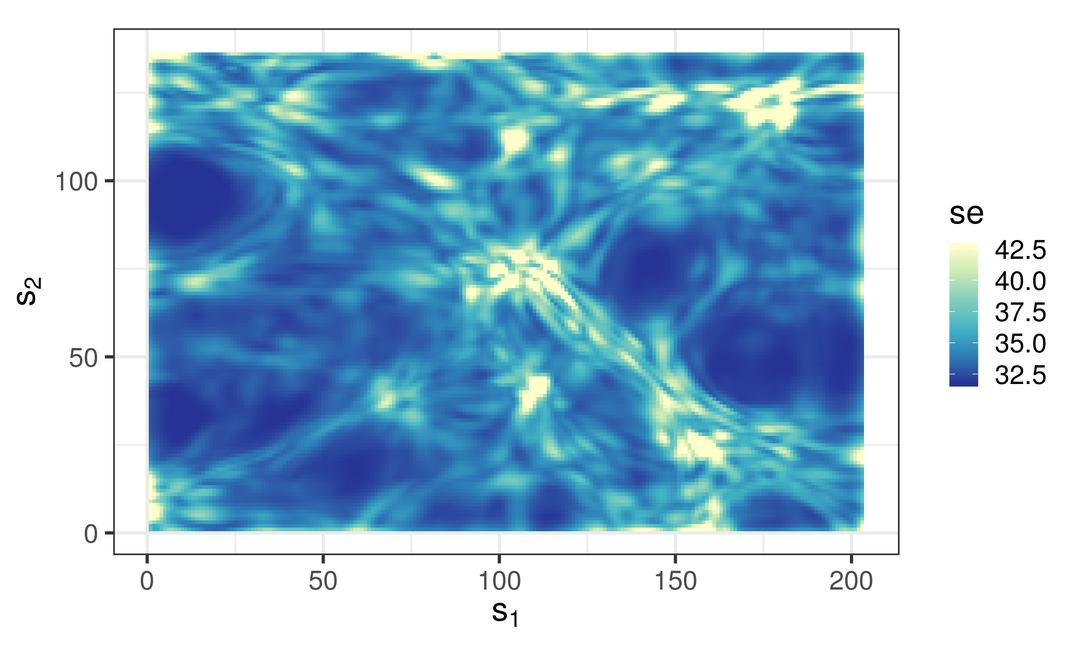}\\
  \includegraphics[width=0.47\textwidth]{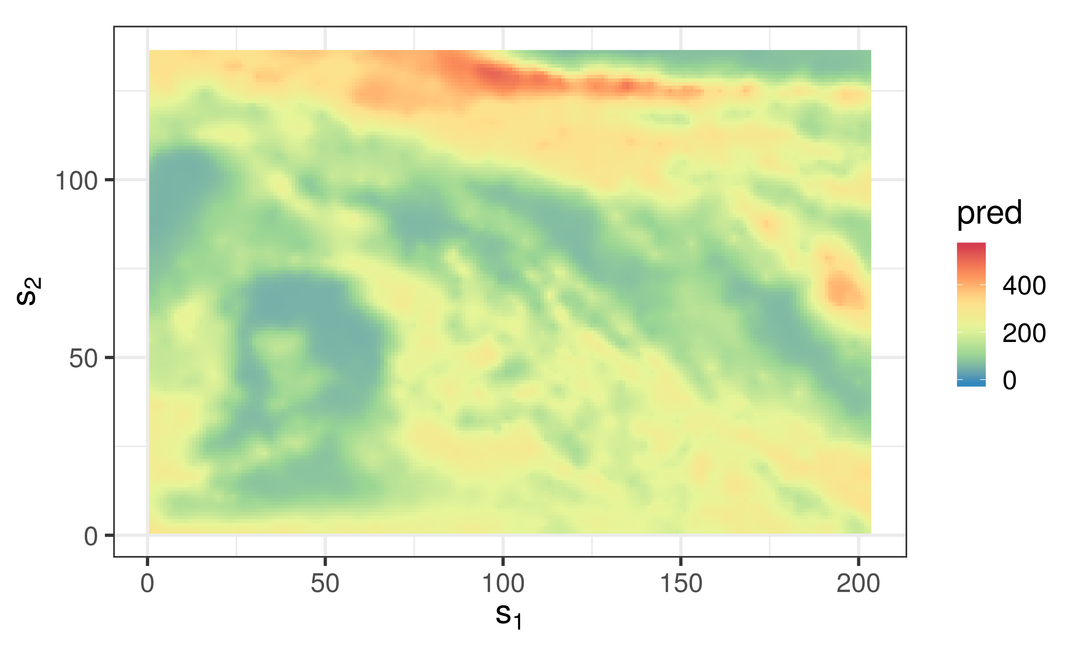}
  \includegraphics[width=0.47\textwidth]{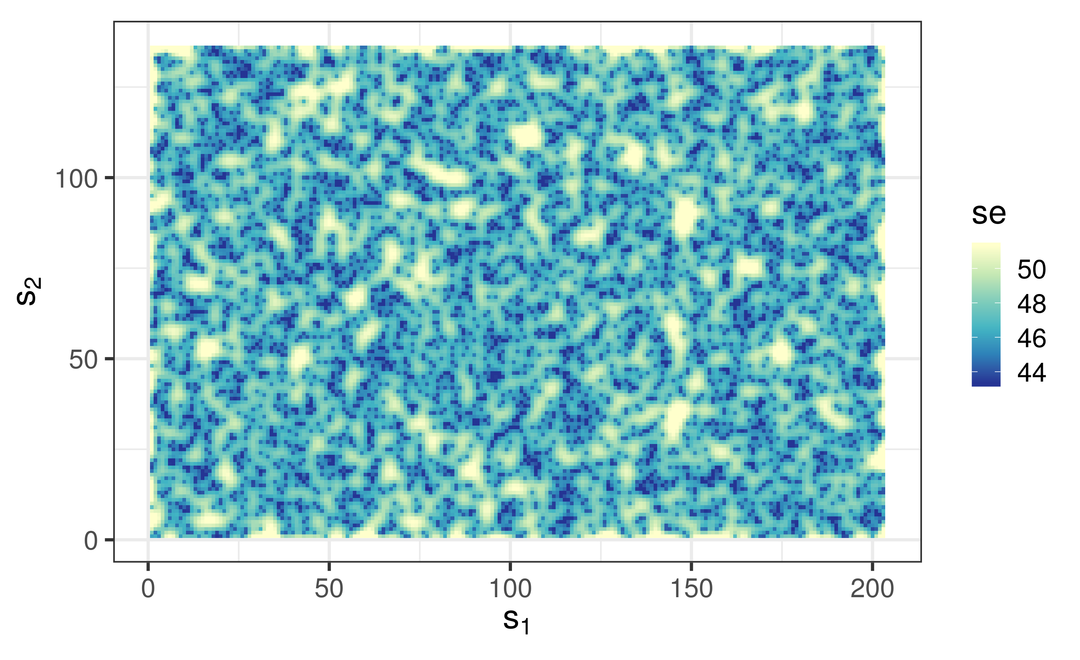}
  \else
  \includegraphics[width=0.47\textwidth]{../deep_spat_src_blinded_for_submission/MODIStest/img/MODIS2truth.png}
  \includegraphics[width=0.47\textwidth]{../deep_spat_src_blinded_for_submission/MODIStest/img/MODIS2obs.png}\\
  \includegraphics[width=0.47\textwidth]{../deep_spat_src_blinded_for_submission/MODIStest/img/MODIS2pred_SDSP.png}
  \includegraphics[width=0.47\textwidth]{../deep_spat_src_blinded_for_submission/MODIStest/img/MODIS2se_SDSP.png}\\
  \includegraphics[width=0.47\textwidth]{../deep_spat_src_blinded_for_submission/MODIStest/img/MODIS2pred_krige.png}
  \includegraphics[width=0.47\textwidth]{../deep_spat_src_blinded_for_submission/MODIStest/img/MODIS2se_krige.png}
  \fi
  \end{center}
  \caption{Results corresponding to the scene shown in the right panel of Figure~\ref{fig:L1}. (Top-left) The scene regridded to a lower resolution, with black squares marking insets C and D shown in more detail in the bottom two rows of Figure~\ref{fig:insets}. (Top-right) The observations, uniformly sampled from the gridded data. (Centre) The prediction (left) and prediction standard error (right) when modelling using the SDSP. (Bottom) The prediction (left) and prediction standard error (right) when using Gaussian-process regression with stationary, isotropic exponential covariance function. All values shown are in units of W/m$^2$/\SI{}{\micro\metre}/st. \label{fig:MODIS2}}
\end{figure}

\begin{table}[t!]
  \caption{Diagnostic Results for the Two MODIS Scenes.%Diagnostic results for the two scenes. The mean absolute prediction error (MAPE), root-mean-squared prediction error (RMSPE), continuous-ranked probability score (CRPS) and interval score at the 5\% level (IS) of the various models listed in the main text. The diagnostics are computed from the out-of-sample validation data and the predictive distributions associated with these data.\vspace{0.1in}
    \label{tab:MODIS}}
  \begin{center}
\begin{tabular}{clcccc}
Scene & Model & MAPE & RMSPE & CRPS & IS \\ 
  \hline

1 & FRK & 17.17 & 29.08 & 14.72 & 187.16 \\ 
  & GP & 15.66 & 26.95 & 13.77 & 171.78 \\ 
  & SDSP & 15.36 & 25.98 & 12.55 & 156.72 \\ 
  & SDSP-RBF & 16.34 & 27.79 & 13.97 & 178.87  \vspace{0.1in}\\       

2 & FRK & 29.20 & 40.62 & 21.94 & 225.11 \\ 
  & GP & 27.76 & 39.10 & 21.47 & 224.53 \\ 
  & SDSP & 26.53 & 36.40 & 19.65 & 195.72 \\ 
  & SDSP-RBF & 27.80 & 38.38 & 20.81 & 209.19 

\end{tabular}
\end{center}
\end{table}

In Figure~\ref{fig:MODIS1} we show the full data set, the observations that were used for making inference, and the predictions and prediction standard errors from the SDSP and Gaussian process. The SDSP adapts to the different regimes of ice and cloud, providing an almost constant prediction over the ice coupled with very low prediction variance, and spatially-varying anisotropy over the region containing clouds. The inferred spatially-varying anisotropy is apparent in regions of high variability from the map of prediction standard errors. Gaussian process regression, as expected, smooths out most of the salient features that could be extracted from the data shown in the top-right panel. Furthermore, the prediction standard-error maps are reflective of the stationarity assumption, with no distinction made between the ice--cloud regimes. Similar conclusions can be drawn from Figure~\ref{fig:MODIS2}, where complex spatially-varying anisotropy predominant in the second scene is correctly captured by the SDSP.

Diagnostics for the two scenes and the models we considered are shown in Table~\ref{tab:MODIS}. For these two scenes, the improvement in prediction accuracy and uncertainty quantification of the SDSP over the GP, using all the diagnostics we considered, is on the order of 5--10\%. The improvement of the SDSP over the FRK model is even greater, suggesting that the inclusion of the warping layer in several of the low-rank models used in spatial statistics may be especially beneficial.

\begin{figure}[t!]
  \begin{center}
  \ifarxiv
  \begin{tabular}{m{1cm}m{4cm}m{4cm}m{4cm}}
  A &
  \includegraphics[width=0.25\textwidth]{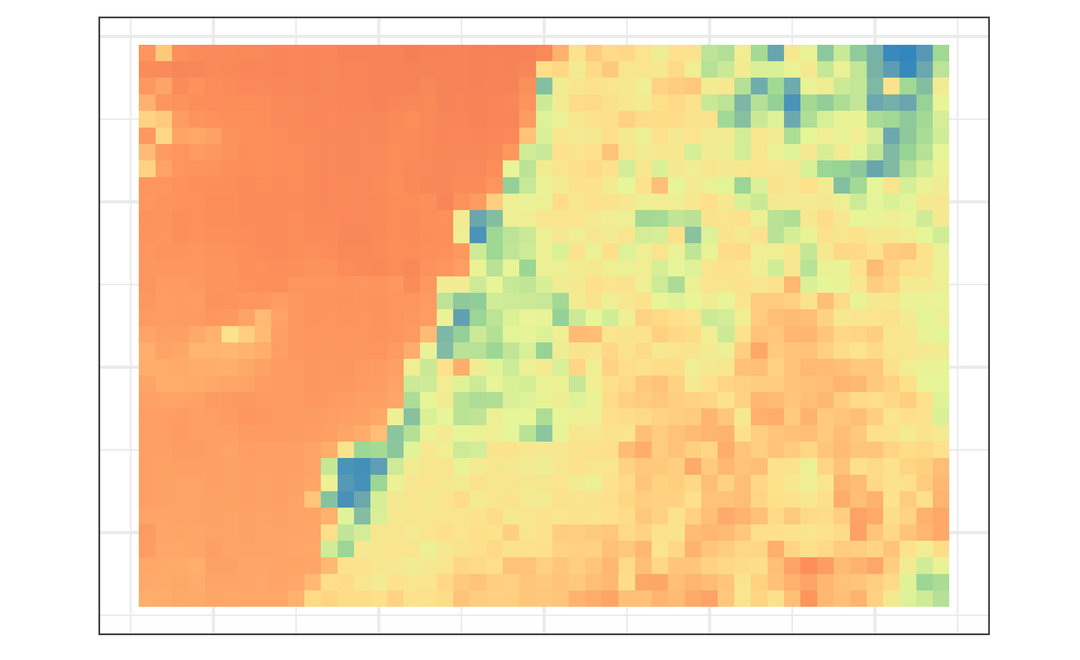}&
  \includegraphics[width=0.25\textwidth]{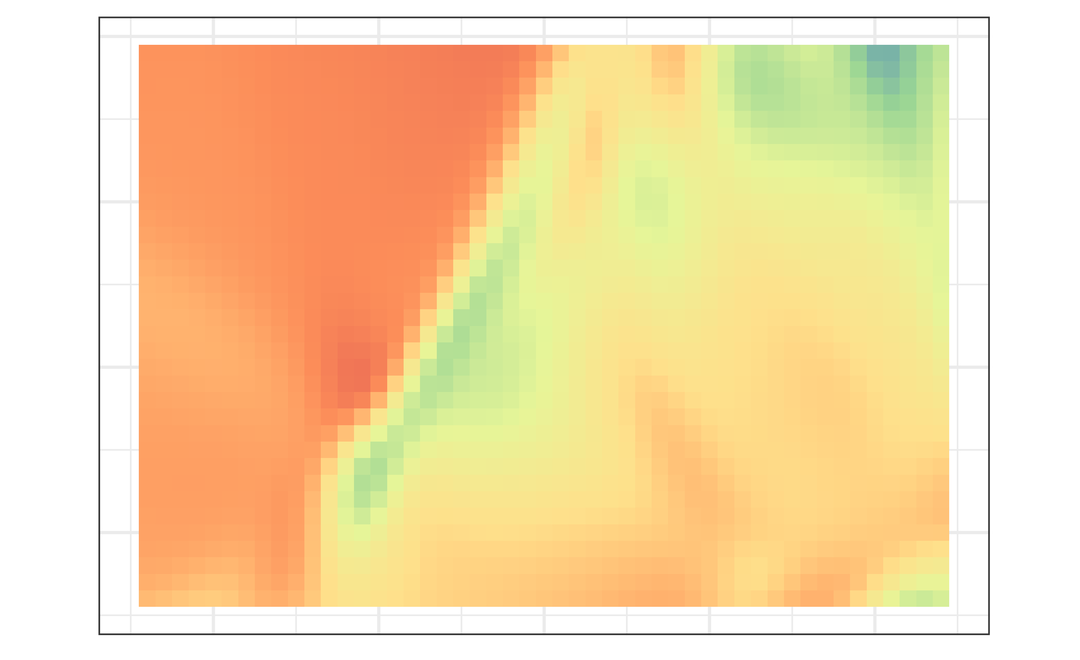}&
  \includegraphics[width=0.25\textwidth]{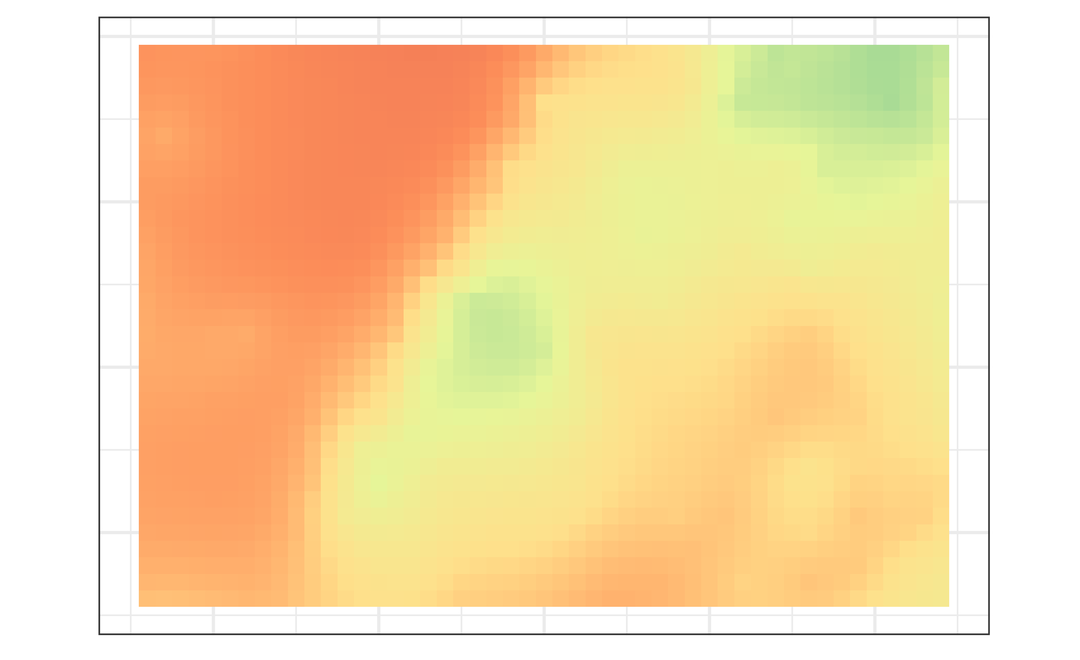}\\
  B &
  \includegraphics[width=0.25\textwidth]{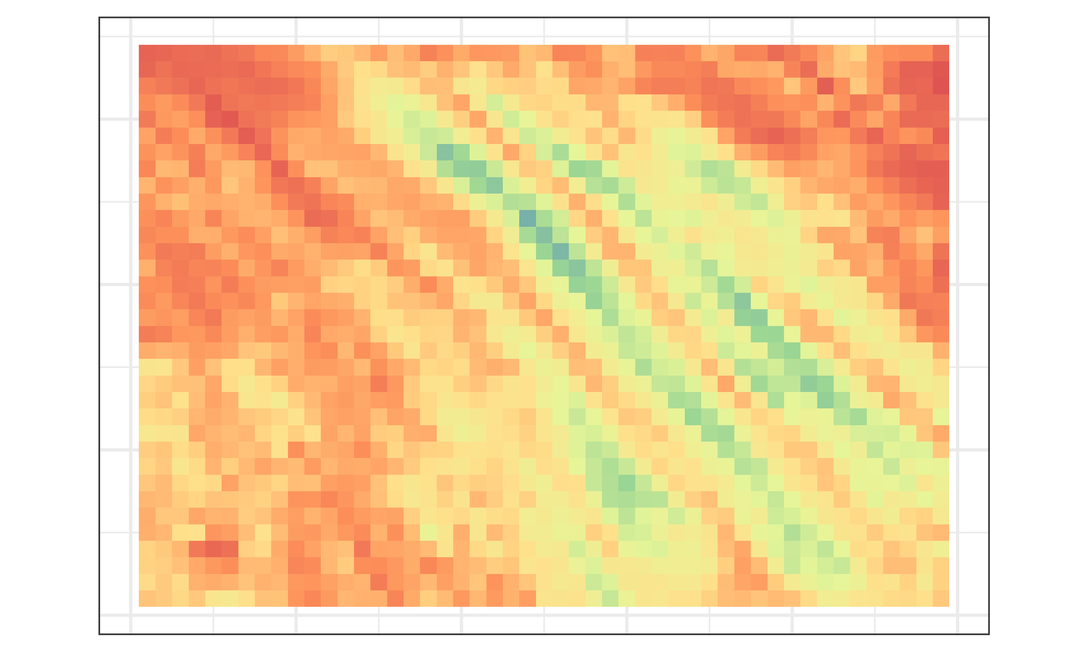}&
  \includegraphics[width=0.25\textwidth]{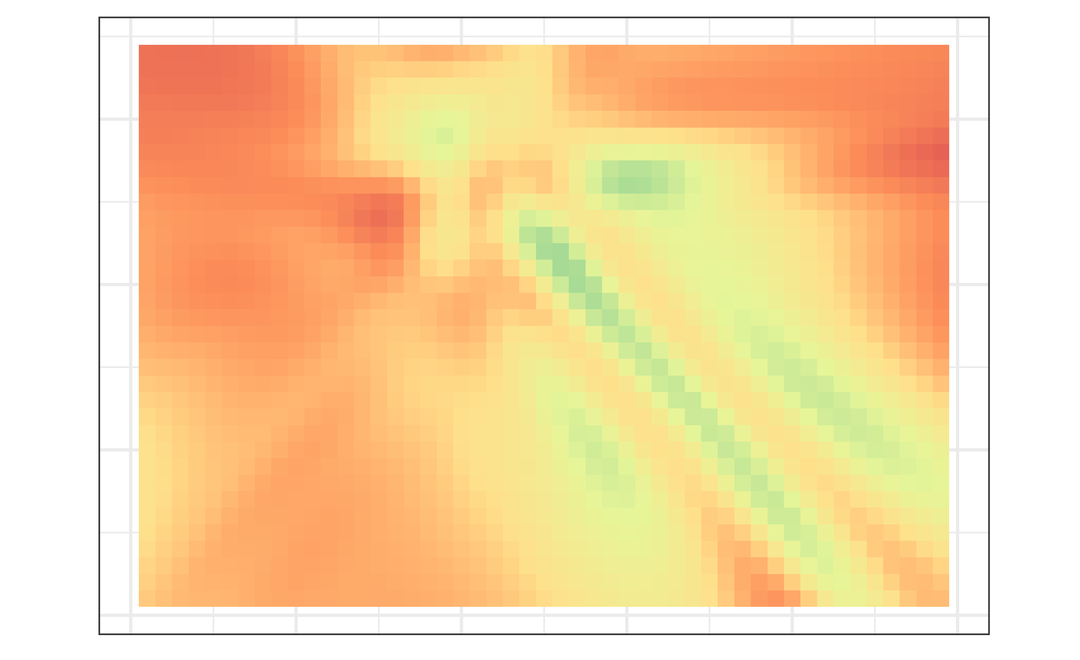}&
  \includegraphics[width=0.25\textwidth]{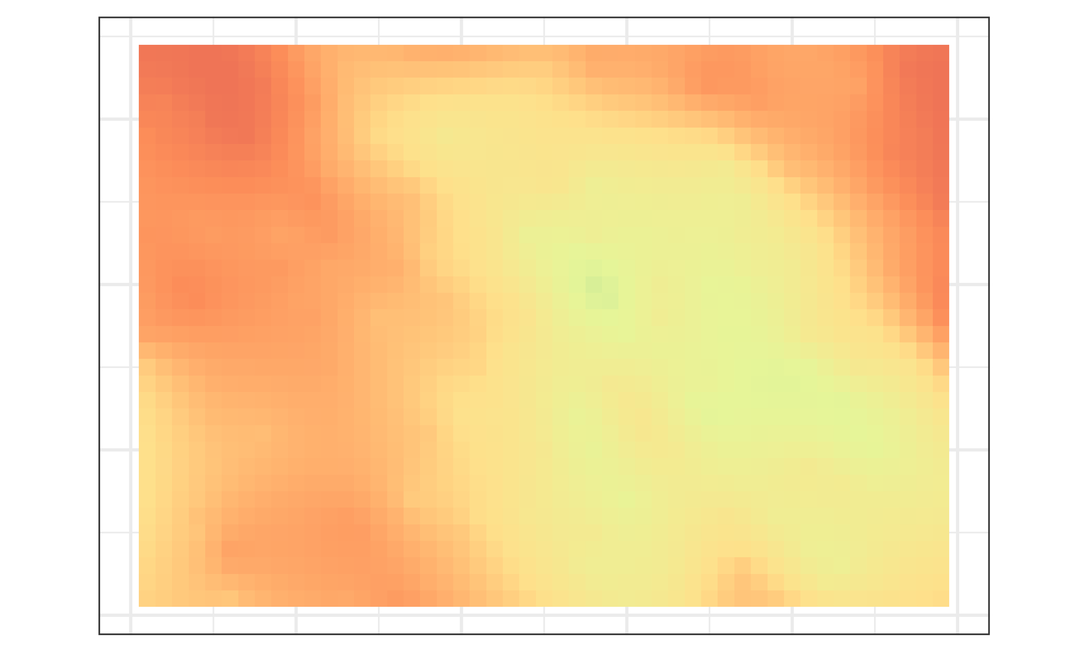}\\
  C &
  \includegraphics[width=0.25\textwidth]{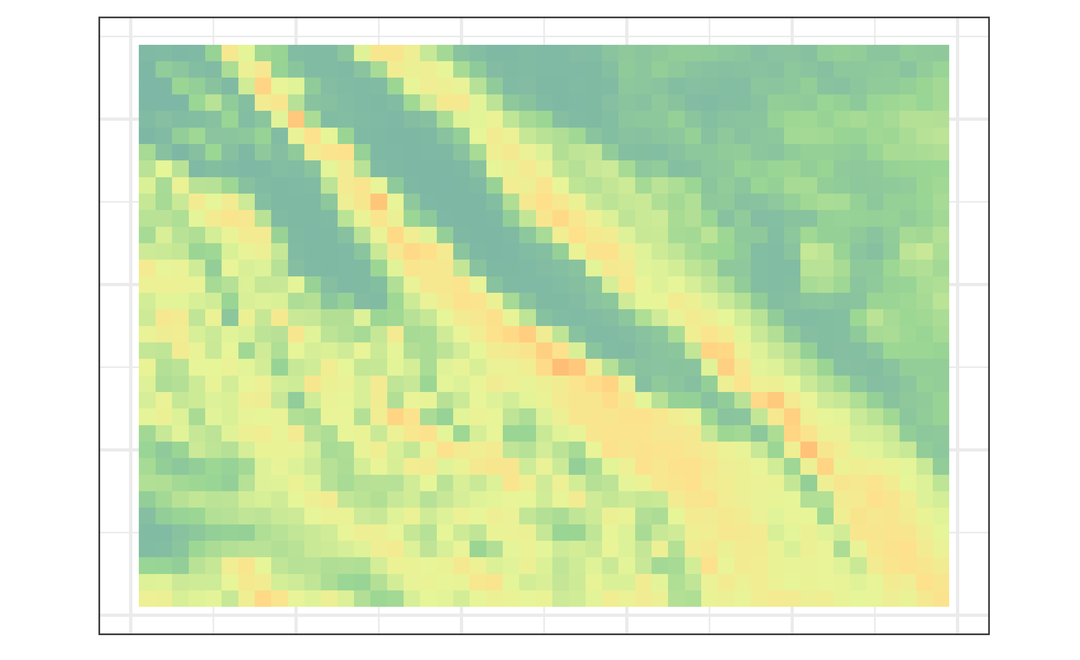}&
  \includegraphics[width=0.25\textwidth]{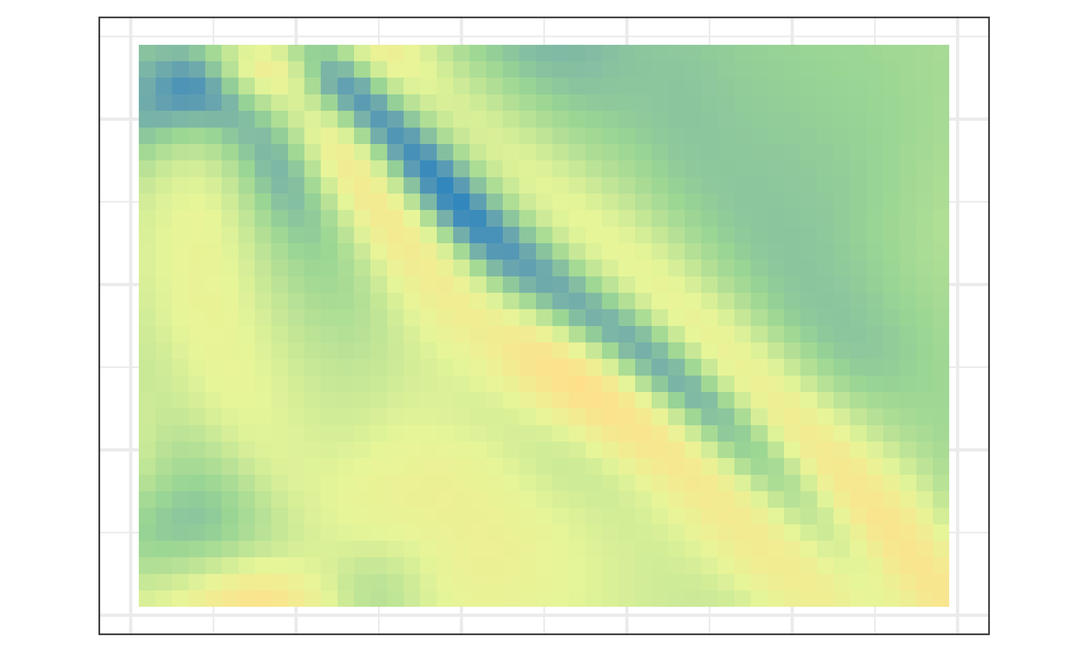}&
  \includegraphics[width=0.25\textwidth]{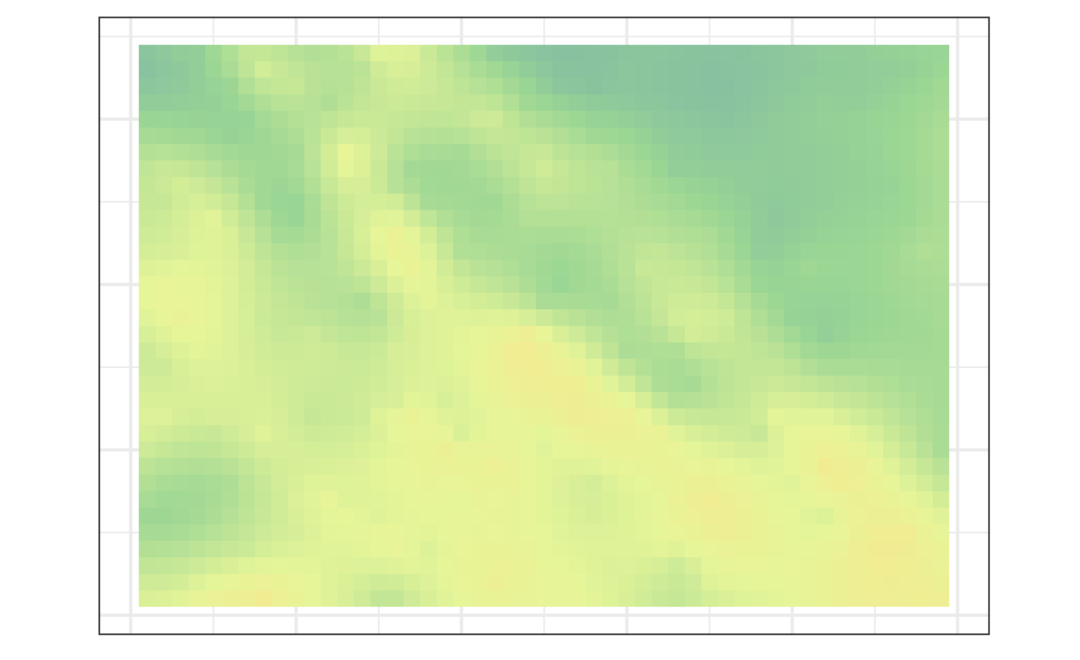}\\
  D &
  \includegraphics[width=0.25\textwidth]{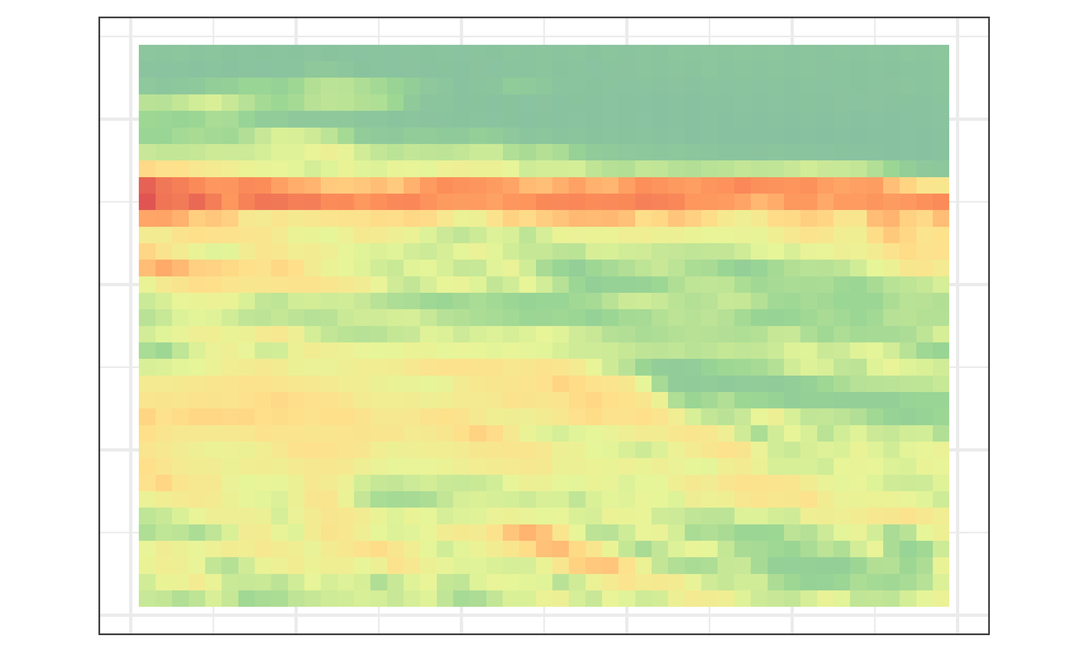}&
  \includegraphics[width=0.25\textwidth]{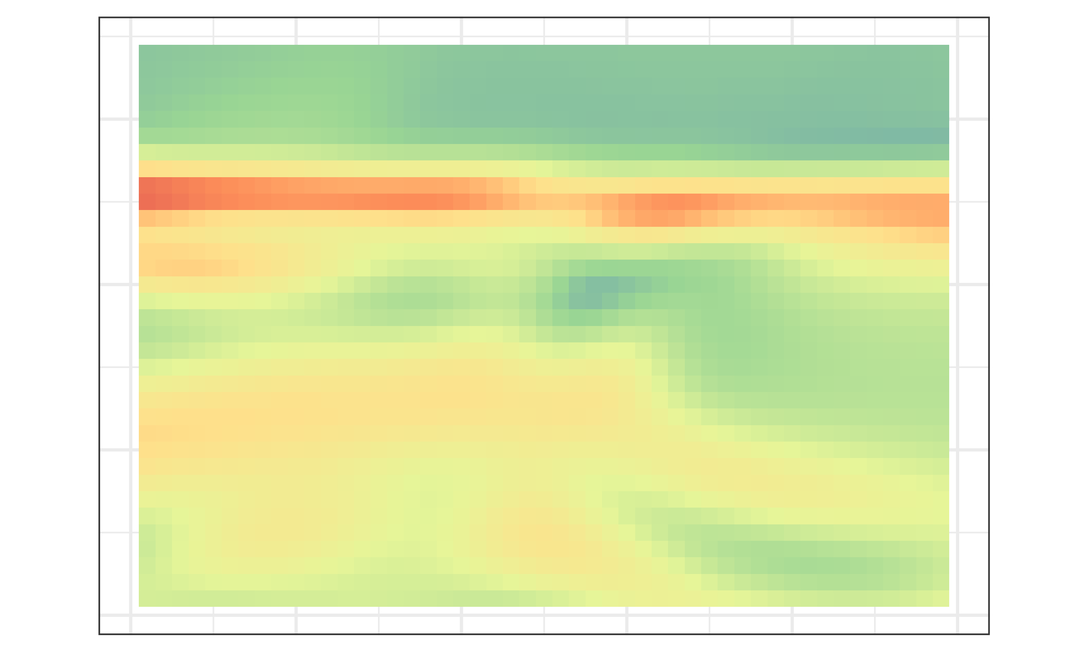}&
  \includegraphics[width=0.25\textwidth]{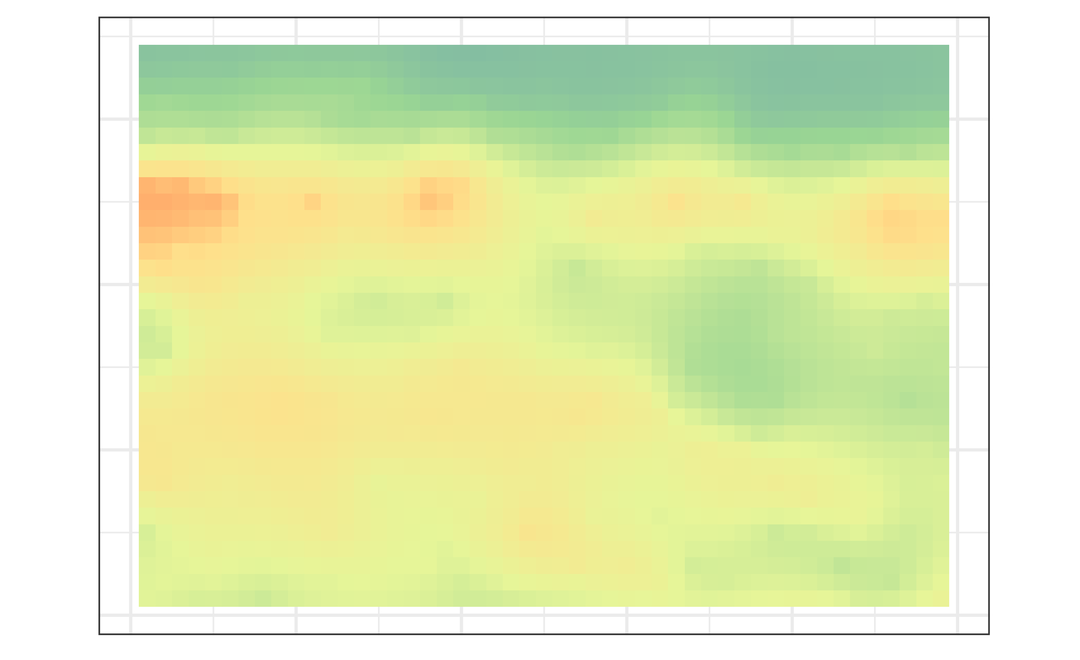}
  \end{tabular}
  \else
    \begin{tabular}{m{1cm}m{4cm}m{4cm}m{4cm}}
  A &
  \includegraphics[width=0.25\textwidth]{../deep_spat_src_blinded_for_submission/MODIStest/img/MODIS1zoom1_true.png}&
  \includegraphics[width=0.25\textwidth]{../deep_spat_src_blinded_for_submission/MODIStest/img/MODIS1zoom1_SDSP.png}&
  \includegraphics[width=0.25\textwidth]{../deep_spat_src_blinded_for_submission/MODIStest/img/MODIS1zoom1_krige.png}\\
  B &
  \includegraphics[width=0.25\textwidth]{../deep_spat_src_blinded_for_submission/MODIStest/img/MODIS1zoom2_true.png}&
  \includegraphics[width=0.25\textwidth]{../deep_spat_src_blinded_for_submission/MODIStest/img/MODIS1zoom2_SDSP.png}&
  \includegraphics[width=0.25\textwidth]{../deep_spat_src_blinded_for_submission/MODIStest/img/MODIS1zoom2_krige.png}\\
  C &
  \includegraphics[width=0.25\textwidth]{../deep_spat_src_blinded_for_submission/MODIStest/img/MODIS2zoom1_true.png}&
  \includegraphics[width=0.25\textwidth]{../deep_spat_src_blinded_for_submission/MODIStest/img/MODIS2zoom1_SDSP.png}&
  \includegraphics[width=0.25\textwidth]{../deep_spat_src_blinded_for_submission/MODIStest/img/MODIS2zoom1_krige.png}\\
  D &
  \includegraphics[width=0.25\textwidth]{../deep_spat_src_blinded_for_submission/MODIStest/img/MODIS2zoom2_true.png}&
  \includegraphics[width=0.25\textwidth]{../deep_spat_src_blinded_for_submission/MODIStest/img/MODIS2zoom2_SDSP.png}&
  \includegraphics[width=0.25\textwidth]{../deep_spat_src_blinded_for_submission/MODIStest/img/MODIS2zoom2_krige.png}
  \end{tabular}
  \fi
  \end{center}
  \caption{Insets corresponding to those marked in Figure~\ref{fig:MODIS1} (top two rows) and in Figure~\ref{fig:MODIS2} (bottom two rows), labelled A--D. The first column of images shows true values, the second column shows the prediction using the SDSP, and the third column the prediction using Gaussian process regression with stationary, isotropic, exponential covariance function.\label{fig:insets}}
 \end{figure}

Despite these improvements in the diagnostics we consider, the greater utility of the SDSP lies in its ability to predict spatial features that regular Gaussian-process models can not. We show such an example in Figure~\ref{fig:insets}, which zooms into four regions, two from each scene (as marked in the top-left panels of Figure~\ref{fig:MODIS1} and \ref{fig:MODIS2}, respectively). In the top row we see some rocky outcrops delineating the land boundary of Antarctica correctly reproduced by the SDSP, while in the remaining rows, we see sharp boundaries in cloud cover being predicted.

One other way to quantify the improvement in prediction is through a field comparison method \citep[][Chapter 6]{Wikle_2019} such as the threat score \citep[TS,][Chapter 7]{Wilks_2006} which, for a given binary classification of each pixel in an image, is defined as the number of true positives divided by the sum of true positives and incorrect classifications. The TS is bounded from above by one, and a higher TS constitutes a better prediction. We construct a binary version of our `true' process by thresholding at $Z^\thr_\obs = 160$ W/m$^2$/\SI{}{\micro\metre}/st; anything below this threshold is deemed to not be cloud or ice and classified as a positive. We then take the predictions from each of our models, and threshold them using  thresholds, $Z^\thr_\pred$, ranging between 50 and 250 W/m$^2$/\SI{}{\micro\metre}/st. For each of the resulting binary predictions we then compute the TS with respect to the original thresholded image.

\begin{figure}[t!]
  \begin{center}
  \ifarxiv
  \includegraphics[width=0.45\textwidth]{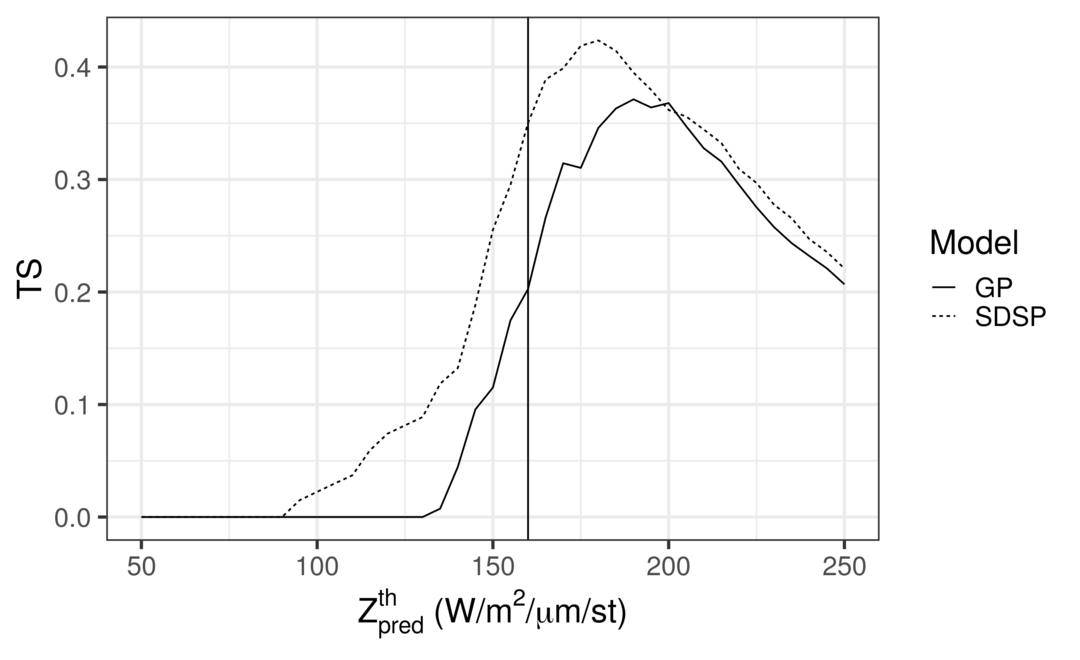}
  \includegraphics[width=0.45\textwidth]{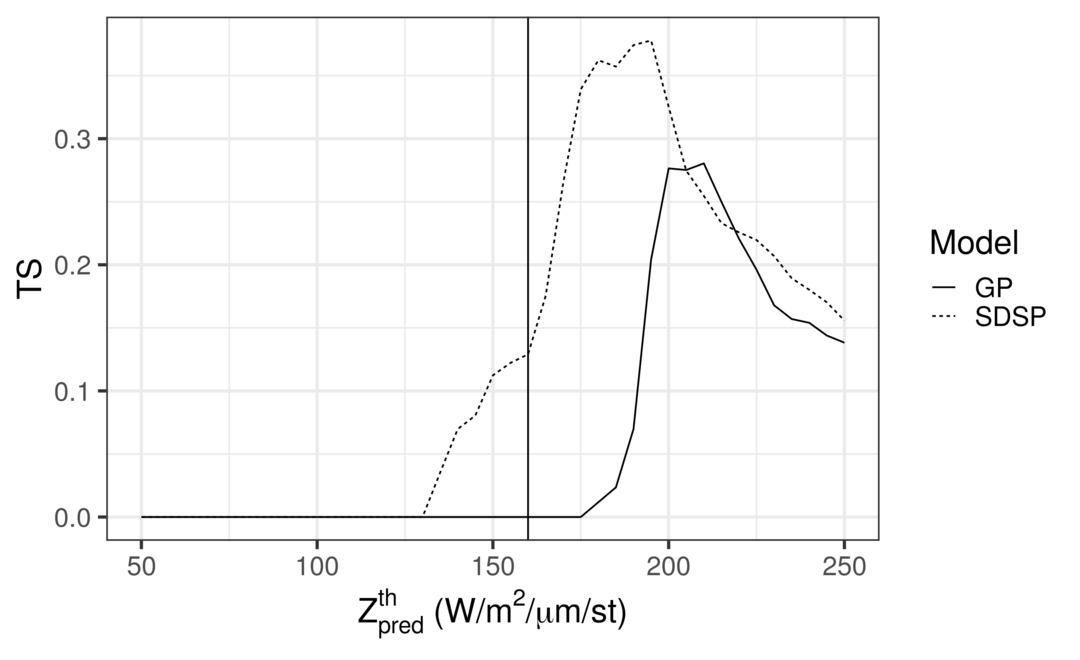}
  \includegraphics[width=0.45\textwidth]{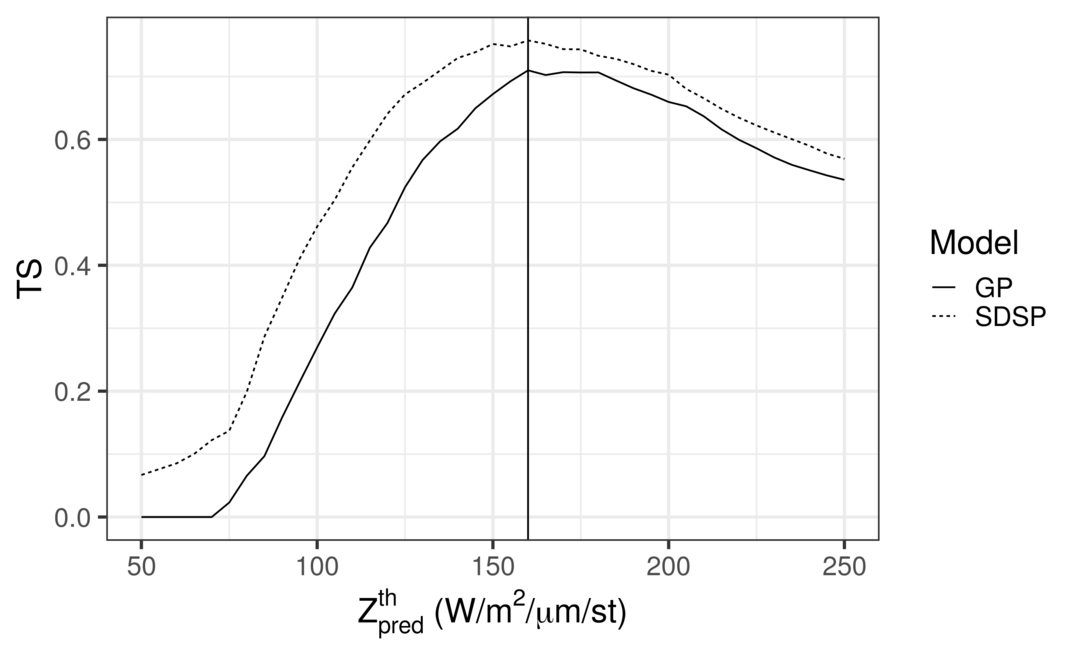}
  \includegraphics[width=0.45\textwidth]{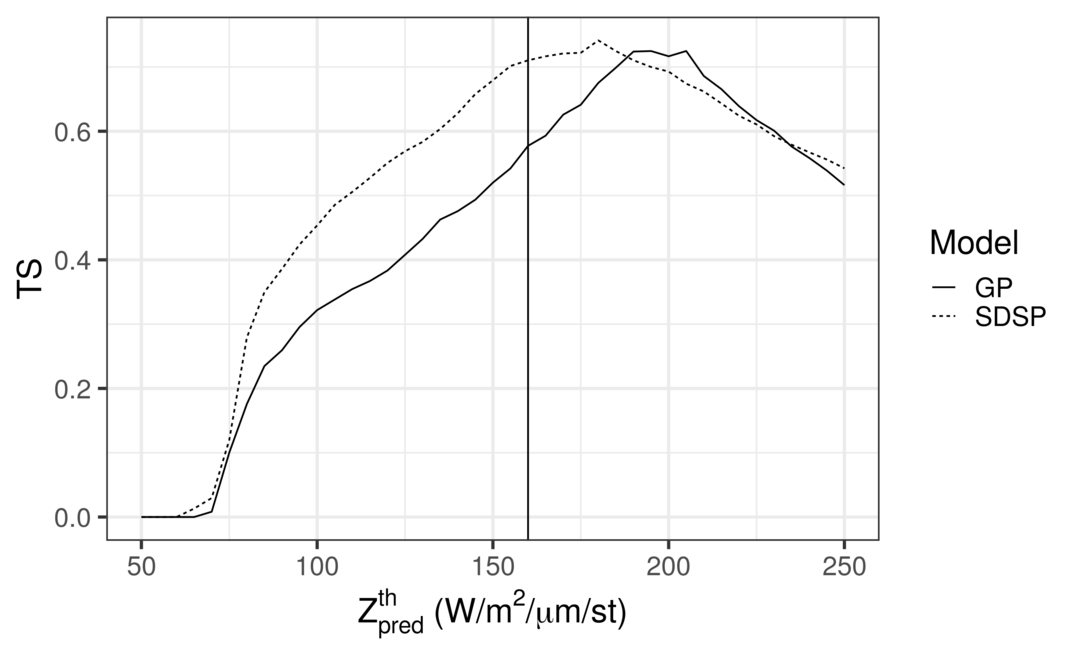}
  \else
    \includegraphics[width=0.45\textwidth]{../deep_spat_src_blinded_for_submission/MODIStest/img/MODIS1zoom1_TS.png}
  \includegraphics[width=0.45\textwidth]{../deep_spat_src_blinded_for_submission/MODIStest/img/MODIS1zoom2_TS.png}
  \includegraphics[width=0.45\textwidth]{../deep_spat_src_blinded_for_submission/MODIStest/img/MODIS2zoom1_TS.png}
  \includegraphics[width=0.45\textwidth]{../deep_spat_src_blinded_for_submission/MODIStest/img/MODIS2zoom2_TS.png}
  \fi
  \end{center}
  \caption{Threat scores as a function of the threshold $Z^\thr_\pred$ when using the SDSP (dotted line) and the GP (solid line) for Insets A (top-left), B (top-right), C (bottom-left) and D (bottom-right) shown in Figure~\ref{fig:insets}. The vertical lines denote the threshold used with the true values, $Z_\obs^\thr = 160$ W/m$^2$/\SI{}{\micro\metre}/st.\label{fig:TS}}
 \end{figure}

The resulting TSs for Insets A--D shown in Figure~\ref{fig:insets} are displayed in Figure~\ref{fig:TS}. The highest TS is generally not obtained at $Z^\thr_\obs$, however the SDSP gives considerably higher TSs for nearly all $Z^\thr_\pred$. Of particular note is Inset B in the first scene, where the GP recorded a TS of 0 when $Z^\thr_\pred = Z^\thr_\obs$. This is a consequence of the GP oversmoothing the salient features in the data, a drawback which the SDSP, which is highly adaptive to such features, has the ability to remedy.

\section{Conclusion}\label{sec:Conclusion}

SIWGPs and SDSPs are deep-learning models that are able to model processes with highly-complex nonstationary and anisotropic covariance structure. The smooth injective constraint inherent in their construction restricts the class of warpings and avoids the notorious problem of `space-folding.' \Copy{Interpretation}{Moreover, since the warping functions are low-dimensional, they can be visualised and are relatively intuitive. The codomain of a warping function also has a natural interpretation as a `warped geographic domain;' it is hard to obtain similar insights on the learned nonlinear mapping from conventional deep-learning models, which are non-injective in general.}

Deep-learning software frameworks such as \texttt{TensorFlow}, and GPU computation, facilitate the implementation of algorithms for making inference and predicting with deep compositional spatial models. Results on both simulated and real data show the huge potential of both the SIWGP and the SDSP in applications of geostatistics.  \Copy{Overfit}{The results also suggest that, whenever computationally feasible, SDSPs should be used, as they appear to be robust against over-fitting, and provide posterior distributions for the warping weights that may be useful for uncertainty quantification. The SIWGP, however, is a good alternative whenever GPU memory or computing time is limited, provided one is aware that the model may over-fit the data if no regularisation is used. As part of this work we have developed an \texttt{R} package, \texttt{deepspat} (online supplementary material), that renders the  implementation of both the SIWGP and the SDSP straightforward.}

This article has presented SIWGPs and SDSPs that are relatively simple, and there are a number of avenues that could be explored next to render the models more widely applicable. First, the process model is currently  a low-rank model, and the number of basis functions needs to be small (on the order of one or two thousand). %We have seen in Section~\ref{sec:MODIS} that, without warping, this low-rank model can do worse than standard kriging, but can outperform kriging when the warping function is estimated.
  Several studies \citep[e.g.,][]{Heaton_2019} show that, for large datasets, higher-rank models are required to obtain good prediction accuracy. Such models are available with the use of sparse precision or covariance matrices; however sparse linear algebraic operations on GPUs tend to be considerably slower than their dense counterpart, and it is not clear at this stage whether such process models could be used within an SIWGP or SDSP. The use of a full-rank Gaussian process and composite-likelihood in the top layer \citep{Eidsvik_2014} is an attractive way forward for large datasets. Alternatively, one might explore the use of minibatch stochastic gradient descent for the SIWGP/SDSP; we give a brief description of how this could be done in Section~\ref{sec:SGD} \ifarxiv (supplementary material)\else of the online supplementary material\fi.

Second, we have only considered some smooth, injective warpings when constructing the deep architectures. \citet{Perrin_1999} considered other RBFs, while one can envisage others based on twists and spirals. A potentially useful unit not considered in this article is the \emph{dimension-expansion} unit. As noted by \citet{Bornn_2012}, if $\fvec(\svec) = (s_1, s_2, \tilde f(\svec))'$, where $\tilde f(\cdot)$ is some unknown map, then there is a trivial one-to-one mapping between $\svec = (s_1, s_2)'$ and $\fvec(\svec)$. New dimensions can be used to capture important features in the data, and can also be modelled using low-rank representations. Dimension expansion was recently used to warp space and time by \citet{Shand_2017}.

Third, we have kept the parameters in the AWUs and RBFs at each layer fixed. Inference might improve if some of these are also estimated; for example, in Section~\ref{sec:1D} we let the AWU steepness parameter $\theta_{1} = 200$, while in Section \ref{sec:L1} we found that $\theta_{1} = 20$ gave more sensible warping functions. In higher-dimensional applications (for example spatio-temporal), covering the domain with RBFs would be challenging; a better strategy in this case might be to only use a small set of RBFs and instead estimate their centroids and apertures. In the case of spatio-temporal data, using a parsimonious variant based on recurrent networks, as explored by \citet{McDermott_2019}, might be advantageous.

Fourth, we have not formally investigated the implication of the chosen warping-function architectures in our studies, that is, the choice of units and their ordering. %Such considerations include the choice and number of activation functions, the use of recurrent connections or otherwise, and the number of hidden layers.
Such considerations are important for our models. % -- an axial warping followed by a M{\"o}bius transformation is very different from a M{\"o}bius transformation followed by an axial warping. 
\Copy{ModelSelection}{One may fit several models with different architectures and then use model averaging or model selection or, if this is infeasible, just use a single very deep architecture containing several units. Preliminary investigations of the latter case (Section~\ref{sec:different-architectures} and Section~\ref{sec:1D-stat-experiment} in the online supplementary material) show that the SDSP prediction quality does not drastically deteriorate when architectures that are overly complex are used. The use of sparsity-inducing priors to further encourage turning off warping units that are not needed, may also be considered. The horseshoe prior, for example, has been used with some success within a variational inference scheme that is similar to the one we present here \citep{Ghosh_2019}.}

Finally, in this article we have considered the SIWGP and SDSP in isolation of other process explanatory variables. We have also only considered Gaussian likelihood functions and point-referenced data. However, our modelling and inferential framework can be extended in a relatively straightforward manner to handle explanatory variables and non-Gaussian areal data if needed.

%% \ifblind
%% \else 
%% \section*{Acknowledgments}

%% The Aqua/MODIS Level 1B Calibrated Radiances (500m) data sets were acquired from the Level-1 and Atmosphere Archive \& Distribution System (LAADS) Distributed Active Archive Center (DAAC), located in the Goddard Space Flight Center in Greenbelt, Maryland (\url{https://ladsweb.nascom.nasa.gov/}). We thank Simone Rossi for running the code for the random Fourier features model and Matt Moores, Aidan Sims and S{\'e}bastien Marmin for helpful discussions. Dr.~Zammit-Mangion's research was supported by an Australian Research Council (ARC) Discovery Early Career Research Award, DE180100203.
%% \fi

\ifarxiv
\section*{Acknowledgements}
\myacktwo
\else
\fi

\ifarxiv
\else

\bigskip

%% \begin{center}
%% {\large\bf SUPPLEMENTARY MATERIAL}
%% \end{center}

\section*{Supplementary Material}

\begin{description}

\item[Sections~\ref{sec:1D-details}--\ref{sec:SGD}:] PDF file containing supplementary text. Section \ref{sec:1D-details} contains further implementation details of the models compared to in Section~\ref{sec:1D}, Section \ref{sec:1D-stat-experiment} describes an experiment that illustrates the behaviour of the SIWGP and SDSP when the underlying process is stationary, and Section \ref{sec:SGD} details the equations that would be required to implement gradient descent using minibatches. (.pdf file)

\item[\texttt{R} package for the DGPsparse model:] \texttt{R} package \texttt{deepGP} containing code to implement the DGPsparse model in Section~\ref{sec:1D}. (GNU zipped tar file)

\item[\texttt{R} package for the SIWGP/SDSP models:] \texttt{R} package \texttt{deepspat} containing code to implement the SIWGP and SDSP models used in Sections~\ref{sec:1D}--\ref{sec:L1}. (GNU zipped tar file)

\item[Source code and data:] \texttt{R} code, \texttt{Python} code and datasets required to reproduce the results in Sections~\ref{sec:1D}--\ref{sec:L1}. (GNU zipped tar file)
  
\end{description}

\fi

\appendix

\section{Inference and prediction with the SIWGP and the SDSP}\label{sec:Inference}

\subsection{Inference for SIWGPs}\label{sec:SIWGP}

In SIWGPs, all unknowns in the set of transformed weights $\widetilde\Wmat_{1:n} \equiv \{\widetilde\Wmat_1,\dots, \widetilde\Wmat_{n}\}$ and the set of parameters $\Thetab_{1:n} \equiv \{\Thetab_1,\dots, \Thetab_{n}\}$ appearing in the warping model, as well as the process-model parameters $\taub_{n+1}$ and the measurement-model noise variance $\sigma^2_{\epsilon}$, are estimated using maximum likelihood. (Recall that several of the components of $\Thetab_{1:n}$ are in fact known and fixed by assumption). We collect the warping weights and parameters into the set $\Lambdamat \equiv \{\widetilde\Wmat_{1:n}, \Thetab_{1:n}\}$. Inference needs to be made on the process $Y(\cdot)$ through $\Wmat_{n+1} = \wvec_{n+1}$ (which is random) by conditioning on the noisy data. 

Let $\Zvec \equiv (Z_1, \dots, Z_N)'$ denote the observed data and $\Yvec \equiv (Y(\svec_1),\dots, Y(\svec_N))'$ the latent process at the locations $\svec_1, \dots, \svec_N \in G$. Then $\Zvec = \Amat_\Lambda\wvec_{n+1} + \epsilonb$, where $\Amat_\Lambda \equiv (\phib_{n+1}(\fvec(\svec_i; \Lambdamat); \Thetab_{n+1}): i = 1,\dots,N)'$ and $\epsilonb \equiv (\epsilon_i: i = 1,\dots, N)'$. Recall that $\fvec(\cdot) \equiv \fvec_n \,\circ\, \fvec_{n-1}\, \circ\,\cdots\,\circ\, \fvec_1(\cdot)$, and it therefore depends on all of $\Lambdamat$. Omitting the dependence on $\Thetab_{n+1}$, which is fixed by assumption, the  marginal, or integrated, likelihood is
$$
p(\Zvec \mid \Lambdamat, \taub_{n+1}, \sigma^2_\epsilon) = \int p(\Zvec \mid \wvec_{n+1}, \Lambdamat,\sigma^2_\epsilon) p(\wvec_{n+1} \mid \taub_{n+1})\intd \wvec_{n+1},
$$
which can be written out as
\begin{align}
  \log p(\Zvec \mid \Lambdamat, \taub_{n+1}, \sigma^2_\epsilon) = \textrm{const.~} - &\frac{N}{2}\log{\sigma^2_\epsilon}  - \frac{1}{2}\log{|\Sigmamat_{\tau_{n+1}}|} - \frac{1}{2}\log{\left|\frac{1}{\sigma^2_\epsilon}\Amat_\Lambda'\Amat_\Lambda + \Sigmamat_{\tau_{n+1}}^{-1}\right|} - \nonumber \\
  & \frac{1}{2}\Zvec'\left[\frac{1}{\sigma^2_\epsilon}\Imat - \frac{1}{\sigma^4_\epsilon}\Amat_\Lambda\left(\frac{1}{\sigma^2_\epsilon}\Amat_\Lambda'\Amat_\Lambda + \Sigmamat_{\tau_{n+1}}^{-1}\right)^{-1}\Amat_\Lambda'\right]\Zmat. \label{eq:marglik}
\end{align}
%In a single layer model with no hidden layers (i.e., when $n = 0$ and $\Lambdamatu \equiv \emptyset$), the marginal likelihood and its gradient with respect to $\Thetab_{n+1}$ can be evaluated efficiently in under a second on a modern desktop when using optimised sparse-matrix algebra routines, and when $N$ and the number of tent functions, $r_{n+1}$, are each on the order of 10,000. In practice, since the hidden layers are very low-dimensional compared to the $(n+1)$th (output) layer, the time required to evaluate the likelihood is dominated by $N$ and $r_{n+1}$, and not by the number of hidden layers. 

Estimates of $\Lambdamat, \taub_{n+1},$ and $\sigma^2_\epsilon$ ($\widehat\Lambdamat, \widehat\taub_{n+1},$ and $\widehat\sigma^2_\epsilon$, respectively), can be found using gradient-based optimisation. Note that \eqref{eq:scaling} is not differentiable everywhere with respect to the weights on the original scale; specifically, it is not differentiable along a finite number of hyperplanes that satisfy $\wvec_i^{(k)'}\vvec = 0$ where $\vvec$ depends on the chosen knots. Since these hyperplanes occupy an infinitesimally small volume in the vector space of $\wvec_i^{(k)}$, they are almost certainly never going to be coincident with an estimate of $\wvec_i^{(k)}$ in a gradient descent. Differentiable approximations to the maximum and minimum functions are available if desired \citep[e.g.,][]{Lange_2014}. 

\subsubsection*{Prediction}

Once $\Lambdamat$, $\taub_{n+1}$, and $\sigma^2_\epsilon$ are estimated, they are used for prediction. Specifically, for a set of $N^*$ prediction locations on $G$, $\svec^*_j, j = 1,\dots,N^*$, define $\Yvec^* \equiv (Y(\svec_j^*): j = 1,\dots,N^*)'$. The prediction and prediction variance of $\Yvec^*$ are
\begin{align}
  E(\Yvec^* \mid \Zmat, \widehat\Lambdamat, \widehat\taub_{n+1}, \widehat\sigma^2_\epsilon) &= \Amat_{\hat\Lambda}^*\muvec_{n+1}^*, ~~\textrm{and} \label{eq:predYstar}\\
  Var(\Yvec^* \mid \Zmat, \widehat\Lambdamat,\widehat\taub_{n+1}, \widehat\sigma^2_\epsilon) &= \Amat_{\hat\Lambda}^*\Sigmamat_{n+1}^{*}\Amat_{\hat\Lambda}^{*'}, \label{eq:predvarYstar}
\end{align}
respectively, where $\Amat_{\hat\Lambda}^* \equiv (\phib_{n+1}(\fvec(\svec_j^*; \widehat\Lambdamat); \Thetab_{n+1}): j = 1,\dots,N^*)'$, and
\begin{align}
  \Sigmamat_{n+1}^{*^{-1}} &=  \frac{1}{\widehat{\sigma}^2_\epsilon}\Amat_{\hat\Lambda}'\Amat_{\hat\Lambda} + \Sigmamat_{\hat\tau_{n+1}}^{-1}, \\
  \muvec_{n+1}^* &=  \frac{1}{\widehat{\sigma}^2_\epsilon}\Sigmamat_{n+1}^*\Amat_{\hat\Lambda}'\Zvec,\label{eq:Qmu}
\end{align}
are the precision and expectation of the weights $\wvec_{n+1}$ when conditioned on the data $\Zvec$ and the estimated quantities $\widehat\Lambdamat, \widehat\taub_{n+1},$ and $\widehat\sigma^2_\epsilon$. %One can use computationally-efficient sparse-linear algebraic methods for \eqref{eq:predvarYstar} if only the marginal prediction variances are sought \citep{Zammit_2018b}.

\subsection{Inference for SDSPs}\label{sec:SDSP}

Some of the notation used here is introduced in Appendix~\ref{sec:SIWGP}. Here we describe our variational Bayes (VB) approach for finding an approximation to the posterior distribution $p(\widetilde\Wmat_{1:n} \mid \Zvec, \Thetab_{1:n}, \taub_{n+1}, \sigma^2_\epsilon)$, which we denote as $q(\widetilde\Wmat_{1:n})$ and  whose form we specify later. In VB, the marginal likelihood is first bounded, and the lower bound is then maximised with respect to the parameters appearing in $q(\widetilde\Wmat_{1:n})$, as well as $\Thetab_{1:n}, \taub_{n+1},$ and $\sigma^2_\epsilon$. 

Consider the set of spatial locations of the data on $G$, $\Smat \equiv (\svec_1, \dots, \svec_N) \equiv \Fmat_0$. The function outputs at the first layer from the inputs $\Smat$ are $\Fmat_1 = (\fvec_1(\svec_1), \dots, \fvec_1(\svec_N))$, which, recall, we write as $\fvec_1(\Smat)$ for conciseness.  Similarly, the outputs at the $i$th layer are $\Fmat_i = \fvec_i(\Fmat_{i-1})$. Collect all these warped variables into $\Fmat_{1:n} \equiv \{\Fmat_1,\dots, \Fmat_{n}\}$ and assume that the matrix $\Smat^\alpha$ consists of some or all of the columns of $\Smat$, so that $\Fmat_i^\alpha$ also consists of some or all of the columns of $\Fmat_i$, $i = 1,\dots,n$. (This latter assumption removes the need for defining separate variational distributions over the knots.) Our marginal likelihood is (again, ignoring the dependence on $\Thetab_{n+1}$, which is fixed by assumption)
\begin{align}
  \log &~p(\Zvec \mid \Thetab_{1:n}, \taub_{n+1}, \sigma^2_\epsilon) \nonumber \\
  &= \log \int~p(\Zvec \mid \Fmat_{1:n}, \widetilde\Wmat_{1:n},  \Thetab_{1:n}, \taub_{n+1}, \sigma^2_\epsilon)p(\Fmat_{1:n}, \widetilde\Wmat_{1:n} \mid \Thetab_{1:n}) \intd \Fmat_{1:n} \intd \widetilde\Wmat_{1:n}\nonumber \\
&= \log \int~q(\Fmat_{1:n}, \widetilde\Wmat_{1:n}\mid\Thetab_{1:n})\frac{p(\Zvec \mid \Fmat_{1:n}, \widetilde\Wmat_{1:n},  \Thetab_{1:n}, \taub_{n+1}, \sigma^2_\epsilon)p(\Fmat_{1:n}, \widetilde\Wmat_{1:n} \mid \Thetab_{1:n})}{q(\Fmat_{1:n}, \widetilde\Wmat_{1:n}\mid\Thetab_{1:n})} \intd \Fmat_{1:n} \intd \widetilde\Wmat_{1:n}\nonumber\\
&\ge E_{q(\Fmat_{1:n}, \widetilde\Wmat_{1:n}\mid\Thetab_{1:n})}\left[\log \frac{p(\Zvec \mid \Fmat_{1:n}, \widetilde\Wmat_{1:n},  \Thetab_{1:n}, \taub_{n+1}, \sigma^2_\epsilon)p(\Fmat_{1:n}, \widetilde\Wmat_{1:n} \mid \Thetab_{1:n})}{q(\Fmat_{1:n}, \widetilde\Wmat_{1:n}\mid\Thetab_{1:n})}\right] \equiv \mathcal{E},\label{eq:E}
\end{align}
\noindent by Jensen's inequality. Now, $\Zvec$ is conditionally independent of $\widetilde\Wmat_{1:n}$, $\Fmat_{1:(n-1)}$, and $\Thetab_{1:n}$  when conditioned on $\Fmat_{n}$, and hence the likelihood $p(\Zvec \mid \Fmat_{1:n}, \widetilde\Wmat_{1:n},  \Thetab_{1:n}, \taub_{n+1}, \sigma^2_\epsilon)= p(\Zvec \mid \Fmat_{n},  \taub_{n+1}, \sigma^2_\epsilon)$. This term is identical to \eqref{eq:marglik} with $\Amat_{\Lambda}$ replaced with $\phib_{n+1}(\Fmat_{n};\Thetab_{n+1})'$.

%So far we have not specified a form for $q(\Fmat_{n-1}, \Wmatu_{n-1})$.
Unfortunately, free-form variational optimisation (where we do not specify the functional form of $q(\Fmat_{1:n}, \widetilde\Wmat_{1:n}\mid\Thetab_{1:n})$) is not analytically tractable. Following the approach first used for the latent-variable Gaussian process \citep{Titsias_2010} and subsequently for sparse DGPs \citep{Damianou_2013}, we constrain $q(\Fmat_{1:n}, \widetilde\Wmat_{1:n}\mid\Thetab_{1:n})$ to take the form
\begin{align*}
  q(\Fmat_{1:n}, \widetilde\Wmat_{1:n}\mid\Thetab_{1:n}) &=  p(\Fmat_{n} \mid \widetilde\Wmat_{n}, \Fmat_{n-1}, \Thetab_{n})\cdots p(\Fmat_1 \mid \widetilde\Wmat_1, \Thetab_1) q(\widetilde\Wmat_{n})\cdots q(\widetilde\Wmat_1).
\end{align*}
We further let $q(\widetilde\Wmat_i) = \prod_{k = 1}^{d_i} q(\tilde\wvec_i^{(k)})$ where $q(\tilde\wvec_i^{(k)}) = \Gau(\mvec_i^{(k)}, \Vmat_i^{(k)}(\etab_i^{(k)})),~i = 1,\dots,n.$ The covariance matrix of the variational distribution, $\Vmat_i^{(k)}(\etab_i^{(k)}),$ is parameterised through its lower Cholesky factor. That is, $\Vmat_i^{(k)}(\etab_i^{(k)}) = (\Lmat_i^{(k)}(\etab_i^{(k)}))(\Lmat_i^{(k)}(\etab_i^{(k)}))'$, where
$$
\Lmat_i^{(k)}(\etab_i^{(k)}) = \begin{pmatrix}
  e^{\eta_{i,11}^{(k)}} \\
  \eta_{i,21}^{(k)} & e^{\eta_{i,22}^{(k)}} \\
  \eta_{i,31}^{(k)} & \eta_{i,32}^{(k)} & e^{\eta_{i,33}^{(k)}} \\
  \vdots & \vdots & \vdots & \ddots 
 \end{pmatrix},
$$
and where the exponential terms on the diagonal ensure that $\Vmat_i^{(k)}(\etab_i^{(k)})$ is positive-definite. This Cholesky factor can be made sparse if desired \citep[e.g.,][]{Tan_2018}. Note that $\Vmat_i^{(k)}(\etab_i^{(k)})$ is an approximate \emph{posterior} covariance matrix. Therefore, a simple parameterisation based on a spatial covariance model (one with an exponential covariance function, say) is not appropriate in this case.

Substituting our choice of $q(\Fmat_{1:n}, \widetilde\Wmat_{1:n}\mid\Thetab_{1:n})$ into \eqref{eq:E} we obtain an expression for the lower bound
\begin{align*}
  \mathcal{E} &= E_{q(\Fmat_{n}\mid\Thetab_{1:n})}\left[\log p(\Zvec \mid \Fmat_{n},  \taub_{n+1}, \sigma^2_\epsilon)\right] - \sum_{i=1}^n KL(q(\widetilde\Wmat_{i}) \| p(\widetilde\Wmat_{i})) \equiv  \mathcal{E}_1 - \mathcal{E}_2,
\end{align*}
where $\mathcal{E}_1$ is the expected marginal log-likelihood (where the expectation is taken under the variational posterior distribution of the warped locations, defined below in \eqref{eq:Fn1}), and $\mathcal{E}_2$ is the sum of Kullback--Leibler divergences between the variational posterior distributions over $\{\widetilde\Wmat_{i}\}$ and the respective prior distributions. This latter term can be calculated analytically since both $q(\widetilde\Wmat_{i})$ and $p(\widetilde\Wmat_{i})$ are multivariate Gaussian distributions for $i = 1,\dots,n$.

The term $\mathcal{E}_1$ cannot be evaluated analytically. However, since it is an expectation, it can be approximated using Monte Carlo:
$$
\mathcal{E}_1 \approx \frac{1}{N_{MC}}\sum_{l=1}^{N_{MC}} \log p(\Zvec \mid \Fmat_{n}^{(l)}, \taub_{n+1}, \sigma^2_\epsilon),
$$
where $N_{MC}$ is the number of Monte Carlo samples used in the approximation and $\Fmat_{n}^{(l)} \sim q(\Fmat_{n}\mid\Thetab_{1:n})$. Since the term $\mathcal{E}_1$ is approximated through Monte Carlo, this type of variational inference is often referred to as stochastic variational inference.

A sample $\Fmat_n^{(l)}$ can be obtained easily by noting that $q(\Fmat_{n} \mid \Thetab_{1:n})$ can be expressed as the marginalisation
\begin{align}
  q(\Fmat_{n}\mid\Thetab_{1:n}) = \int & p(\Fmat_{n}\mid \Fmat_{n-1},\widetilde\Wmat_{n}, \Thetab_{n})p(\Fmat_{n-1}\mid \Fmat_{n-2},\widetilde\Wmat_{n-1},  \Thetab_{n-1})\cdots \nonumber\\
  &\times p(\Fmat_1 \mid \widetilde\Wmat_1,  \Thetab_1)q(\widetilde\Wmat_{n})\cdots q(\widetilde\Wmat_1)\intd\Fmat_{1:(n-1)}\intd\widetilde\Wmat_{1:n}.\label{eq:Fn1}
\end{align}
In our case, the distributions $p(\Fmat_{i}\mid \Fmat_{i-1},\widetilde\Wmat_{i},  \Thetab_{i}), i = 1,\dots,n,$ are degenerate at $\Fmat_{i} = \gvec_i(\Wmat_{i}\phib_{i}(\Fmat_{i-1}; \Thetab_{i}); \Fmat^\alpha_{i-1}),$
%% and hence we can rewrite \eqref{eq:Fn1} as
%% \begin{align*}
%%   q(\Fmat_{n}\mid\Thetabu_n) = \int \delta_{\Fmat_{n}}(\gvec_n(\Wmat_{n}&\phib_{n}(\gvec_{n-1}(\Wmat_{n-1}\phib_{n-1}(\cdots \gvec_1(\Wmat_1\phib_1(\Smat; \Thetab_1); \widetilde\Smat)\cdots ; \Thetab_{n-1}); \widetilde\Fmat_{n-2}); \Thetab_n); \widetilde\Fmat_{n-1}))\\
%%   & \times q(\Wmat_{n})\cdots q(\Wmat_1)\intd \Wmatu_{n},
%% \end{align*}
%% where $\delta_x(\cdot)$ is the Dirac delta function on $x$
where recall that $\Wmat_i \equiv h_i^{-1}(\widetilde\Wmat_i), i = 1,\dots, n$. Sampling thus proceeds by first sampling $\widetilde\Wmat_{1:n}^{(l)}$ from the variational distributions, back-transforming layer-wise to obtain $\Wmat_{1:n}^{(l)}$,  and then computing $\Fmat_{n}^{(l)}$ through
\begin{equation}
\Fmat_{n}^{(l)} = \gvec_n(\Wmat_{n}^{(l)}\phib_{n}(\gvec_{n-1}(\Wmat_{n-1}^{(l)}\phib_{n-1}(\cdots \gvec_1(\Wmat_1^{(l)}\phib_1(\Smat;\Thetab_1); \Smat^\alpha)\cdots;\Thetab_{n-1}); \Fmat_{n-2}^{\alpha^{(l)}});\Thetab_{n});\Fmat_{n-1}^{\alpha^{(l)}}).\label{eq:Fmatnm1}
\end{equation}
%where we have notated an explicit dependence on $\Smat$, $\Thetabu_n$ and the knots (which, recall, are dependent on the weights and therefore different for each MC sample) to facilitate our discussion on prediction later on in this section.
Equation \eqref{eq:Fmatnm1} shows that all that is needed to sample $\Fmat_{n}$ is to (deterministically) propagate $\Smat$ and $\Smat^\alpha$ through the layers and rescaling functions with the weights fixed to the back-transformed sample $\Wmat_{1:n}^{(l)}$. 

Now, in variational Bayes one sets out to find the variational parameters (in our case the mean and Cholesky-factor elements) that maximise the lower bound, but these parameters no longer appear explicitly inside the partial objective $\mathcal{E}_1$ due to the use of the Monte Carlo samples. However, since $q(\tilde\wvec_i^{(k)})$ is Gaussian, a sample from $q(\tilde\wvec_i^{(k)})$ is also a sample from $\mvec_i^{(k)} + \Lmat_i^{(k)}(\etab_i^{(k)})\evec_i^{(k)},$ where $\evec_i^{(k)} \sim \Gau(\zerob,\Imat)$. This so-called re-parameterisation trick \citep{Kingma_2014, Xu_2018} ensures that the set of variational parameters still explicitly appear within $\mathcal{E}_1$ despite the use of a Monte Carlo approximation. % Therefore, one need only generate $N_{MC}$ samples from $\evec_i^{(k)}$ once for each $i$ and $k$; these are subsequently kept fixed.

The optimisation problem reduces to the following. Let $\Mmat_i \equiv (\mvec_i^{(1)}, \dots, \mvec_i^{(d_i)})$ and $\Mmat_{1:n} \equiv \{\Mmat_1,\dots,\Mmat_{n}\}$. Similarly, let $\Gammamat_i \equiv (\etab_i^{(1)}, \dots, \etab_i^{(d_i)})$ and $\Gammamat_{1:n} = \{\Gammamat_1,\dots,\Gammamat_{n} \}$. Then
\begin{align}
  (\widehat\Mmat_{1:n}, \widehat\Gammamat_{1:n},\widehat\Thetab_{1:n}, \widehat\taub_{n+1}, \widehat\sigma^2_\epsilon) = &\argmax_{\Mmat_{1:n}, \Gammamat_{1:n}, \Thetab_{1:n},\taub_{n+1},\sigma^2_\epsilon} \frac{1}{N_{MC}}\sum_{l=1}^{N_{MC}}\log p(\Zvec \mid \Fmat_{n}^{(l)}, \taub_{n+1}, \sigma^2_\epsilon,) \nonumber \\
  & \qquad\qquad- \sum_{i=1}^n KL(q(\widetilde\Wmat_{i}) \| p(\widetilde\Wmat_{i})),\label{eq:varoptim}
\end{align}
where the dependence of $\Fmat_{n}^{(l)}$ on  $\Mmat_{1:n}, \Gammamat_{1:n},$ and $\Thetab_{1:n}$ is given through \eqref{eq:Fmatnm1} and application of the re-parameterisation trick to $\widetilde\Wmat_{1:n}$.

\subsubsection*{Prediction}

The variational prediction distribution for $\Yvec^*$ is given by
\begin{align}
  p(\Yvec^* \mid \Zvec, \widehat\Thetab_{1:n}, \widehat\taub_{n+1}, \widehat\sigma^2_\epsilon) = \int &p(\Yvec^* \mid \wvec_{n+1}, \Fmat^*_n,)p(\wvec_{n+1} \mid \Fmat_n, \Zvec, \widehat\taub_{n+1}, \widehat\sigma^2_\epsilon) \nonumber \\
  & \times q(\Fmat_n, \Fmat^*_n \mid \widehat\Thetab_{1:n}) \intd\wvec_{n+1}\intd\Fmat_n\intd\Fmat^*_n,\label{eq:varpred}
\end{align}
where $q(\Fmat_n, \Fmat^*_n\mid\widehat\Thetab_{1:n})$ is the (joint) variational posterior distribution over $\Fmat_n$ and $\Fmat^*_n$. Samples from this joint distribution can be generated by noting that
\begin{align}
  q(\Fmat_{n}, \Fmat^*_n\mid\Thetab_{1:n}) = &\int  p(\Fmat^*_n \mid \Fmat^*_{n - 1}, \Fmat^\alpha_{n-1},\widetilde\Wmat_n, \Thetab_n)p(\Fmat_{n}\mid \Fmat_{n-1},\widetilde\Wmat_{n}, \Thetab_{n}) \nonumber \\
  &\times p(\Fmat^*_{n-1} \mid \Fmat^*_{n - 2}, \Fmat^\alpha_{n-2},\widetilde\Wmat_{n-1}, \Thetab_{n-1})p(\Fmat_{n-1}\mid \Fmat_{n-2},\widetilde\Wmat_{n-1},  \Thetab_{n-1})\cdots \nonumber \\
  &\times p(\Fmat^*_{1} \mid \widetilde\Wmat_{1}, \Thetab_{1})p(\Fmat_1 \mid \widetilde\Wmat_1,  \Thetab_1)q(\widetilde\Wmat_{n})\cdots q(\widetilde\Wmat_1)\intd\Fmat^*_{1:(n-1)}\intd\Fmat_{1:(n-1)}\intd\widetilde\Wmat_{1:n}.\nonumber
\end{align}
%$p(\Fmat_i, \Fmat^*_i \mid \Wmat_i, \Fmat_{i - 1}, \Fmat^*_{i - 1}) = p(\Fmat^*_i \mid \Fmat^*_{i - 1}, \widetilde\Fmat_{i-1},\Wmat_i)p(\Fmat_i \mid \Fmat_{i - 1}, \Wmat_i)$. %, that is, the data locations and the prediction locations at each layer are conditionally independent given the weights in that layer and the inputs to that layer (which, recall, include the knot locations).
\noindent Hence, as when fitting the model, one need only generate samples from $q(\widetilde\Wmat_{1:n})$ and back-transform them layer-wise; these are then used to simultaneously generate samples (jointly) of $\Fmat_{n}$  and $\Fmat_n^*$. Specifically, the warping $\Fmat_n^{(l)}$ is found from \eqref{eq:Fmatnm1} and $\Fmat_n^{*^{(l)}}$ is found from
$$
{\small
\Fmat_{n}^{*^{(l)}} = \gvec_n(\Wmat_{n}^{(l)}\phib_{n}(\gvec_{n-1}(\Wmat_{n-1}^{(l)}\phib_{n-1}(\cdots \gvec_1(\Wmat_1^{(l)}\phib_1(\Smat^*;\Thetab_1); \Smat^\alpha)\cdots;\Thetab_{n-1}); \Fmat^{\alpha^{(l)}}_{n-2});\Thetab_{n});\Fmat_{n-1}^{\alpha^{(l)}}),}
$$
\noindent where $\Smat^* = (\svec^*_1,\dots,\svec^*_{N^*})$ and recall that $\Fmat_i^{\alpha^{(l)}}$ is a submatrix of, or identical to, $\Fmat_i^{(l)}, i = 1,\dots, n-1$. Note that the samples of the transformed weights are obtained from the optimised variational distributions. That is, a sample from $q(\tilde\wvec_i^{(k)})$, for $i = 1,\dots,n$ and $k = 1,\dots, d_i$, is a sample from  $\Gau(\hat\mvec_i^{(k)}, \Vmat_i^{(k)}(\hat\etab_i^{(k)}))$, where $\hat\mvec_i^{(k)}$ and $\hat\etab_i^{(k)}$ are obtained from \eqref{eq:varoptim}. Therefore, for each sample of weights, the warped prediction locations and knots at each layer are found by simply (deterministically) propagating $\Smat^*$ and $\Smat^\alpha$ through the layers, respectively, with the weights fixed to the sample $\Wmat_{1:n}^{(l)}$.

The resulting approximation to \eqref{eq:varpred} is the Gaussian mixture
$$
p(\Yvec^* \mid \Zvec, \widehat\Thetab_{1:n}, \widehat\taub_{n+1}, \widehat\sigma^2_\epsilon) \approx \frac{1}{N_{MC}}\sum_{l = 1}^{N_{MC}}\int p(\Yvec^* \mid \wvec_{n+1}, \Fmat^{*^{(l)}}_n)p(\wvec_{n+1} \mid \Fmat_n^{(l)}, \Zvec, \widehat\taub_{n+1}, \widehat\sigma^2_\epsilon)\intd \wvec_{n+1},
$$
where each Gaussian mixture component has mean and covariance matrix defined through \eqref{eq:predYstar}--\eqref{eq:Qmu} with $\Amat_{\hat\Lambda}$ replaced with $\phib_{n+1}(\Fmat_{n}^{(l)}; \Thetab_{n+1})'$ and $\Amat_{\hat\Lambda}^*$ replaced with $\phib_{n+1}(\Fmat_{n}^{*^{(l)}}; \Thetab_{n+1})'$.  Since each Gaussian distribution in the mixture has equal weighting, we suggest sampling a small amount of samples (say 100) from each component, and combining them to obtain an empirical approximation of $p(\Yvec^* \mid \Zvec, \widehat\Thetab_{1:n}, \widehat\taub_{n+1}, \widehat\sigma^2_\epsilon)$. In contrast to the SIWGP, note that these prediction distributions can be highly non-Gaussian. 

%\red{KILLER PROOF: Show that all bijective functions in $L_2$ can be represented using an infinite composition of bijective mappings? To rebut the criticism of Iovleff and Perrin, 2000} 

\subsection{Computational complexity}\label{sec:Complexity}

The objective function used for parameter estimation with the SIWGP is the marginal likelihood \eqref{eq:marglik}. Evaluating this function requires the warping of the data locations in $G$ via $\fvec(\cdot)$.  A warping operation in any given layer is a multiplication of a weights matrix of size $d_i \times r_i$ with a matrix constructed by evaluating $r_i$ basis functions at $N$ locations. Since we fix $d_i = 2, i = 1,\dots,n$, the time complexity of each individual warping layer is $O(Nr_i)$. The remaining operations for evaluating the marginal likelihood have time complexity $O(Nr_{n+1}^2 + r_{n+1}^3)$. The space complexity for the SIWGP is $O(\sum_{i=1}^n Nr_i + Nr_{n+1} + r_{n+1}^2)$. In practice, $r_{n+1} \gg \sum_i r_i$, and therefore $\textrm{max}\{Nr_{n+1}, r_{n+1}^2\}$ needs to be in the tens or low hundreds of millions for the available memory space on conventional GPUs to be sufficient.

The objective function used for inference with the SDSP \eqref{eq:varoptim} requires the evaluation of a Kullback--Leibler divergence, which needs negligible computing time (since our prior and variational posterior distributions are Gaussian and have factored representations), and the marginal likelihood function evaluated $N_{MC}$ times, where $N_{MC}$ is the number of Monte Carlo samples used to approximate expectations in the VB algorithm. The time complexities associated with making inference with our VB algorithm are hence $O(N_{MC}Nr_i)$ for each warping layer, and $O(N_{MC}(Nr_{n+1}^2 + r_{n+1}^3))$ for the remaining operations. The space complexity is $O(N_{MC}(\sum_{i=1}^n Nr_i + Nr_{n+1} + r_{n+1}^2))$.  \texttt{TensorFlow} implements AD in reverse accumulation mode, so that the time and memory required to compute and store the derivatives for the SIWGP and the SDSP are the same as those for the corresponding objective function, multiplied by factors in the single digits, in what is known as the cheap gradient principle \citep[][Section 3.3]{Griewank_2008}. We note that the gradient-based methods we adopt, both for the SIWGP and the SDSP, are guaranteed to converge to local maxima.

\bibliography{Bibliography}

\ifarxiv

\def\spacingset#1{\renewcommand{\baselinestretch}%
{#1}\small\normalsize}

\spacingset{1.5} % DON'T change the spacing!
\ifarxiv \spacingset{1}
\else \fi

\ifarxiv
\else
\ifblind \author{} \else 
\author{Andrew Zammit-Mangion, Tin Lok James Ng, \\ Quan Vu \& Maurizio Filippone} \fi
\date{}


\maketitle
\fi

\renewcommand{\thesection}{S\arabic{section}}
\renewcommand{\thefigure}{S\arabic{figure}}
\ifarxiv \setcounter{section}{0} \fi

\numberwithin{equation}{section} 

\makeatletter
\@addtoreset{equation}{section}
\makeatother

\ifarxiv
\newpage
\section*{Supplementary Material}
\fi

\section{Details on the 1D Simulation Experiment}\label{sec:1D-details}

In this section we provide implementation details of the models we compared the SIWGP/SDSP to in Section \ref{sec:1D}. Details for DGPfull are presented in Section~\ref{sec:DGPfull}; DGPRFF in Section~\ref{sec:DGPRFF}; DGPsparse in Section~\ref{sec:DGPsparse}; GP in Section~\ref{sec:GP}; and SDSP-MCMC in Section~\ref{sec:SDSP-MCMC}. Note that in this section the notation differs slightly from that of the main text when describing the various methods; terms are  explicitly defined, where appropriate, for the reader's benefit.

\subsection{DGPfull}\label{sec:DGPfull}

The DGPfull model we implemented is the following two-layer (i.e., one hidden layer) Gaussian process. Let $\Zvec \equiv (Z_1,\dots,Z_N)'$ denote the data; $\Yvec \equiv (Y_1,\dots,Y_N)'$ the process at the data locations $\Smat \equiv (s_1,\dots,s_N)'$; and $\Fmat_1 = (f_{11},\dots,f_{1N})'$ the warped locations. We let
\begin{align*}
  \Zvec \mid \Yvec &\sim\Gau(\Yvec, \sigma^2_\epsilon\Imat), \\
  \Yvec \mid \Fmat_1 &\sim\Gau(\zerob, \Kmat_2(\Fmat_1)), \\
  \Fmat_1 &\sim \Gau(\zerob, \Kmat_1(\Smat)),
\end{align*}
\noindent where
\begin{align}
  \Kmat_2(\Fmat_1) &\equiv (\sigma^2_2\exp(-(f_{1i} - f_{1j})^2/\alpha_2) : i,j = 1,\dots N),\label{eq:Kmat2}\\
  \Kmat_1(\Smat) &\equiv (\sigma^2_1\exp(-(s_{i} - s_{j})^2/\alpha_1) : i,j = 1,\dots N), \label{eq:Kmat1}
\end{align}
are covariance matrices constructed using the squared-exponential covariance function. Since estimating parameters in the DGPfull model is difficult, the parameters $\sigma^2_\epsilon$, $\sigma^2_i,$ and $\alpha_i$, for $i = 1,2$, were estimated by fitting a DGPRFF model with squared-exponential covariance functions and one hidden layer. Specifically, for the first case study  we fixed $\log \sigma^{2}_{\epsilon} = -4.336$, $\log \sigma^{2}_1 = 0.713$, $\log \sigma^{2}_2 = -0.067$, $\log \alpha_1 = -2.194$, and $\log \alpha_2 = -1.406 $,  while for the second case study we fixed   $\log \sigma^{2}_{\epsilon} = -3.830$, $\log \sigma^{2}_1 = 0.270$, $\log \sigma^{2}_2 = 1.408$, $\log \alpha_1 = -1.645$, and $\log \alpha_2 = -1.032 $.

Let $\Yvec^* \equiv (Y^*_1,\dots,Y^*_{N^*})'$ and $\Fmat_1^{*} \equiv (f^*_{11},\dots,f^*_{1N^*})'$ denote the processes and hidden layer variables at the $N^*$ prediction locations. The conditional distribution of $\Yvec$, when conditioned on the data $\Zvec$, is not available in closed form, and was hence approximated using a Gibbs sampling Markov chain Monte Carlo (MCMC) scheme. Specifically, we iteratively sampled from the distributions $p(\Yvec \mid \Fmat_1, \Zvec)$ and $p(\Fmat_1 \mid \Yvec)$. While the former of these is Gaussian and hence easy to sample from, the latter is not. As in \citet{Cutajar_2016} we used  elliptical slice sampling to sample from this conditional distribution. The full conditional distributions of $\Yvec^*$ and $\Fmat_1^*$, namely $p(\Yvec^* \mid \Yvec, \Fmat_1^*)$ and $p(\Fmat_1^* \mid \Fmat_1)$,  are  Gaussian and available in closed form. Thus, sampling from these distributions proceeds through sampling by composition: First samples of $\Fmat^*_1 \mid \Fmat_1$ are generated followed by samples of  $\Yvec^* \mid \Yvec, \Fmat_1^*$. 

The MCMC scheme was implemented in \texttt{Python}. The number of iterations and burn-in samples were set to 5000 and 100, respectively. The elliptical slice sampler took, on average, about 85 s to generate a single sample from the conditional distribution $p(\Fmat_1 \mid \Yvec)$. Convergence was assessed by visually inspecting trace plots of $\Yvec^{*}$ at a small number of randomly-selected prediction locations. 

\subsection{DGPRFF}\label{sec:DGPRFF}

We fit DGPRFF models with one, two, and five hidden layers. Of these, the model with two hidden layers provided the best predictive performance and, therefore, this is the model presented in the main text.

The DGPRFF model with two hidden layers (and a single-dimensional output at each layer) is the hierarchical model given by
\begin{align*}
  \Zvec \mid \Yvec &\sim\Gau(\Yvec, \sigma^2_\epsilon\Imat), \\
  \Yvec \mid \Fmat_2 &\sim\Gau(\zerob, \Kmat_3(\Fmat_2)), \\
  \Fmat_2 \mid \Fmat_1 &\sim \Gau(\zerob, \Kmat_2(\Fmat_1)),\\
  \Fmat_1 &\sim \Gau(\zerob, \Kmat_1(\Smat)),
\end{align*}
where $\Kmat_3, \Kmat_2,$ and $\Kmat_1$ are constructed from squared-exponential covariance functions (see Section~\ref{sec:DGPfull}). By Bochner's Theorem, one can represent the squared-exponential correlation function as an expectation of sums and products of trigonometric functions, where the expectation is taken with respect to a Gaussian distribution (in the spectral domain) that is fully determined by the length scale parameter $\alpha$. One can therefore approximate this expectation (and, hence, the correlation function) through Monte Carlo to obtain a set of trigonometric basis functions that can be used to reconstruct the squared-exponential function in expectation; see \citet{Cutajar_2016} for details.

In summary, dimension-reduction in the DGPRFF is achieved by modelling $\Kmat_i(\Fmat_{i-1}) = \Phimat_i(\Fmat_{i-1}; \Thetab_i)\Phimat_i(\Fmat_{i-1}; \Thetab_i)'$, for $i = 1,2,3,$ and for $\Smat \equiv \Fmat_0$, where $\Phimat_i$ are sine and cosine functions evaluated at the data/warped data locations. The ensuing weight-space view of the DGPRFF model is
\begin{align*}
  \Zvec &= \Yvec + \epsilonb, \\
  \Yvec &= \Phimat_3(\Fmat_2; \Thetab_3)\wvec_3, \\
  \Fmat_2 &= \Phimat_2(\Fmat_1; \Thetab_2)\wvec_2, \\
  \Fmat_1 &= \Phimat_1(\Smat; \Thetab_1)\wvec_1, 
\end{align*}
\noindent where $\wvec_i \sim \Gau(\zerob, \Imat), i = 1,2,3,$ and $\epsilonb \sim \Gau(\zerob, \sigma^2_\epsilon\Imat)$. The parameters vector $\Thetab_i$ contains the length-scale parameter $\alpha_i$ and the variance $\sigma^2_i$, $i = 1,2,3$. In our implementation we sampled the spectral frequencies associated with $\alpha_i$ once and adjusted them for each step when optimising $\alpha_i$; see the procedure \texttt{PRIOR-FIXED} in \citet{Cutajar_2016}. The parameters in the variatonal distribution over $\wvec_i, i = 1,2,3,$ were found using stochastic gradient descent, while expectations taken with respect to $\wvec_i$ were approximated using Monte Carlo (similar to what we do in Appendix \ref{sec:SDSP}); specifically, we used $N_{MC} = 25$ Monte Carlo samples.

At each layer we let the number of Fourier features (i.e., sine and cosine basis functions) equal 256 and ran stochastic gradient descent for 50,000 iterations. For the first 10,000 iterations we kept the covariance-function parameters fixed, and optimised the variational parameters for $w_i, i = 1,2,3$. For the remaining 40,000 iterations we optimised both the weights and covariance-function parameters simultaneously. Each iteration took on the order of a tenth of a second to complete.

\subsection{DGPsparse}\label{sec:DGPsparse}

The sparse deep Gaussian process model of \citet{Damianou_2013} was fitted using the \texttt{deepGP} package in \texttt{R}, which we provide as supplementary material. We used one hidden layer; specifically, we employed the following hierarchical model,
\begin{align*}
  \Zvec \mid \Yvec &\sim\Gau(\Yvec, \sigma^2_\epsilon\Imat), \\
  \Yvec \mid \Hmat_1 &\sim\Gau(\zerob, \Kmat_2(\Hmat_1)), \\
  \Hmat_1 \mid \Fmat_1 &\sim \Gau(\Fmat_1, \sigma^2_\xi\Imat), \\
  \Fmat_1 &\sim \Gau(\zerob, \Kmat_1(\Smat)),
\end{align*}
where $\Hmat_1$ can be seen as jittered versions of the smoothly warped locations $\Fmat_1$, and $\Kmat_2$ and $\Kmat_1$ are given by \eqref{eq:Kmat2} and \eqref{eq:Kmat1}, respectively. 
Sparsity is introduced into the model through inducing points and variables for both $\Yvec$ and $\Fmat_1$. Denote the inducing points for $\Fmat_1$ as $\bar\Smat_0$, and those for $\Yvec$ as $\bar\Smat_1$, and denote the corresponding inducing variables at these points as $\bar\Fmat_1$ and $\bar\Yvec$, respectively. Then, omitting the dependence on the inputs and the inducing points,
\begin{align*}
  p(\Zvec, \Yvec, \Hmat_1, \Fmat_1, \bar\Yvec, \bar\Fmat_1) &= p(\Zvec \mid \Yvec)p(\Yvec \mid \Hmat_1, \bar\Yvec)p(\bar\Yvec) \times \\
  &~~~~ p(\Hmat_1 \mid \Fmat_1) p(\Fmat_1 \mid \bar \Fmat_1)  p(\bar\Fmat_1).
\end{align*}
Variational Bayes is used to make inference with the deep sparse GP. Specifically, similar to Appendix \ref{sec:SDSP}, the marginal likelihood is lower-bounded, and the variational distribution is constrained to take the form
\begin{align*}
  q(\Yvec, \Hmat_1, \Fmat_1, \bar\Yvec, \bar\Fmat_1) &= p(\Yvec \mid \Hmat_1, \bar\Yvec) q(\Hmat_1)q(\bar\Yvec) \times \\
  & ~~~~p(\Fmat_1 \mid \bar \Fmat_1)q(\bar\Fmat_1).
\end{align*}
The package \texttt{deepGP} implements the approach of \citet{Damianou_2013}, that is, it finds closed-form expressions for $q(\bar\Fmat_1)$ and $q(\bar\Yvec)$ using free-form optimisation, and constrains $q(\Hmat_1)$ using a mean-field approach, so that
$$
q(\Hmat_1) = \prod_{j=1}^N q(h_{1j}) = \prod_{j=1}^N \Gau(m_{1j}, \sigma^2_{1j}).
$$
The $\{m_{1j}\}$ and $\{\sigma^2_{1j}\}$ are variational parameters that need to be optimised concurrently with the inducing-point locations $\bar\Smat_0$ and $\bar\Smat_1$, and the model parameters $\alpha_1, \alpha_2, \sigma^2_1, \sigma^2_2, \sigma^2_\epsilon,$ and $\sigma^2_\xi$.

%% and
%% \begin{align}
%%   \fvec_{i} &\sim \Gau(\zerob, C_{D_{i}}(\cdot); \Thetab_{i}), \quad i = 1,\dots, n, \\
%%   \epsilonb_{i} &\sim \Gau(\zerob, \sigma^2_{\epsilonb_i} \Imat), \quad i = 1,\dots, n.
%% \end{align}

In early attempts to fit the DGPsparse models, we found difficulty optimising the inducing point locations. We hence resorted to fixing these inducing points such that they are equally spaced in the domain on which they lie, much in the same way as our injective warpings are rescaled in the SDSP (see \eqref{eq:scaling}). Such a choice rendered the gradient descent optimisation procedure stable and easy to tune. The number of inducing points in each layer can also affect the predictive performance of DGPsparse. In the first case study, we set the number of inducing points in $\bar\Smat_0$ and $\bar\Smat_1$ to 10 and 3, respectively. In the second case study,  the number of inducing points in $\bar\Smat_0$ and $\bar\Smat_1$ were set to 35 and 25, respectively. These numbers reflect those that gave us the best out-of-sample predictive performance after several attempts.  We optimised the other parameters concurrently for 5000 iterations, with adaption following the 500th iteration (adaption was carried out by halving the learning rate associated with a parameter every time an associated gradient step decreased the lower-bound, rather than increased it).

After the variational and model parameters were estimated, we computed the predictions and prediction variances of the true process at  % Using the predictions of the previous layer, the predictions at one layer can be calculated by the law of iterated expectation and variance. Recursively, the predictions at the top layer can be computed easily.
a set of $N^*$ prediction locations, $\Smat^*_j \equiv (s_j^*: j = 1,\dots,N^*)'$. Let $\Hmat_1^* \equiv (h^*_{1j}: j = 1,\dots,N^*)'$ denote the jittered warped inputs at the prediction locations, and $\Yvec^* \equiv (Y_j^*: j = 1,\dots,N^*)'$ as the process of interest at these locations. The variational prediction distribution for $\Yvec^*$ is given by

\begin{equation}\label{eq:predsparseGP}
p(\Yvec^* \mid \Zmat) = \int p(\Yvec^* \mid \Hmat_1, \Hmat_1^*, \Zvec) p(\Hmat_1^* \mid \Hmat_1) q(\Hmat_1)  \intd \Hmat_1^* \intd \Hmat_1.
\end{equation}

\noindent The first term of the integrand of \eqref{eq:predsparseGP} is Gaussian, and hence the Monte Carlo approximation
$$
p(\Yvec^* \mid \Zmat) \approx \sum_{l=1}^{N_{MC}}p(\Yvec^* \mid \Hmat_1^{(l)}, \Hmat_1^{*^{(l)}}, \Zvec), \quad (\Hmat_1^{(l)}, \Hmat_1^{*^{(l)}})\sim q(\Hmat_1, \Hmat_1^*),
$$
is a Gaussian mixture. Note that $q(\Hmat_1, \Hmat_1^*)\equiv p(\Hmat_1^* \mid \Hmat_1) q(\Hmat_1)$ is also Gaussian, and hence easy to sample from. In our implementation we obtained the approximate variational predictive distribution through $N_{MC} = 100$ Monte Carlo samples.

%% This suggests the following Monte Carlo approximatio

%% Using the posterior mean and posterior variance of $\gvec_1^*$, we can calculate the prediction and prediction variance of $\Yvec^*$ using law of total expectation and variance as follows.
%% \begin{align}
%% E(\Yvec^* | \Yvec, \svec, \svec^*) &= E_{\gvec_1^* | \bmu_1}[E(\Yvec^*| \Yvec, \bmu_1, \gvec_1^*)] \\
%% Var(\Yvec^* | \Yvec, \svec, \svec^*) &= E_{\gvec_1^* | \bmu_1}[Var(\Yvec^*| \Yvec, \bmu_1, \gvec_1^*)] + Var_{\gvec_1^* | \bmu_1}[E(\Yvec^*| \Yvec, \bmu_1, \gvec_1^*)]
%% \end{align}

\subsection{GP}\label{sec:GP}
Standard Gaussian process regression was carried out using the \texttt{Python} package \texttt{GPflow} \citep{Matthews_2017}. We considered zero-mean Gaussian processes with Mat{\'e}rn covariance functions, with smoothness parameters $\nu = 1/2, 3/2$, and $5/2$. For both case studies best results were obtained with $\nu = 3/2$, and hence only results using models with this smoothness parameter are discussed in Section \ref{sec:1D}.  In \texttt{GPflow}, the length-scale and variance parameters in the Mat{\'e}rn covariance function, as well as the noise variance,  are estimated using maximum likelihood, while the prediction and prediction standard errors are obtained using standard Gaussian-process regression equations \citep[e.g.,][Chapter 2]{Rasmussen_2006} with the estimated parameters plugged in. Maximum likelihood estimation with GPs with only a few data points is known to be quick. For both case studies optimisation required under a second to complete on our setup.

\subsection{SDSP-MCMC}\label{sec:SDSP-MCMC}

The SDSP-MCMC model is the same SDSP model used in Section \ref{sec:1D}, but with inference made using MCMC. As in Section~\ref{sec:DGPfull}, since parameters are difficult to estimate in deep compositional models using MCMC,  we fixed the parameters appearing at the top layer and in the observation model to those estimated using variational Bayes. Specifically, the parameters $\sigma^2$, $l$ (appearing in $\taub_{n+1}$), and $\sigma^2_{\epsilon}$ were fixed to $0.442, 0.704$ and $0.010132$, respectively, in the first case study, and to $0.525, 0.312,$ and $0.010146$, respectively, in the second case study.

MCMC was used to determine the posterior distributions over the weights $\wvec_2$ and $\wvec_1$ using \texttt{Stan} \citep{Carpenter_2017}.  Ten thousand samples were generated, and the first 1000 were discarded as burn-in, requiring just over an hour of computation time in total. The predictions and prediction standard errors of $\Yvec^*$ were then obtained through sampling by composition. 

\section{1D simulation experiment on a stationary process}\label{sec:1D-stat-experiment}

In this section we evaluate the SIWGP and SDSP on noisy data generated from a Gaussian process with a stationary Mat{\'e}rn covariance function with smoothness parameter 3/2,
$$
\mathcal{M}(\hvec) \equiv \sigma^2\left(1 + \sqrt{3}\|\hvec\|/\rho\right)\exp\left(-\sqrt{3}\|\hvec\|/\rho\right),
$$
and with $\sigma^2 = 1$ and $\rho = 0.05$. We use the same prediction grid, number of observations, and measurement-error variance as in Section~\ref{sec:1D}. To assess the impact of the warping complexity on the predictions, we computed prediction diagnostics using SIWGPs and SDSPs with an AWU comprising different numbers of sigmoid basis functions (from 10 to 100, in steps of 10). We used 100 bisquare functions to model the process at top layer, $Y(\cdot)$.

In Figure~\ref{fig:stat_setup} we show the true process, the generated data, as well as the predictions when using the true model with parameters estimated using maximum likelihood. In Figure~\ref{fig:stat_diags} we show the RMSPE, MAPE, CRPS, and IS for the SIWGPs and SDSPs as a function of the number of basis functions in the AWU, $r_1$ (dots), and compare these values to those given by the true model (horizontal lines). We see that when $r_1$ is low, both the SIWGP and the SDSP give predictions that are in line with what can be expected from the true model which, we recall, is stationary in this example. As $r_1$ increases, the SIWGP begins to overfit, and this has a deleterious effect on the prediction performance. The SDSP appears to be robust to over-fitting; this is likely due to the use of prior distibutions that promote zero warping, as well as the use of what is in effect a stochastic optimisation algorithm that tends to avoid spurious modes in the lower bound (Appendix~\ref{sec:SDSP}).

\begin{figure}[t!]
  \ifarxiv
  \includegraphics[width=\textwidth]{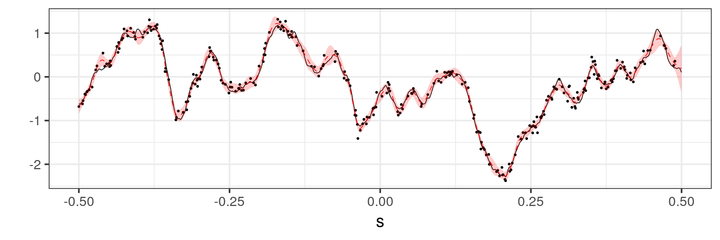}
  \else
  \includegraphics[width=\textwidth]{../deep_spat_src_blinded_for_submission/Supplementary_1Dtest/img/SimStudy.png}
  \fi
  \caption{True process (solid black line) and data (black dots) used in the experiment of Section~\ref{sec:1D-stat-experiment}. The posterior mean (red, dashed line) and the 95\% posterior prediction intervals (red shading) using the true model (a Gaussian process with zero mean and a Mat{\'e}rn 3/2 covariance function), with the parameters estimated using maximum likelihood, are also shown. \label{fig:stat_setup}}
\end{figure}

\begin{figure}[ht!]
  \ifarxiv
  \includegraphics[width=\textwidth]{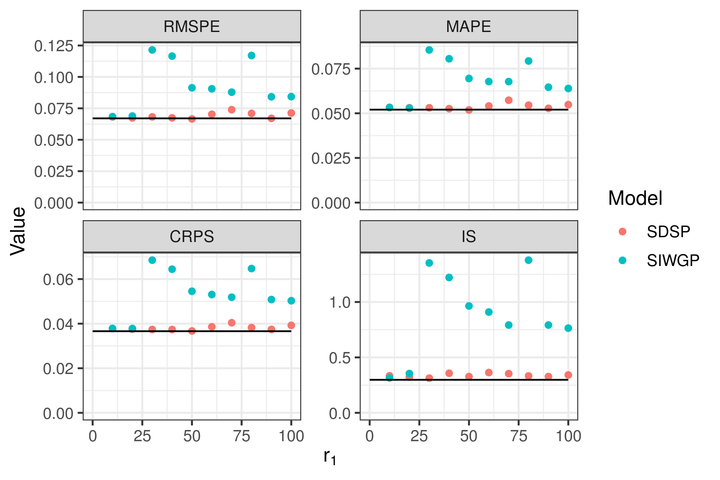}
  \else
  \includegraphics[width=\textwidth]{../deep_spat_src_blinded_for_submission/Supplementary_1Dtest/img/DiagResults.png}
  \fi
  \caption{Predictive diagnostics for the SIWGPs and SDSPs with various values of $r_1$ when fitted to the data shown in Figure~\ref{fig:stat_setup}. Predictive diagnostics obained using the true model (a Gaussian process with zero mean and a Mat{\'e}rn 3/2 covariance function), with the parameters estimated using maximum likelihood, are shown as horizontal black lines.\label{fig:stat_diags}} 
\end{figure}

\section{Minibatch Stochastic Gradient Descent for Large Datasets}\label{sec:SGD}

Although all inverses and log-determinants in \eqref{eq:marglik} are relatively small in size and easy to compute, the matrix multiplication  $\Amat_\Lambda'\Amat_\Lambda$ may become infeasible when the dataset is huge. In such cases, it might be reasonable to instead consider the log-likelihood without $\wvec_{n+1}$ integrated out which, due to conditional independence, then reduces to a sum over data points, that is,
\begin{equation}
    \log p(\Zvec \mid \wvec_{n+1}, \Lambdamat, \sigma^2_\epsilon) = \sum_{j=1}^{N} \log p(Z_j \mid \wvec_{n+1}, \Lambdamat, \sigma^2_\epsilon),\label{eq:sumloglik}
\end{equation}
where
$$
Z_j \mid \wvec_{n+1}, \Lambdamat, \sigma^2_\epsilon \sim \Gau(\avec^{(j)'}_\Lambda\wvec_{n+1}, \sigma^2_\epsilon),
$$
and recall that $\avec^{(j)'}_\Lambda$ denotes the $j$th row of $\Amat_\Lambda$. Now, consider the log-likelihood contribution of a single data point $Z$, where $Z$ is selected uniformly at random from $\Zvec$. Then $P(Z = Z_j) = 1/N$ for $j = 1,\dots,N$, and
\begin{align*}
  E(\log p(Z \mid \wvec_{n+1}, \Lambdamat, \sigma^2_\epsilon)) &= \sum_{j=1}^N\log p(Z_j \mid \wvec_{n+1}, \Lambdamat, \sigma^2_\epsilon) P(Z = Z_j) \\
  &= \frac{1}{N}\sum_{j=1}^N\log p(Z_j \mid \wvec_{n+1}, \Lambdamat, \sigma^2_\epsilon) \\
  &= \frac{1}{N}\log p(\Zvec \mid \wvec_{n+1}, \Lambdamat, \sigma^2_\epsilon),
\end{align*}
from \eqref{eq:sumloglik}. Therefore $N\log p(Z \mid \wvec_{n+1}, \Lambdamat, \sigma^2_\epsilon)$ is an unbiased estimator of $\log p(\Zvec \mid \wvec_{n+1}, \Lambdamat, \sigma^2_\epsilon)$, and $N{\boldsymbol\nabla}\log p(Z \mid \wvec_{n+1}, \Lambdamat, \sigma^2_\epsilon)$ is an unbiased estimator of ${\boldsymbol\nabla}\log p(\Zvec \mid \wvec_{n+1}, \Lambdamat, \sigma^2_\epsilon)$. Using the unbiased estimate $N{\boldsymbol\nabla}\log p(Z_j \mid \wvec_{n+1}, \Lambdamat, \sigma^2_\epsilon)$ (where $Z_j$ is randomly sampled from $\Zvec$) instead of the true gradient based on $\Zvec$ when doing gradient descent results in a stochastic gradient descent algorithm.

Similar arguments apply for when grouping the individual data points into minibatches so that\begin{equation*}
    \log p(\Zvec \mid \wvec_{n+1}, \Lambdamat, \sigma^2_\epsilon) = \sum_{j=1}^{N_b} \log p(\Zvec^m_j \mid \wvec_{n+1}, \Lambdamat, \sigma^2_\epsilon),
\end{equation*}
where $N_b$ is the number of minibatches, $\Zvec^m_j$ is a minibatch of size $m_b \ll N$, and where we have assumed for convenience that $m_b = N / N_b$ is an integer. In this case 
$$
\Zvec^m_j \mid \wvec_{n+1}, \Lambdamat, \sigma^2_\epsilon \sim \Gau(\Amat^{(j)}_\Lambda\wvec_{n+1}, \sigma^2_\epsilon\Imat),
$$
where $\Amat^{(j)}_\Lambda$ contains the rows of $\Amat_\Lambda$ corresponding to the data in $\Zvec^m_j$. An unbiased estimator of $\log p(\Zvec \mid \wvec_{n+1}, \Lambdamat, \sigma^2_\epsilon)$ is then given by $N_b\log p(\Zvec^{m} \mid \wvec_{n+1}, \Lambdamat, \sigma^2_\epsilon)$ where $\Zvec^m$ is a random sample of size $m_b$ from $\Zvec$.

%% In this appendix we provide the implementation of minibatch for SIWGP and SDSP.

%% For SIWGP, because of the assumption of independence, the marginal log-likelihood can be written as a sum of terms from individual data points as follows.
%% \begin{equation}
%%     \log p(\Zvec \mid \wvec_{n+1}, \Lambdamat, \sigma^2_\epsilon) = \sum_{j=1}^{N} \log p(Z(\svec_j) \mid \wvec_{n+1}, \Lambdamat, \sigma^2_\epsilon). \label{eq:SIWGP}
%% \end{equation}
%% since

%% \begin{align}
%%     &\log p(\Zvec \mid \wvec_{n+1}, \Lambdamat, \taub_{n+1}, \sigma^2_\epsilon) \nonumber \\ &= \textrm{const.~} - \frac{N}{2}\log{\sigma^2_\epsilon} - \frac{1}{2\sigma^2_\epsilon}(\Zvec - \Amat_\Lambdau \wvec_{n+1})'(\Zvec - \Amat_\Lambdau \wvec_{n+1}) \nonumber \\
%%     &= \textrm{const.~} - \frac{N}{2}\log{\sigma^2_\epsilon} -\frac{1}{2\sigma^2_\epsilon}\left[ \sum_{i=1}^{N} Z(\svec_i)^2 - 2 \sum_{i=1}^{N} Z(\svec_i) [\phib_{n+1}(\fvec(\svec_i, \Lambdau))]' \wvec_{n+1} + \sum_{i=1}^{N} [[\phib_{n+1}(\fvec(\svec_i, \Lambdau))]' \wvec_{n+1}]^2 \right]
%% \end{align}

Minibatches can also be used when modelling the data using an SDSP, except that now $\wvec_{n+1}$ and $\Fmat_{n+1} \equiv \Yvec$ are not integrated out and instead equipped with a variational distribution $q(\Fmat_{n+1},\wvec_{n+1} \mid \Fmat_n) \equiv p(\Fmat_{n+1} \mid \wvec_{n+1}, \Fmat_n)q(\wvec_{n+1})$ where $p(\Fmat_{n+1} \mid \wvec_{n+1}, \Fmat_n)$ is degenerate and $q(\wvec_{n+1}) = \Gau(\mvec_{n+1}, \Vmat_{n+1}(\etab_{n+1}))$, where $\Vmat_{n+1}$ is constrained to be diagonal. Omitting details, for this model we have that
$$
\mathcal{E}_1 \approx \frac{1}{N_{MC}}\sum_{l=1}^{N_{MC}} \log p(\Zvec \mid \Fmat_{n+1}^{(l)}, \sigma^2_\epsilon),
$$
where $\Fmat_{n+1}^{(l)}$ is sampled akin to \eqref{eq:Fmatnm1}. As shown earlier in this section, it is easy to see that $N_b \log p(\Zvec^m \mid \Fmat_{n+1}^{(l)}, \sigma^2_\epsilon)$ is an unbiased estimator of $\log p(\Zvec \mid \Fmat_{n+1}^{(l)}, \sigma^2_\epsilon)$ and that hence one can write
$$
\mathcal{E}_1 \approx \frac{N_b}{N_{MC}}\sum_{l=1}^{N_{MC}} \log p(\Zvec^m_j \mid \Fmat_{n+1}^{(l)}, \sigma^2_\epsilon),
$$
which can be subsequently used for minibatch stochastic gradient descent.

%% \begin{align}
%%     \log p(\Zvec \mid \wvec_{n+1}, \Thetabu_{n}, \sigma^2_\epsilon) &\geq E_{q(\Fmatu_{n}, \Wmatu_{n}\mid\Thetabu_n)}\left[\log \frac{p(\Zvec \mid \Fmatu_{n}, \Wmatu_{n}, \wvec_{n+1}, \Thetabu_{n}, \sigma^2_\epsilon)p(\Fmatu_{n}, \Wmatu_{n} \mid \Thetabu_{n})}{q(\Fmatu_{n}, \Wmatu_{n}\mid\Thetabu_n)}\right] \nonumber \\
%%     &=E_{q(\Fmat_{n}\mid\Thetabu_n)}\left[\log p(\Zvec \mid \Fmat_{n},  \wvec_{n+1}, \sigma^2_\epsilon)\right] - \sum_{i=1}^n KL(q(\Wmat_{i}) \| p(\Wmat_{i})) \nonumber \\
%%     &=\mathcal{E}_1 - \mathcal{E}_2.
%% \end{align}
%% The term $\mathcal{E}_1$ can be approximated using Monte Carlo as follows.
%% \begin{equation}
%% \mathcal{E}_1 \approx \frac{1}{N_{MC}}\sum_{l=1}^{N_{MC}} \log p(\Zvec \mid \Fmat_{n}^{(l)}, \wvec_{n+1}, \sigma^2_\epsilon). \label{eq:MC}
%% \end{equation}
%% Because of the assumption of independence, equation \eqref{eq:MC} can be re-written as
%% \begin{equation}
%%     \mathcal{E}_1 \approx \frac{1}{N_{MC}}\sum_{l=1}^{N_{MC}}  \sum_{j=1}^{N} \log p(Z(\svec_j) \mid f_{n}^{(l)}(\svec_j), \wvec_{n+1}, \sigma^2_\epsilon). \label{eq:SDSP}
%% \end{equation}
%% As shown in equations \eqref{eq:SIWGP} and \eqref{eq:SDSP}, the marginal log-likelihood can be written in terms of individual data points. This allows us to implement minibatch stochastic gradient descent to optimise the parameters. %Note that now in both models we are optimising with respect to w_{n+1} and not tau_{n+1}.

\ifarxiv
\else
\bibliography{Bibliography}
\fi

\fi

\end{document}